\documentclass[a4paper,fleqn,usenatbib]{mnras}

\usepackage[T1]{fontenc}
\usepackage{ae,aecompl}
\usepackage{graphicx}	
\usepackage{amsmath}	
\usepackage{amssymb}	
\newcommand{\ie}{$i.e.,\;$}
\newcommand{\eg}{$e.g.,\;$}

\title[On the nature of infrared-faint radio sources]{On the nature of infrared-faint radio sources in the SXDF and VLA-VVDS fields}
\author[Singh et al.]{Veeresh Singh$^{1,2}$\thanks{E-mail: veeresh@prl.res.in}, Yogesh Wadadekar$^{3}$, C. H. Ishwara-Chandra$^{3}$,  
\newauthor Sandeep Sirothia$^{3,4,5}$, Jonathan Sievers$^{2}$, Alexandre Beelen$^{6}$ and Alain Omont$^{7}$
\\
$^{1}$Astronomy \& Astrophysics Division, Physical Research Laboratory, Ahmedabad 380009, India \\
$^{2}$Astrophysics and Cosmology Research Unit, School of Chemistry and Physics, University of KwaZulu-Natal, Durban 4041, South Africa \\
$^{3}$National Centre for Radio Astrophysics, TIFR, Post Bag 3, Ganeshkhind, Pune 411007, India\\ 
$^{4}$Square Kilometre Array South Africa, 3rd Floor, The Park, Park Road, 7405, Pinelands, South Africa \\
$^{5}$Department of Physics and Electronics, Rhodes University, PO Box 94, Grahamstown 6140, South Africa \\
$^{6}$Institut d'Astrophysique Spatiale, B$\hat{\rm a}$t. 121, Universit{\'e} Paris-Sud, 91405 Orsay Cedex, France \\
$^{7}$CNRS, UMR 7095, Institut d'Astrophysique de Paris, 75014, Paris, France \\
}

\date{Accepted XXX. Received YYY; in original form ZZZ}

\pubyear{2017}

\begin{document}
\label{firstpage}
\pagerange{\pageref{firstpage}--\pageref{lastpage}}
\maketitle

\begin{abstract}
Infrared-Faint Radio Sources (IFRSs) are an unusual class of objects that are relatively bright at radio wavelengths but
have faint or undetected infrared counterparts even in deep surveys. 
We identify and investigate the nature of IFRSs using deep radio (S$_{\rm 1.4~GHz}$ $\sim$ 100 $\mu$Jy beam$^{-1}$ at 5$\sigma$), 
optical (m$_{\rm r}$ $\sim$ 26 -- 27.7 at 5$\sigma$), and near-IR (S$_{\rm 3.6~{\mu}m}$ $\sim$ 1.3 -- 2.0 $\mu$Jy beam$^{-1}$ at 5$\sigma$) data available 
in two deep fields namely the Subaru X-ray Deep Field (SXDF) and the Very Large Array - VIMOS VLT Deep Survey (VLA-VVDS) field. 
In 1.8 deg$^{2}$ of the two fields we identify a total of nine confirmed and ten candidate IFRSs. 
We find that our IFRSs are high-redshift radio-loud AGN, with 12/19 sources having redshift estimates in the range of $z$ $\sim$ 1.7 -- 4.3, while 
a limit of $z$ $\geq$ 2.0 is placed for the remaining seven sources. 
Notably, our study finds, for the first time, IFRSs with measured redshift $>$ 3.0, and also, 
the redshift estimates for IFRSs in the faintest 3.6 $\mu$m flux regime {\ie}S$_{\rm 3.6~{\mu}m}$ $<$ 1.3 ${\mu}$Jy.
Radio observations show that our IFRSs exhibit both compact unresolved as well as 
extended double-lobe morphologies, and have predominantly steep radio spectra 
between 1.4 GHz and 325 MHz. The non-detection of all but one IFRSs in the X-ray band and the 
optical-to-MIR colour (m$_{\rm r}$ - m$_{\rm 24~{\mu}m}$) suggest that a significant fraction of IFRSs are likely to be hosted 
in dusty obscured galaxies.    
\end{abstract}

\begin{keywords}
galaxies: active -- galaxies: high redshift -- galaxies: radio -- infrared: galaxies 
\end{keywords}



\section{Introduction}
\label{sec:Intro}
In recent years, systematic searches for the optical and infrared counterparts of radio sources in deep fields have yielded a new class of radio sources 
named as Infrared-Faint Radio Sources (IFRSs). As the name suggests, IFRSs constitute radio sources with faint or undetected 
near-IR (NIR) counterparts \citep{Norris06,Norris11,Herzog14}. 
These sources were first discovered by \cite{Norris06} who reported 22 radio sources in 3.7 deg$^{2}$ of the 1.4 GHz 
Australia Telescope Large Area Survey (ATLAS) in the {\it Chandra} Deep Field South (CDFS), with 
no detected 3.6 $\mu$m counterparts in the {\it Spitzer} Wide-area Infrared Extragalactic Survey (SWIRE; \citealt{Lonsdale03}). 
Further, using the ATLAS and SWIRE data \cite{Middelberg08} found 31 IFRSs in 3.6 deg$^{2}$ of the European Large Area IR space observatory Survey
South 1 (ELAIS-S1) field.  
The 1.4 GHz flux density of IFRSs, found in these deep fields, ranges from 
few hundred $\mu$Jy to few tens of mJy, while non-detection in the SWIRE puts an upper flux limit of 5 $\mu$Jy in the 3.6 $\mu$m band. 
Thus, the ratio of radio-to-infrared flux densities for IFRSs ranges from several hundred to several thousands, similar to that for 
High-redshift Radio Galaxies (H{\it z}RGs). However, unlike H{\it}zRGs, the faintness of IFRSs in the optical and IR bands makes 
it difficult to unveil their true nature.
Notably, all 53 IFRSs found in the ATLAS also lacked optical counterparts, thus suggesting them 
to be extreme counterparts of Optically Invisible Radio Sources (OIRSs) identified by \cite{Higdon05}. 
The OIRSs are compact radio sources with no detected optical counterparts up to R-band magnitude of 25.7. 
\cite{Higdon08} reported that a substantial fraction (34 per cent) of OIRSs also lacked counterparts in 3.6 $\mu$m band, 
and suggested them to be possible candidates for H{\it z}RGs ($z$ $>$ 2).  
\par
Given the faintness of optical and IR counterparts of IFRSs, the initial investigations on IFRSs were limited to 
radio wavelengths. For example, many IFRSs have been detected with Very Long Baseline Interferometry (VLBI) that inferred high brightness temperature 
(T$_{\rm B}$) $\sim$ 10$^{6}$ K \citep[see][]{Norris07,Middelberg08,Herzog15a}. 
The IFRSs were also found to exhibit radio spectra steeper ($\alpha$ $<$ -1) than the general radio population \citep{Middelberg11}. 
Several IFRSs also showed polarisation in radio, thus suggesting them to be AGN rather than star-forming galaxies (SFGs) \citep{Banfield11,Middelberg11,Collier14}. 
In an attempt to detect IR counterparts of IFRSs, \cite{Huynh10} used ultra-deep {\it Spitzer} 
imaging in the extended {\it Chandra} Deep Field South (eCDFS) and reported the detection of only two IFRSs having 3.6 $\mu$m fluxes 
5.5$\pm$0.3 $\mu$Jy and  6.6$\pm$0.3 $\mu$Jy. The fainter IFRS of these two also showed an optical counterpart with $V_{\rm AB}$ = 26.27 and 
$z_{\rm AB}$ = 25.62 in the optical Advanced Camera for Surveys (ACS; \citealt{Giavalisco04}) images. 
Using one of the deepest {\it Spitzer} imaging surveys, the {\it Spitzer} Extragalactic Representative Volume Survey (SERVS; \citealt{Mauduit12}) 
with 3$\sigma$ flux limit of 1.5 $\mu$Jy beam$^{-1}$, \cite{Norris11} reported candidate detections of only three ATLAS IFRSs with the possibility of 
being them spurious detections due to noise. The stacking of SERVS image cut-outs at the radio positions of IFRSs resulted in a 
median flux density of $\leq$ 0.2 $\mu$Jy beam$^{-1}$, which, in turn, suggests the extreme ratio of radio-to-3.6 $\mu$m flux densities 
{for IFRSs}. 
\par 	
One of the most important aspects of the study of IFRSs is to estimate their redshifts. 
However, in most cases, it has been difficult to estimate the redshifts of IFRSs due to the faintness of their optical and IR counterparts. 
The lack of redshift causes a major hindrance in understanding their nature, including 
basic properties such as distance, physical size, luminosity, and cosmological epoch. 
We note that \cite{Zinn11} presented a catalogue of 55 IFRSs in four deep fields ({\ie}CDFS, ELAIS-S1, FLS, and COSMOS), but 
without redshift estimates. 
Furthermore, using the Unified Radio Catalogue (URC; \citealt{Kimball08}) and the Wide-Field Infrared Survey Explorer (WISE; \citealt{Wright10}) 
data, \cite{Collier14} presented a catalogue of 1317 IFRSs 
that are suggested to be closer version of IFRSs found in the deep surveys, but only 18/1317 (1.3 per cent) IFRSs have 
spectroscopic redshifts (2 $<$ $z$ $<$ 3) from the Sloan Digital Sky Survey (SDSS). 
In order to obtain redshifts of IFRSs \cite{Herzog14} carried out spectroscopic observations with the Very Large Telescope 
(VLT) for four IFRSs selected by their relatively bright optical counterparts (m$_{\rm r}$ $\sim$ 22.0 -- 24.1 in Vega) 
and found spectroscopic redshifts of only three IFRSs. 
\par
Indeed, despite several attempts, the nature of IFRSs population remains unclear mainly due to their non-detection in the optical 
and IR bands. 
Therefore, obtaining more sensitive optical and IR observations of radio surveyed fields is key to understand the nature of IFRSs.  
Subaru X-ray Deep Field (SXDF) and Very Large Array - VIMOS VLT Deep Survey (VLA-VVDS) fields possess 
deep multiwavelength ({\ie}radio, optical, NIR, mid-IR (MIR), far-IR (FIR) and X-ray) data, and therefore, both the two fields 
are advantageous for investigating the nature of IFRSs. Also, the availability of deep 1.4~GHz (5$\sigma$ $=$ 100 $\mu$Jy beam$^{-1}$) and 3.6 $\mu$m 
(5$\sigma$ $\simeq$ 1.3 -- 2.0 $\mu$Jy beam$^{-1}$) data in both the fields allow us to identify the IFRSs population at fainter flux levels. 
In this paper, we identify and investigate the nature of IFRSs, in the relatively fainter flux regime, in the SXDF and VLA-VVDS field.
\\
Details about the existing multiwavelength data in the two fields are given in Section \ref{sec:Data}. 
The selection criteria and the sample of our IFRSs are described in Section \ref{sec:Identification}. 
In Section \ref{sec:Redshifts}, we present redshift estimates of our IFRSs. 
Section \ref{sec:Comparison} is devoted to a comparison between our sample and previously reported IFRSs samples. 
The radio properties of our IFRSs are discussed in Section \ref{sec:RadioProp}, and the properties of multiwavelength counterparts of IFRSs 
are discussed in Section \ref{sec:MWCounterparts}. 
The results and conclusions of our study are summarised in Section \ref{sec:Summary}.     
The cosmological parameters used in this paper are H$_{0}$ = 71 km s$^{-1}$ Mpc$^{-1}$, 
${\Omega}_{\rm M}$ = 0.27 and ${\Omega}_{\rm {\Lambda}}$ = 0.73.
In the $\Lambda$CDM cosmology, the linear scale spans a limited range between 4 kpc arcsec$^{-1}$ 
to 8.5 kpc arcsec$^{-1}$ in the redshift range of 0.5 $\leq$ ${\it z}$ $\leq$ 12.   
All magnitudes are listed in the AB system unless mentioned otherwise.
\section{Multiwavelength Data}
\label{sec:Data}
Both the SXDF (centred at RA = 02$^{\rm h}$ 18$^{\rm m}$ 00$^{\rm s}$ and DEC = -05$^{\circ}$ 00$^{\prime}$ 00$^{{\prime}{\prime}}$ 
with a total area of 1.3 deg$^{2}$; \citealt{Furusawa08}) and the VLA-VVDS 
(centred at RA = 02$^{\rm h}$ 26$^{\rm m}$ 00$^{\rm s}$ and DEC = -04$^{\circ}$ 30$^{\prime}$ 00$^{{\prime}{\prime}}$ with a total area of 1.0 deg$^{2}$; 
\citealt{LeFevre04}) 
fields are defined on the basis of area covered by the deep optical surveys from Subaru and VLT, respectively. 
Both the optically-surveyed fields have also been fully or partially surveyed at other wavelengths ranging from radio, IR to X-ray.
The multiwavelength data available in the two fields are listed in Table~\ref{table:MWData}, and the footprints 
of these multiwavelength surveys are shown in 
Fig.~\ref{fig:Footprints}. 
Since the identification of IFRSs is based on the search for 3.6 $\mu$m counterparts of 1.4 GHz radio sources (see Section~\ref{sec:Identification}), we 
begin with 1.4 GHz VLA radio source catalogues that cover 0.8/1.3 deg$^{2}$ in the central region of the SXDF and full 1.0 deg$^{2}$ in the VLA-VVDS. 
The search for our IFRSs is further limited to the common area (0.8 deg$^{2}$ in the SXDF and 0.96 deg$^{2}$ in the VLA-VVDS) 
covered by both the radio and 3.6 $\mu$m surveys. 
The multiwavelength counterparts of IFRSs are searched from the deepest available data in the overlapped regions. 
\\
In the following sub-sections we provide brief details about the available multiwavelength data. 
The footprints of some surveys ({\eg}1.4 GHz, 610 MHz, X-ray) are matched with the optical coverage, 
and therefore, in such cases, we do not provide the survey co-ordinates repeatedly.
\subsection{Radio data}
\subsubsection{1.4 GHz VLA surveys}
{\bf SXDF} : Full 1.3 deg$^{2}$ area of the SXDF was surveyed at 1.4 GHz using VLA `BnC' configuration \citep{Simpson06}. 
The mosaiced 1.4 GHz radio map, made from 14 overlapping pointings arranged in a hexagonal pattern, has nearly uniform rms noise 
of 20 $\mu$Jy beam$^{-1}$ in the central 0.8 deg$^{2}$ while it increases in the peripheral regions. 
Based on these data, \cite{Simpson06} presented a catalogue of 512 radio sources with a detection limit of 5$\sigma$ = 100 $\mu$Jy beam$^{-1}$ 
in the central 0.8 deg$^{2}$. 
The VLA radio map has a synthesized beam-size of 5$\arcsec$.0 $\times$ 4$\arcsec$.0 with position angle (PA) 170$^{\circ}$.
 \\
{\bf VLA-VVDS} : \cite{Bondi03} carried out 1.4 GHz VLA `B' survey of full 1.0 deg$^{2}$ of the VLA-VVDS field. 
Their nine pointing mosaiced radio map of the full field has nearly uniform rms noise of 17 $\mu$Jy beam$^{-1}$ 
and a resolution of 6$\arcsec$.0.
The survey resulted in a total of 1054 radio sources detected above the 5$\sigma$ limit.  

\subsubsection{610 MHz GMRT survey}
The VLA-VVDS field has also been surveyed at 610 MHz with the Giant Metrewave Radio Telescope (GMRT), 
although there are no deep 610 MHz observations in the SXDF. 
The 610 MHz GMRT survey of the VLA-VVDS covers full 1.0 deg$^{2}$ with a sensitivity limit  
of 250 $\mu$Jy beam$^{-1}$ at the 5$\sigma$ level and a resolution of 6$\arcsec$.0 \citep{Bondi07}. 
There are a total of 514 radio sources detected at $\geq$ 5$\sigma$ in 1.0 deg$^{2}$. 
\subsubsection{325 MHz GMRT survey}
Both the SXDF and VLA-VVDS fields also lie within the 325 MHz GMRT survey 
(centred at RA = 02$^{\rm h}$ 21$^{\rm m}$ 00$^{\rm s}$ and DEC = -04$^{\circ}$ 30$^{\prime}$ 00$^{{\prime}{\prime}}$) 
of 12 deg$^{2}$ covered with 16 pointings, in the XMM-LSS field (Wadadekar et al., in preparation). 
The 325 MHz GMRT mosaiced image, with synthesized beam-size of 9$\arcsec$.4 $\times$ 7$\arcsec$.4, has an average rms noise 
of 150 $\mu$Jy beam$^{-1}$ and it reaches down to 120 $\mu$Jy beam$^{-1}$ in the central region. 
The 325 MHz GMRT survey is the deepest low-frequency radio survey in the XMM-LSS field, 
and detects a total of 3304 radio sources at $\geq$ 5$\sigma$ with a noise cut-off $\leq$ 300 $\mu$Jy. 
We have already used these 325 MHz GMRT data in combination with the existing multiwavelength data to study 
ultra steep spectrum radio sources \citep{Singh14}, the radio-FIR correlation for distant SFGs \citep{Basu15}, and 
a giant relic radio galaxy at $z\sim 1.3$ \citep{Tamhane15}.

\begin{figure}
\includegraphics[angle=0,width=8.5cm,trim={2.0cm 1.8cm 3.0cm 0.6cm},clip]{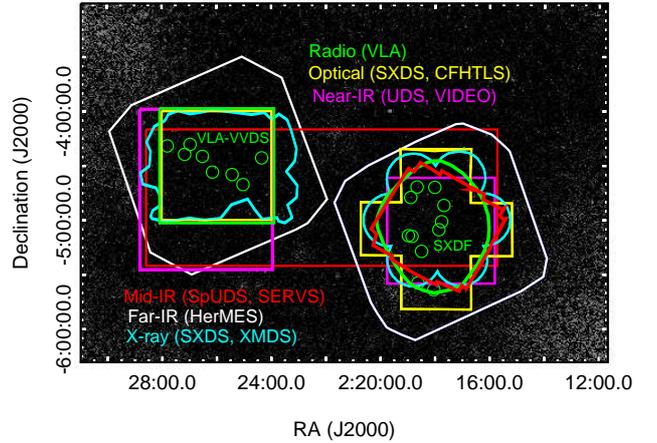}
\caption{The footprints of deep multiwavelength surveys ({\ie} 1.4 GHz radio in green, optical in yellow, NIR in magenta, MIR in red, FIR in white 
and X-ray in cyan) in the SXDF (right side) and VLA-VVDS (left side) over-plotted on to a part of 325 MHz GMRT image. 
The footprints of large area surveys ({\ie}SWIRE and 325 MHz GMRT) that cover both the deep fields are not shown here. 
The positions of our IFRSs are indicated by green circles.}
\label{fig:Footprints} 
\end{figure}

\subsection{Optical data}
{\bf SXDF} : The optical data are obtained from the Subaru {\it XMM-Newton} Deep Survey (SXDS) which is 
carried out with the Suprime-Cam on the Subaru telescope \citep{Furusawa08}. 
The SXDS provides photometric data in $B$, $V$, $Rc$, $i^{\prime}$, and $z^{\prime}$ bands to {the 5$\sigma$} depths of 28.4, 27.8, 
27.7, 27.7, and 26.6 mag, respectively. 
Using the SXDS data \cite{Simpson06} present the optical identifications of 1.4 GHz radio sources. 
\\
{\bf VLA-VVDS} : Full 1.0 deg$^{2}$ of the VLA-VVDS was covered by the VLT VIMOS Deep Survey (VVDS) and
the Canada-France-Hawaii Telescope Legacy Survey D1 field (CFHTLS D1). 
The CFHTLS D1 provides photometric observations in $u^{\star}$, $g^{\prime}$, $r^{\prime}$, $i^{\prime}$, and 
$z^{\prime}$ bands to the 5$\sigma$ depths of 26.5, 26.4, 26.1, 25.9 and 25.0 mag, respectively \citep{Ilbert06}. 
The VVDS carried out with the CFH12K wide-field mosaic camera yields optical photometric data in B, V, R and I bands to the 5$\sigma$ limiting magnitudes 
(50 per cent completeness for point sources) of $B = 26.5$, $V = 26.2$, $R = 25.9$ 
and $I = 25.0$ \citep[see][]{McCracken03}. 
The 5$\sigma$ limiting magnitudes at 90 per cent completeness are nearly one magnitude brighter ($I = 24.0$).
\subsection{NIR data}
{\bf SXDF} : 0.63 deg$^{2}$ of the SXDF is covered by the Ultra Deep Survey (UDS) 
from the UKIRT Infrared Deep Sky Survey (UKIDSS; \citealt{Lawrence07}). 
The UDS is centred at RA = 02$^{\rm h}$ 17$^{\rm m}$ 48$^{\rm s}$ and DEC = -05$^{\circ}$ 06$\arcmin$ 00$\arcsec$ (J2000) and covers 
a total area of 0.77 deg$^{2}$. 
The magnitude limits of the UDS DR 11 at 5$\sigma$ level are $J = 25.6$, $H = 25.1$, and $K = 25.3$ \\
{\bf VLA-VVDS} : The NIR VISTA Deep Extragalactic Observations (VIDEO; \citealt{Jarvis13}) survey covers full 1.0 deg$^{2}$ of the VLA-VVDS field. 
In the XMM-LSS field, the VIDEO survey covers a total area of 4.5 deg$^{2}$ that is subdivided into three tiles each covering 1.5 deg$^{2}$. 
The VIDEO survey provides photometric observations in $Z, Y, J, H,$ and $Ks$ bands 
with 5$\sigma$ magnitude limits of 25.7, 24.5, 24.4, 24.1, and 23.8, respectively. 
\begin{table*}
\hspace{-1.0cm}
\begin{minipage}{140mm}
\caption{ Summary of the multiwavelength data}
\scalebox{0.85}{
\begin{tabular}{@{}cccccccccccccc@{}}
\hline
              & \multicolumn{4}{c}{SXDF}   &  \multicolumn{4}{c}{VLA-VVDS}   \\  \cline{2-5} \cline{6-9}
   Band       & Survey             & 5$\sigma$ Depth  &  Area        & Reference  & Survey    & 5$\sigma$ Depth &  Area        & Reference  \\
              &                    & (unit)       & (deg$^{2}$)  &            &           & (unit)       & (deg$^{2}$)  &            \\  \hline
{\it Radio}   &                    & ($\mu$Jy beam$^{-1}$) &           &            &           &  ($\mu$Jy beam$^{-1}$) &              &            \\         
 1.4 GHz      & VLA                &   100        &  0.8         & 1          & VLA        &    80       & 1.0          &  2          \\  
 610 MHz      &                    &              &              &            & GMRT      &     250      & 1.0          &  3           \\   
 325 MHz      &  GMRT              &   750        & 0.8 (12.0)   & 4          & GMRT      &     750      & 1.0 (12.0)   &  4          \\   
{\it Optical} &                    & (mag)        &              &            &           & (mag)        &              &             \\
    B         &      SXDS          & 28.4         &  0.8         & 5          & VVDS      & 26.5         &  1.0         &  6          \\
    V         &      SXDS          & 27.8         &  0.8         & 5          & VVDS      & 26.2         &  1.0         &  6          \\
    R         &      SXDS          & 27.7         &  0.8         & 5          & VVDS      & 25.9         &  1.0         &  6          \\   
    I         &                    &              &              &            & VVDS      & 25.0         &  1.0         &  6          \\         
 u$^{\star}$  &                    &              &              &            & CFHTLS D1 & 26.5         &  1.0         &  7          \\         
 g$^{\prime}$ &                    &              &              &            & CFHTLS D1 & 26.4         &  1.0         &  7          \\         
 r$^{\prime}$ &                    &              &              &            & CFHTLS D1 & 26.1         &  1.0         &  7          \\         
 i$^{\prime}$ &      SXDS          & 27.7         & 0.8          & 5          & CFHTLS D1 & 25.9         &  1.0         &  7          \\         
 z$^{\prime}$ &      SXDS          & 26.6         & 0.8          & 5          & CFHTLS D1 & 25.0         &  1.0         &  7          \\         
{\it NIR} &                    &              &              &            &           &              &              &             \\         
     Z        &                    &              &              &            & VIDEO     & 25.7         &  1.0 (4.5)   &  9          \\         
     Y        &                    &              &              &            & VIDEO     & 24.5         &  1.0 (4.5)   &  9          \\         
     J        & UDS                &      25.6    &   0.6 (0.8)  & 8          & VIDEO     & 24.4         &  1.0 (4.5)   &  9          \\         
     H        & UDS                &      25.1    &   0.6 (0.8)  & 8          & VIDEO     & 24.1         &  1.0 (4.5)   &  9          \\         
     K        & UDS                &      25.3    &   0.6 (0.8)  & 8          & VIDEO     & 23.8         &  1.0 (4.5)   &  9          \\         
{\it MIR}  &                    &              &              &            &           &              &              &             \\         
 3.6 $\mu$m   & SpUDS              & 24.0         &  0.6 (1.0)   & 10         &           &              &              &              \\  
              & SERVS              & 23.1         &  0.7 (4.5)   & 11         & SERVS     & 23.1         & 0.8 (4.5)    &  11           \\
              &                    & ($\mu$Jy beam$^{-1}$) &              &            &           & ($\mu$Jy beam$^{-1}$) &              &             \\         
              & SWIRE              & 7.3          &  0.8 (9.0)   & 12         & SWIRE     &  7.3         & 0.9 (9.0)    &  12          \\
              &                    & (mag)        &              &            &           & (mag)        &              &             \\         
 4.5 $\mu$m   & SpUDS              & 23.9         &  0.6 (1.0)   & 10         & SERVS     & 23.1         &  0.8 (4.5)   &  11          \\         
              &                    & ($\mu$Jy beam$^{-1}$) &              &            &           & ($\mu$Jy beam$^{-1}$) &              &             \\         
 5.8 $\mu$m   & SpUDS              &   8.5        &  0.6 (1.0)   & 10         & SWIRE     &   27.5       &  0.9 (8.7)   &  12          \\         
 8.0 $\mu$m   & SpUDS              &  10.5        &  0.6 (1.0)   & 10         & SWIRE     &   32.5       &  0.9 (8.7)   &  12          \\    
{\it FIR}  &                    & (mJy)        &              &            &           & (mJy)        &              &             \\         
 24 $\mu$m    & SpUDS              & 0.1          &  0.6 (1.0)   & 10         & SWIRE     &  0.45        &  0.9 (9.0)   &  12         \\         
 70 $\mu$m    &                    &              &              &            &           &  2.75        &  0.9 (9.0)   &  12          \\         
 110 $\mu$m   & HerMES             & 11.2         &  0.8 (4.2)   &  13        & HerMES    & 28.8         &  1.0 (6.2)   &  13         \\         
 160 $\mu$m   & HerMES             & 21.4         &  0.8 (4.2)   &  13        & HerMES    & 54.9         &  1.0 (6.2)   &  13         \\         
 250 $\mu$m   & HerMES             & 11.2         &  0.8 (4.2)   &  13        & HerMES    & 11.2         &  1.0 (6.2)   &  13         \\         
 350 $\mu$m   & HerMES             & 9.3          &  0.8 (4.2)   &  13        & HerMES    & 9.3          &  1.0 (6.2)   &  13         \\         
 500 $\mu$m   & HerMES             & 13.4         &  0.8 (4.2)   &  13        & HerMES    & 13.4         &  1.0 (6.2)   &  13         \\  
 {\it X-ray}  &       & (ergs cm$^{-2}$ s$^{-1}$) &              &            &           & (ergs cm$^{-2}$ s$^{-1}$) &   &            \\     
0.5~--~2.0 keV  & SXDS             & 6 $\times$ 10$^{-16}$ & 0.8 (1.1) & 14  & XMDS      & 1 $\times$ 10$^{-15}$ & 1.0 (3.0) & 15     \\   
2.0~--~10 keV   & SXDS             & 3 $\times$ 10$^{-15}$ & 0.8 (1.1) & 14  & XMDS      & 7 $\times$ 10$^{-15}$ & 1.0 (3.0) & 15     \\     
 \hline     
\end{tabular}}
\label{table:MWData} 
\\
Notes - The total area of surveys having coverages outside the deep fields are given within brackets. 
The values for area and sensitivity of different surveys are rounded-off to the first  decimal place. 
Radio surveys from VLA and GMRT have no specific names. 
The optical surveys SXDS, VVDS and CFHTLS-D1 are carried out with the Suprime-Cam/Subaru, CFH12K/VLT and MegaCam/CFHT, respectively.
The UDS and VIDEO NIR surveys are based on the UKIRT Wide Field Camera (WFCAM) and VISTA IR Camera (VIRCAM), respectively.
Spitzer surveys (SpUDS, SERVS, SWIRE) in MIR (3.6 $\mu$m, 4.5 $\mu$m, 5.0 $\mu$m and 8.0 $\mu$m) and FIR (24 $\mu$m and 70 $\mu$m) 
bands are from IRAC and MIPS, respectively. 
In HerMES, 110 $\mu$m and 160 $\mu$m observations are from PACS, while 250 $\mu$m, 350 $\mu$m and 500 $\mu$m observations are from SPIRE. In X-ray, SXDS and XMDS surveys 
are carried out with the {\it XMM-Newton}.
References, 1 : \cite{Simpson06}; 2 : \cite{Bondi03}; 3 : \cite{Bondi07}; 4 : Wadadekar et al. (in preparation); 5 : \cite{Furusawa08}; 6 : \cite{Ciliegi05}; 
7 : \cite{Ilbert06}; 8 : \cite{Lawrence07}; 9 : \citep{Jarvis13}; 10 : \citep{Dunlop07}; 11 : \cite{Mauduit12}; 12 : \cite{Lonsdale03}
13 : \cite{Oliver12}; 14 : \cite{Ueda08}; 
15 : \cite{Chiappetti05}. \\
\end{minipage}
\end{table*}

\subsection{MIR data}
{\bf SXDF} : The MIR data in the SXDF are available from the {\it Spitzer} Public Legacy survey of the UKIDSS Ultra Deep Survey 
(SpUDS; \citealt{Dunlop07}), which was carried out with all four IRAC bands (3.6 $\mu$m, 4.5 $\mu$m, 5.8 $\mu$m and 8.0 $\mu$m), and one 
MIPS band (24 $\mu$m). 
The SpUDS covers full 0.8 deg$^{2}$ of the UDS and 0.6 deg$^{2}$ of the SXDF with 5$\sigma$ detection limits of 
[3.6] = 24.0 (1.3 $\mu$Jy beam$^{-1}$) and [4.5] = 23.9 (1.7 $\mu$Jy beam$^{-1}$). \\
{\bf VLA-VVDS} : Nearly 0.82 deg$^{2}$ out of the total 1.0 deg$^{2}$ of the VLA-VVDS field is covered by 
the {\it Spitzer} Extragalactic Representative Volume Survey (SERVS; \citealt{Mauduit12}) which is 
a medium deep survey at 3.6 $\mu$m and 4.5 $\mu$m with a 5$\sigma$ sensitivity limit of 23.1 mag (2.0 $\mu$Jy beam$^{-1}$) in both the bands. 
The SERVS covers a total area of 4.5 deg$^{2}$ in the XMM-LSS field 
and is centred at RA = 02$^{\rm h}$ 20$^{\rm m}$ 00$^{\rm s}$ DEC = -04$^{\circ}$ 48$\arcmin$ 00$\arcsec$ (J2000). 
\\
SERVS also covers 0.6/0.8 deg$^{2}$ of the radio coverage in the SXDF.  
Furthermore, both SXDF and VLA-VVDS have substantial overlap with the SWIRE \citep{Lonsdale03}) 
which is centred at RA = 02$^{\rm h}$ 21$^{\rm m}$ 00$^{\rm s}$ DEC = -04$^{\circ}$ 30$\arcmin$ 00$\arcsec$ (J2000) and 
covers a total of 8.7 deg$^{2}$ in all four IRAC bands (3.6 $\mu$m, 4.5 $\mu$m, 5.8 $\mu$m and 8.0 $\mu$m) and 9.0 deg$^{2}$ in MIPS bands (24 $\mu$m, 70 $\mu$m 
and 160 $\mu$m), 
with a 5$\sigma$ sensitivity of 7.3 $\mu$Jy beam$^{-1}$, 9.7 $\mu$Jy beam$^{-1}$, 27.5 $\mu$Jy beam$^{-1}$, and 32.5 $\mu$Jy beam$^{-1}$ in the four IRAC bands, respectively, 
and 0.45 mJy beam$^{-1}$, 2.75 mJy beam$^{-1}$ and 17.5 mJy beam$^{-1}$ in three MIPS bands, respectively.  
\subsection{FIR data}
Both the SXDF and VLA-VVDS fields are fully covered by the {\it Herschel}/SPIRE surveys that are 
part of the Level-4 {\it Herschel} Multi-tiered Extragalactic Survey (HerMES; \citealt{Oliver12}). 
The level-4 HerMES observations carried out at 250 $\mu$m, 350 $\mu$m and 500 $\mu$m bands with the 5$\sigma$ sensitivity limit 
of 11.2 mJy beam$^{-1}$, 9.3 mJy beam$^{-1}$ and 13.4 mJy beam$^{-1}$, respectively. 
These {\it Herschel} surveys are centred at the optically-surveyed fields and cover area of 2.0 deg$^{2}$ in each field.
\subsection{X-ray data}
{\bf SXDF} : The Subaru {\it XMM-Newton} Deep Survey (SXDS) covers a total area of 1.14 deg$^{2}$ centred at RA = 02$^{\rm h}$ 18$^{\rm m}$ 00$^{\rm m}$ 
and DEC = -05$^{\circ}$ 00$\arcmin$ 00$\arcsec$ \citep{Ueda08}. 
It consists of one central 30$\arcmin$ diameter field with a 100 ks exposure time and six flanking fields with 50 ks exposure time. 
The SXDS presents a catalogue of 866, and 645 X-ray sources with sensitivity limits of 6.0 $\times$ 10$^{-16}$ ergs cm$^{-2}$ s$^{-1}$, 
and 3.0 $\times$ 10$^{-15}$ ergs cm$^{-2}$ s$^{-1}$ in the 0.5 -- 2.0 keV and 2.0 -- 10 keV bands, respectively, 
with detection likelihood of $\geq$ 7 (corresponding to a confidence level of 99.9 per cent). \\
{\bf VLA-VVDS} : The {\it XMM} Medium Deep Survey (XMDS) covers the central region of the XMM-LSS field and provides 
a catalogue of 286 sources detected at 4$\sigma$ in 1 deg$^{2}$ area of the VLA-VVDS field \citep{Chiappetti05}. 
The XMDS covers a contiguous area of about 3 deg$^{2}$ over 19 pointings (typical exposure time of $\sim$ 20 -- 25 ks) and reaches down to 
 flux limits of 1.0 $\times$ 10$^{-15}$ erg cm$^{-2}$ s$^{-1}$ in the 0.5 -- 2.0 keV band and 
7.0 $\times$ 10$^{-15}$ erg cm$^{-2}$ s$^{-1}$ in the 2.0 -- 10 keV band at 4$\sigma$ level. 
\section{Identification of IFRSs}
\label{sec:Identification}
\subsection{Selection criteria}
In order to identify IFRSs we adopt the selection criteria proposed by \cite{Zinn11} and select a radio source as an IFRS 
if the following two conditions are satisfied: (i) $\frac{\rm S_{1.4~GHz}}{\rm S_{3.6~{\mu}m}}$ $>$ 500; 
and (ii) S$_{3.6~{\mu}m}$ $<$ 30 $\mu$Jy (including a non-detection). 
The first criterion allows us to mitigate the contamination from foreground stars and SFGs owing to the well known 
radio-IR correlation \citep{Appleton04}, whereas, 
the second criterion helps us to exclude low-redshift radio sources that tend to be relatively bright in IR. 
For instance, Cygnus A ($\frac{\rm S_{1.4~GHz}}{\rm S_{3.6~{\mu}m}}$ $=$ 2 $\times$ 10$^{5}$), a nearby radio galaxy, 
would be selected as an IFRS based on the first criterion, while it would be excluded based on the second criterion. 
It is worth mentioning that the limiting values used in the IFRSs selection criteria are arbitrary but 
select almost all the known IFRSs ranging from faint to bright ones \citep[see][]{Zinn11,Collier14}. 
\\
For radio sources with undetected 3.6~$\mu$m counterparts, we have only a lower limit on the flux ratio of 1.4~GHz-to-3.6~$\mu$m 
($\frac{\rm S_{1.4~GHz}}{\rm S_{3.6~{\mu}m}}$), and such sources can be potential candidates for IFRSs. 
The non-detection of 3.6 $\mu$m counterparts in the SpUDS, SERVS and SWIRE surveys gives S$_{3.6~{\mu}m}$ $<$ 30 $\mu$Jy (see Table~\ref{table:MWData}). 
Therefore, we include candidate IFRSs in our sample and define them as : (i) radio sources with no detected 3.6~$\mu$m counterparts and 
(ii) the lower limit on $\frac{\rm S_{1.4~GHz}}{\rm S_{3.6~{\mu}m}}$ is less than 500. 
A radio source with an undetected 3.6~$\mu$m counterpart but having limiting value of the flux ratio 
of 1.4~GHz-to-3.6~$\mu$m ($\frac{\rm S_{1.4~GHz}}{\rm S_{3.6~{\mu}m}}$) $>$ 500 becomes a confirmed IFRS. 
\\ 
We note that to ensure the reliability of our IFRSs we have excluded faint radio sources with SNR $<$ 8 and 
also discarded multicomponent extended radio sources with the ambiguous centroid position. We thereby ensure that a spurious radio source is not identified as an IFRS.

\subsection{The sample}

With the selection criteria outlined above we find six confirmed and
five candidate IFRSs in the SXDF field, and three confirmed and five
candidate IFRSs in the VLA-VVDS field. In total, we find nine
confirmed and ten candidate IFRSs in the two fields. All our IFRSs and
their basic parameters are listed in Table~\ref{table:IFRSSample}.
The detailed procedure used for the identification of our IFRSs is
given below.  To find 3.6 $\mu$m counterparts of 1.4 GHz radio sources
we give first priority to the deepest available 3.6 $\mu$m survey
({\ie}SpUDS, SERVS and SWIRE in decreasing order of
priority).

\subsubsection{Identification of IFRSs in the SXDF}
To identify IFRSs in the SXDF we first search for 3.6 $\mu$m counterparts of 1.4 GHz radio sources by cross-matching the 3.6 $\mu$m and 1.4 GHz source catalogues. 
There are a total of 512 1.4 GHz radio sources over 0.8 deg$^{2}$, of which 0.63 deg$^{2}$ is covered by the 3.6 $\mu$m SpUDS survey 
that contains only 438 radio sources (see Fig.~\ref{fig:Footprints}). 
We cross-matched 1.4 GHz radio and 3.6 $\mu$m IR sources using a search radius of 5$\arcsec$.0 around the radio positions. 
This exercise yielded 428/438 cross-matched sources, 
where a majority of sources (401/428 $\sim$ 93.7 per cent) were matched within 1$\arcsec$.0. 
Sources with larger positional offset ($>$ 1$\arcsec$.0) are generally fainter in radio as well as in 3.6 $\mu$m and 
seem to suffer from larger positional uncertainties. 
The chosen search radius of 5$\arcsec$.0 corresponds to a projected linear distance of $\leq$ 45 kpc at a redshift ({\it z}) $>$ 0.5, 
which is the typical size of a galaxy. 
We estimate the flux ratio of 1.4~GHz-to-3.6 $\mu$m for all the cross-matched sources, 
and find that only five out of 428 radio sources satisfy our IFRS selection criteria ({\ie}$\frac{\rm S_{1.4~GHz}}{\rm S_{3.6~{\mu}m}}$ $>$ 500, and S$_{3.6~{\mu}m}$ $<$ 30 $\mu$Jy). 
\\
For the 10 radio sources with no cross-matched 3.6 $\mu$m counterparts, we 
visually inspected the 3.6 $\mu$m SpUDS image cut-outs centred at their radio positions, and 
found that five radio sources are falling either on or close to bright extended 
foreground IR sources, which makes the identification of true 3.6 $\mu$m counterparts difficult. 
This leaves us with only five radio sources with undetected 3.6~$\mu$m counterparts (S$_{3.6~{\mu}m}$ $<$ 1.3 $\mu$Jy). 
With the estimated lower limits on the ratio of 1.4~GHz-to-3.6 ${\mu}$m one of the five radio sources turns out to be 
a confirmed IFRS ($\frac{\rm S_{1.4~GHz}}{\rm S_{3.6~{\mu}m}}$ $>$ 537), while remaining four radio sources 
having $\frac{\rm S_{1.4~GHz}}{\rm S_{3.6~{\mu}m}}$ $\simeq$ 84 -- 453, can be considered candidate IFRSs (see Table~\ref{table:IFRSSample}).  
Thus, within the SpUDS coverage we find only six confirmed IFRSs and four candidate IFRSs. 
We note that among 74/512 radio sources falling outside the SpUDS coverage, 28/74 radio sources are covered within the SERVS footprint, 
while remaining 46/74 radio sources are covered with the SWIRE. 
All 28 radio sources falling within the SERVS coverage area have 3.6 $\mu$m counterparts, 
but none of these sources have flux ratio of 
1.4 GHz-to-3.6 $\mu$m ($\frac{\rm S_{1.4~GHz}}{\rm S_{3.6~{\mu}m}}$) $>$ 500. 
Among 46 radio sources covered with the SWIRE, 43/46 radio sources have 3.6 $\mu$m counterparts, 
but none of these satisfies the IFRSs selection criteria.
Two of the three radio sources with no 3.6 $\mu$m counterparts are fairly faint in radio 
({\ie}SNR $\simeq$ 5$\sigma$) and fall below our cut-off limit of SNR $\geq$ 8. 
And, only one among three radio sources meets the criteria of candidate IFRSs with 
S$_{\rm 3.6~{\mu}m}$ $<$ 7.3 $\mu$Jy and $\frac{\rm S_{1.4~GHz}}{\rm S_{3.6~{\mu}m}}$ $>$ 216. 
Therefore, using SpUDS, SERVS and SWIRE data we find, in total, only six confirmed IFRSs and five candidate IFRSs in the SXDF (see Table~\ref{table:IFRSSample}). 
\subsubsection{Identification of IFRSs in the VLA-VVDS}

The VLA-VVDS field contains a total of 1054 1.4 GHz radio sources in 1.0 deg$^{2}$ area, of which only 0.82 deg$^{2}$ containing 865/1054 radio sources overlaps with the SERVS. 
The cross-matching of 1.4 GHz and 3.6 $\mu$m source catalogues using a search radius of 5$\arcsec$.0 around the radio position yielded 
3.6 $\mu$m counterparts for only 789/865 (91.2 per cent) radio sources.  
We find that none of these 789 cross-matched radio sources meets the IFRS selection criteria 
({\ie}$\frac{\rm S_{1.4~GHz}}{\rm S_{3.6~{\mu}m}}$ $>$ 500, and S$_{\rm 3.6~{\mu}m}$ $\leq$ 30 ${\mu}$Jy). 
For the remaining 76/865 radio sources with no cross-matched counterparts we visually inspected 
the 3.6~$\mu$m SERVS image cut-outs centred at the radio positions. 
We excluded all the radio sources that lie on or close to bright extended foreground IR sources which do not allow us to find the true 3.6~$\mu$m counterparts. 
We also excluded faint radio sources with SNR $<$ 8, and multicomponent extended radio sources with ambiguous centroid position. 
This exercise yielded only eight radio sources with no detected 3.6 $\mu$m counterparts. 
Three among these eight radio sources only three are considered confirmed IFRSs with the limiting flux ratio $\frac{\rm S_{1.4~GHz}}{\rm S_{3.6~{\mu}m}}$ $>$ 500, 
while remaining five radio sources can only be considered as candidate IFRSs (see Table~\ref{table:IFRSSample}). 
One among three confirmed IFRSs has 3.6 $\mu$m counterpart just below 5$\sigma$ with S$_{\rm 3.6~{\mu}m}$ $\sim$ 1.8$\pm$0.3 $\mu$Jy.
142 out of 189/1054 radio sources that are not covered by the SERVS fall within the SWIRE coverage, while remaining 
47/189 radio sources are not covered by any 3.6 $\mu$m survey. 
Cross-matching of 142 radio sources with 3.6 $\mu$m SWIRE catalogue yielded 3.6 $\mu$m counterparts for 130 radio sources, but none of these satisfies 
our IFRS selection criteria. 8 out of 12/142 radio sources with no [3.6] counterparts are faint and fall below our SNR cut-off limit (SNR $\geq$ 8). 
Visual inspection of the 3.6 $\mu$m SWIRE image cut-outs shows that the remaining four sources have counterparts at offset $>$ 5.0 arcsec. 
The larger positional offset may be due to the fact that these radio sources are extended. 
Considering the nearest 3.6 $\mu$m counterparts, none of these radio sources meets our IFRSs selection criteria.
Therefore, using SERVS and SWIRE data we find, in total, only three confirmed IFRSs and five candidate IFRSs in the VLA-VVDS field. 
We caution that our IFRSs sample can suffer from incompleteness, in particular, at fainter flux densities.
\begin{table*}
\hspace{-2.5cm}
\begin{minipage}{140mm}
\caption{The IFRSs sample}
\scalebox{0.8}{
\begin{tabular}{@{}ccccccccccc@{}}
\hline
  RA          &   DEC          & S$_{\rm 1.4~GHz}$ & radio size                           & S$_{3.6~{\mu}m}$           & $\frac{\rm S_{1.4~GHz}}{\rm S_{3.6~{\mu}m}}$   & S$_{\rm 610~MHz}$  & S$_{\rm 325~MHz}$ & ${\alpha}_{\rm 325~MHz}^{\rm 1.4~GHz}$ & Redshift & L$_{\rm 1.4~GHz}$ \\
  (h m s)     &   (d m s)      &  (mJy)                &    (arcsec (kpc))                & ($\mu$Jy)                  &           & (mJy)            & (mJy)             &              & ($z$)        & (W Hz$^{-1}$)         \\ \hline
  SXDF        &                &                       &                                  &                            &           &                  &                   &              &                        \\
  02 18 39.55 & -04 41 49.4    & 50.82$\pm$0.07        &      21 (173)                    & 13.63$\pm$0.254$^{\rm a}$  & 3727.4    &   ...            & 250$\pm$3.5      & -1.09$\pm$0.01       & 2.43(s)      & 2.73 $\times$ 10$^{27}$ \\
  02 17 52.12 & -05 05 22.4    & 6.19$\pm$0.05         &   $\leq$ 1.22 (9.6)              & 3.20$\pm$0.254$^{\rm a}$   & 1934.4    &   ...            & 2.65$\pm$0.4     & 0.58$\pm$0.05        & 2.92         & 5.35 $\times$ 10$^{25}$ \\
  02 18 53.63 & -04 47 35.6    & 16.95$\pm$0.07        &   $\leq$ 1.53 (12.6)             & 27.84$\pm$0.254$^{\rm a}$  & 608.8     &   ...            & 22.8$\pm$0.8     & -0.20$\pm$0.02       & 2.47(s)      & 3.12 $\times$ 10$^{26}$ \\
  02 18 51.38 & -05 09 01.6    & 16.01$\pm$0.07        &   $\leq$ 1.11 (9.5)              & 29.82$\pm$0.268$^{\rm a}$  & 536.9     &   ...            & 62.5$\pm$1.0     & -0.93$\pm$0.01       & 1.75         & 3.18 $\times$ 10$^{26}$ \\
  02 18 03.41 & -05 38 25.5    & 8.91$\pm$0.09         & $\leq$ 2.57 (19.0)               & 13.90$\pm$0.97$^{\rm a}$   & 641       &   ...            & 28.4$\pm$0.2     & -0.79$\pm$0.01       & 3.57(s)      & 7.92 $\times$ 10$^{26}$  \\ 
  02 18 38.24 & -05 34 44.2    & 1.58$\pm$0.02         &  $\leq$ 2.19 (18.7)              & $<$ 7.3$^{\rm c}$          & $>$ 216   &   ...            & 9.7$\pm$0.2      & -1.24$\pm$0.02       & 1.68         & 3.85 $\times$ 10$^{25}$  \\
  02 17 40.69 & -04 51 57.3    & 0.526$\pm$0.047       &    14 (96)                       & $<$ 1.3$^{\rm a}$          & $>$ 405   &   ...            &  4.2$\pm$0.5     & -1.42$\pm$0.01       & 4.32         & 2.04 $\times$ 10$^{26}$ \\   
  02 17 45.84 & -05 00 56.4    & 0.589$\pm$0.013       & $\leq$ 1.58 (13.2)               & $<$ 1.3$^{\rm a}$          & $>$ 453   &   ...            &  5.4$\pm$0.4     & -1.52$\pm$0.01       & 2.22         & 4.15 $\times$ 10$^{25}$ \\    
  02 18 01.23 & -04 42 00.8    & 0.109$\pm$0.013       & $\leq$ 4.64 (39.7)               & $<$ 1.3$^{\rm a}$          & $>$ 84    &   ...            & 0.19$\pm$0.12    & -0.38$\pm$0.01       & 1.72         & 1.19 $\times$ 10$^{24}$   \\   
  02 18 30.13 & -05 17 17.4    & 0.187$\pm$0.013       & $\leq$ 2.98 (25.2)               & $<$ 1.3$^{\rm a}$          & $>$ 144   &   ...            &  0.59$\pm$0.13   & -0.79$\pm$0.01       & 2.04         & 4.59 $\times$ 10$^{24}$   \\   
  02 18 59.19 & -05 08 37.8    & 0.698$\pm$0.014       &  $\leq$ 1.58 (11.6)              & $<$ 1.3$^{\rm a}$          & $>$ 537   &   ...            & 0.23$\pm$0.12    &  0.76$\pm$0.02       & 3.60         & 5.90 $\times$ 10$^{24}$  \\ 
  VLA-VVDS    &                &                       &                                  &                            &           &                  &                  &                      &              &                        \\
  02 27 48.26 & -04 19 05.3    & 0.162$\pm$0.017       &   $\leq$ 3.3 (27.9)              & $<$ 2.0$^{\rm b}$          & $>$ 81    &  $<$ 0.25        & 0.67$\pm$0.15    & -0.97$\pm$0.01       &  $>$ 2.0     &  $>$ 4.63 $\times$ 10$^{24}$     \\
  02 25 02.13 & -04 40 26.9    & 0.202$\pm$0.030       & $\leq$ 5.74 (48.6)               & $<$ 2.0$^{\rm b}$          & $>$ 101   &  $<$ 0.25        &   0.54$\pm$0.11  &  -0.67$\pm$0.01      &  $>$ 2.0     &  $>$ 4.16 $\times$ 10$^{24}$      \\
  02 26 58.10 & -04 18 14.9    & 0.217$\pm$0.016       &    $\leq$ 5.06 (42.9)            & $<$ 2.0$^{\rm b}$          & $>$ 109   &  0.68$\pm$0.05   &   2.72$\pm$0.46  & -1.71$\pm$0.01       &  $>$ 2.0     &  $>$ 1.43 $\times$ 10$^{25}$       \\
  02 27 09.90 & -04 23 44.8    & 0.238$\pm$0.016       &    $\leq$ 3.2 (27.1)             & $<$ 2.0$^{\rm b}$          & $>$ 119   &  0.61$\pm$0.05   &   1.51$\pm$0.13  & -1.26$\pm$0.01       &  $>$ 2.0     &  $>$ 9.48 $\times$ 10$^{24}$     \\
  02 26 31.12 & -04 24 53.3    & 0.699$\pm$0.066       &   18 (152.6)                     & $<$ 2.0$^{\rm b}$          & $>$ 350   &  1.71$\pm$0.15   &   4.56$\pm$0.58  & -1.27$\pm$0.01       &  $>$ 2.0     & $>$ 2.81 $\times$ 10$^{25}$     \\
  02 25 26.14 & -04 34 54.4    & 1.392$\pm$0.049       &   28 (237)                       & $<$ 2.0$^{\rm b}$          & $>$ 696   &  1.87$\pm$0.05   &   3.4$\pm$0.5    & -0.60$\pm$0.01       &  $>$ 2.0     &  $>$ 1.78 $\times$ 10$^{25}$    \\
  02 26 09.09 & -04 33 34.7    & 8.643$\pm$0.020       &   $\leq$ 2.76 (22.7)             & $<$ 2.0$^{\rm b}$          & $>$ 4321  &  18.10$\pm$0.05  &  31.8$\pm$0.7    & -0.89$\pm$0.01       & 2.45         & 3.68 $\times$ 10$^{26}$   \\
  02 24 20.96 & -04 25 44.6    & 18.967$\pm$0.025      &   39 (330.5)                     & 1.8$\pm$0.3$^{\rm b}$      &  10537    &  50.56$\pm$0.10  &  108.8$\pm$1.0   & -1.20$\pm$0.01       &   $>$ 2.0    &  $>$ 7.0 $\times$ 10$^{26}$          \\
  \hline
\end{tabular}}
\label{table:IFRSSample} 
\\
Notes - 
$^{\rm a}$ : SpUDS, $^{\rm b}$ : SERVS, $^{\rm c}$ : SWIRE.
The radio luminosities are in the rest-frame with K-correction applied. 
In redshift column, `(s)' denotes a spectroscopic redshift. 
\end{minipage}
\end{table*} 
\section{Redshift estimates}
\label{sec:Redshifts}
To estimate the redshifts of IFRSs is one the most important aspects in their study. 
We attempt to obtain the redshift estimates of our IFRSs by using the existing data available in the literature (see Table~\ref{table:IFRSSample}). 
The details about the redshift estimates in both the fields are given below. \\
\\
{\bf SXDF} : 
The redshift estimates of our sample sources are obtained mainly from \cite{Simpson12} who reported the spectroscopic and/or photometric 
redshifts of 505/512 1.4 GHz radio sources. The spectroscopic redshift estimates of 267/505 radio sources are 
based on the observations carried out with the Visible Multi-Object Spectrograph (VIMOS) on the VLT and 
also include estimates from previous spectroscopic campaigns \citep[{\eg}][]{Geach07,Smail08,vanBreukelen09,Banerji11,Chuter11}. 
The remaining 238/505 radio sources have photometric redshift estimates derived using 
eleven-band ($u^{\star}$, $B$, $V$, $R$, $i^{\prime}$, $z^{\prime}$, $J$, $H$, $K$, [3.6] and [4.5]) photometric data with the code EASY \citep{Brammer08}. 
The accuracy of the photometric redshifts is determined by comparing the photometric and spectroscopic redshifts of 267 sources. 
It is shown that the photometric redshift estimates are fairly accurate with the normalized median absolute deviation $\sim$ 
0.035 for the histogram of normalized photometric redshift errors (z$_{\rm phot}$ - z$_{\rm spec}$)/(1 + z$_{\rm spec}$). 
There are only 07/512 radio sources with no redshift estimates due to the presence of bright foreground objects in their close vicinity 
in the optical and IR images.
Using \citet{Simpson12} measurements we find that, among 11 IFRSs in the SXDF, 
only three have spectroscopic redshift estimates and eight have photometric redshift estimates 
(see Table~\ref{table:IFRSSample}). 
We note that, based on the optical spectra, \cite{Simpson12} classify one of our IFRSs J021839-044149 as a narrow line AGN, 
while two other IFRSs {\ie}J021853-044735 and J021803-053825 are possibly also narrow line AGN, 
although with some degree of ambiguity due to the poor quality of their spectra.
The optical spectra of these IFRSs are presented in \cite{Simpson12}. 
\par
\begin{figure}
\hspace{-0.6cm}
\includegraphics[angle=0,width=9.0cm,trim={0.5cm 0.0cm 1.0cm 1.0cm},clip]{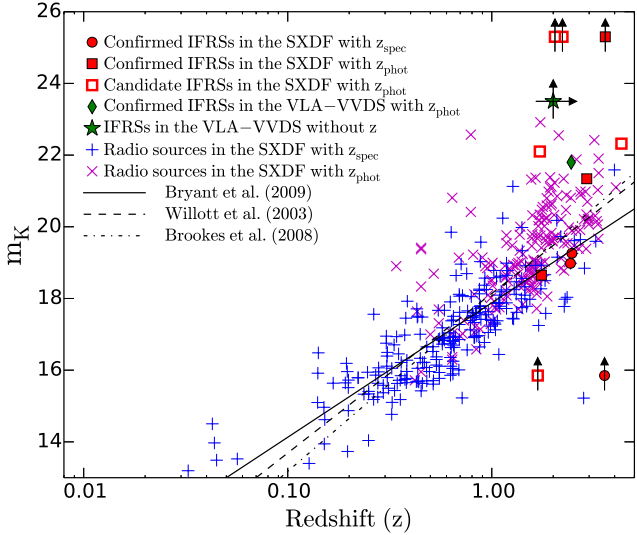}
\caption{The locations of our IFRSs in the K-{\it z} plot displaying the radio sources detected in the SXDF \citep[see][]{Simpson12}. 
The vertical and horizontal arrows indicate lower limits on the K-band magnitudes and redshifts, respectively. 
Two IFRSs fall outside the UDS footprint and the lower limits (m$_{\rm K}$ $\geq$ 15.85) for these two IFRSs are from 2MASS. 
In the VLA-VVDS field, there are seven IFRSs with no redshift and no K-band magnitude, and such sources are placed at {\it z} $>$ 2.0 
based on the trend of K-{\it z} relation and the K-band magnitude lower limits (m$_{\rm K}$ $\geq$ 23.5 from the VIDEO).     
The solid, dashed and dashed-dotted lines represent the K-{\it z} relations for the 
samples of radio galaxies from \citet{Bryant09}, \citet{Willott03} and \citet{Brookes08}, respectively.}
\label{fig:Kzplot} 
\end{figure}

{\bf VLA-VVDS} : 
The redshift estimates of our IFRSs in the VLA-VVDS field are obtained from \cite{McAlpine13} who present photometric redshifts of 942/1054 1.4 GHz radio sources 
using ten-band ($u^{\star}$, $g^{\prime}$, $r^{\prime}$, $i^{\prime}$, $z^{\prime}$, $Z$, $Y$, $J$, $H$, and $Ks$) photometric data. 
The photometric redshifts are derived using code Le Phare6 \citep{Ilbert06} that fits the photometric data with SED templates. 
The accuracy of photometric redshift estimate is assessed by comparing the photometric redshifts with the 
spectroscopic redshifts obtained from the VIMOS VLT Deep Survey (VVDS; \citealt{LeFevre13}).  
There are only 3.8 per cent catastrophic outliers defined as sources with $|$z$_{\rm phot}$ - z$_{\rm spec}$$|$/(1 + z$_{\rm spec}$) $>$ 0.15.  
Owing to the use of more photometric bands the redshift estimates presented in \cite{McAlpine13} supersedes previous 
estimates \citep[e.g.,][]{Ciliegi05,Bardelli09}. 
The procedure followed for deriving the photometric redshifts is described in \cite{Jarvis13}. 
We also use spectroscopic redshift catalogue produced by the VIMOS VLT Deep Survey (VVDS), which is a magnitude limited (17.5 $\leq$ $i$ $\leq$ 24.75) 
spectroscopic redshift survey conducted by the VIMOS multi-slit spectrograph 
at the ESO-VLT \citep{LeFevre13}. There are only 134/1054 1.4 GHz radio sources with spectroscopic redshifts from the VVDS.
Using \citet{McAlpine13} photometric redshift estimates together with the VVDS spectroscopic redshift catalogue 
we find that only one out of eight IFRSs has a photometric redshift estimate 
(see Table~\ref{table:IFRSSample}). \\
Since seven out of eight IFRSs in the VLA-VVDS field do not have redshift estimates, we attempt to constrain
the lower limits of their redshifts by using the empirical relation between $K$-band magnitude and redshift 
{\ie}the $K$-${\it z}$ relation exhibited by radio sources. It has been found that radio galaxies at higher redshifts are systematically 
fainter in the $K$-band \citep[see][]{Willott03,Brookes08,Bryant09}. 
In fact, the radio source population consisting of different kinds of galaxies also follows 
the $K$-{\it z} relation but with a relatively larger dispersion \citep{Simpson12}. 
The $K$-band (centred at 2.2 $\mu$m) is effective in detecting galaxies over a large redshift range (0 $<$ $z$ $<$ 4) as it samples their 
NIR to optical rest-frame emission. 
In Fig.~\ref{fig:Kzplot}, we show the locations of our IFRSs in the $K$-{\it z} plot drawn for the  
1.4 GHz radio sources detected in the SXDF. 
We choose radio sources from the SXDF due to the fact that a large fraction (267/505 = 52.9 per cent) of these 
radio sources have spectroscopic redshifts, and the remaining radio sources have photometric redshifts.
We find that the most of our IFRSs lie at the fainter end of the $K$-{\it z} plot. 
Five out of eleven IFRSs detected in the SXDF do not have $K$-band counterparts, 
and therefore, only lower limits are considered (from the UDS for three IFRSs and from the 2MASS for two IFRSs). 
In the VLA-VVDS, seven IFRSs with no redshift estimates also lack $K$-band counterparts in the VIDEO data. 
Using $K$-band magnitude lower limits (m$_{\rm K}$ $>$ 23.5 at 5$\sigma$) from the VIDEO data and the $K$-{\it z} correlation shown by the 
general radio population, the IFRSs with no redshifts are expected to lie at $z$ $>$ 2.0 (see Fig.~\ref{fig:Kzplot}). 
From the $K$-{\it z} plot it is evident that the lower limit ($z$ $>$ 2.0) on the redshifts of our IFRSs is a fairly conservative 
value, even if we account for the dispersion and non-linearity in the $K$--{\it z} plot, in particular at higher redshifts. 
The lower limit $z$ $>$ 2.0 considered for the IFRSs in the VLA-VVDS is further supported by the fact that the IFRSs with undetected $K$-band counterparts in 
the SXDF have estimated redshifts distributed over $\sim$ 2 -- 4.
We also note that the lower limit on the redshifts is based on the $K$-{\it z} relation for faint radio sources (5$\sigma$ $\simeq$ 0.1 mJy) 
and these IFRSs will lie at much higher redshifts ({\it z} $>$ 5) if we consider the $K$-{\it z} relation for powerful radio galaxies 
(for example, $K$({\it z}) $=$ 17.75 + 3.64 log10({\it z}); \citealt{Bryant09}).
\par
\begin{figure}
\includegraphics[angle=0,width=9.0cm,trim={0.0cm 0.0cm 0.0cm 0.0cm},clip]{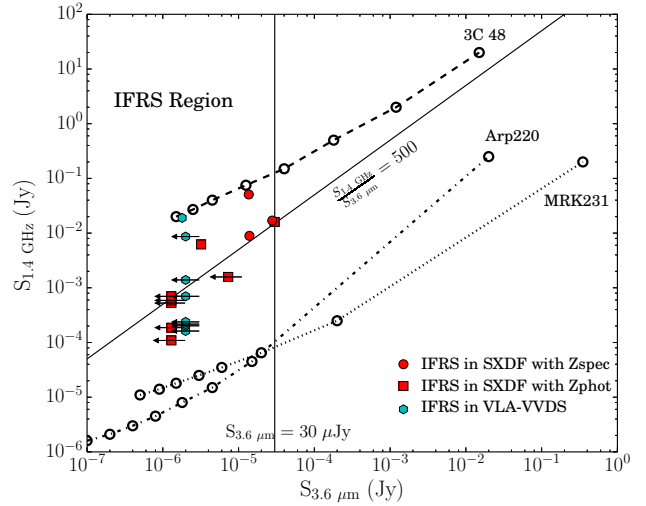}
\caption{The locations of our IFRSs in S$_{\rm 1.4~GHz}$ versus S$_{\rm 3.6~{\mu}m}$ diagnostic plot in which evolutionary tracks 
are shown for 3C 48 (dashed line), Arp 220 (dash-dotted line) and Mrk 231 (dotted line). 
The open circles represent redshifted data points with increasing redshift value from right to left such that the first open circle having the highest flux on any track represents 
the current redshift of the source, second circle represents the source placed at $z$ $=$ 1, third circle represents the source at $z$ $=$ 2 and so on upto $z$ $=$ 7. 
The vertical and diagonal solid lines represent our IFRSs selection criteria S$_{\rm 3.6~{\mu}m}$ $<$ 30 $\mu$Jy 
and $\frac{\rm S_{1.4~GHz}}{\rm S_{3.6~{\mu}m}}$ $\geq$ 500, respectively. 
The sources above $\frac{\rm S_{1.4~GHz}}{\rm S_{3.6~{\mu}m}}$ $=$ 500 line are confirmed IFRSs, 
while the sources below this line are candidate IFRSs.}
\label{fig:RadioVsIRPlot} 
\end{figure}
We also cross-check the accuracy of the lower limit on the redshifts of our IFRSs by using S$_{\rm 1.4~GHz}$ versus S$_{\rm 3.6~{\mu}m}$ 
diagnostic plot (see Fig.~\ref{fig:RadioVsIRPlot}). 
This diagnostic plot proposed by \cite{Zinn11} shows the evolutionary tracks for three different types 
of galaxies {\ie}3C~48 (a radio-loud AGN at $z = 0.37$ and a good IFRS candidate at higher redshift with 
$\frac{\rm S_{1.4~GHz}}{\rm S_{3.6~{\mu}m}}$ $>$ 500), Mrk 231 (a dusty IR bright radio-quiet AGN at $z$ $=$ 0.042) and 
Arp 220 (a star-burst IR luminous galaxy at $z$ $=$ 0.18). 
In order to trace the evolution of 1.4 GHz and 3.6 $\mu$m flux density as the function of redshift, the SED templates of 3C 48, Arp 220
and Mrk 231 were redshifted to $z$ $=$ 1, 2, 3 and so on upto 7, and 1.4 GHz radio and 3.6 $\mu$m flux densities were estimated at respective
redshifts. In the S$_{\rm 1.4~GHz}$ versus S$_{\rm 3.6~{\mu}m}$ diagnostic plot, 
we show the positions of our IFRSs along with the evolutionary tracks of 3C$~$48, Arp$~$220 and Mrk 231. 
The comparison of the positions of our IFRSs having no redshift estimates and the evolutionary tracks of three different types 
of sources suggests that our IFRSs are likely to be at $z$ $>$ 6.0 if they are powerful radio-loud AGN similar to 3C 48, and our 
IFRSs are at $z$ $>$ 4.0 if they are similar 
to Mrk 231, while our IFRS are at $z$ $>$ 2.0 if they are similar to Apr 220. 
Furthermore, the locations of our IFRSs, with known redshifts at $z$ $=$ 1.68 -- 4.3, 
in S$_{\rm 1.4~GHz}$ versus S$_{\rm 3.6~{\mu}m}$ diagnostic plot, suggest that they consist of a mixed population of different types 
of sources ranging from powerful radio-loud AGN to relatively less powerful radio AGN. 
Indeed, the location of a source in the S$_{\rm 1.4~GHz}$ versus S$_{\rm 3.6~{\mu}m}$ plot 
depends on the exact nature of SED which is unknown for most of our IFRSs. 
We note that \cite{Maini16} demonstrated the evolution of 3.6 $\mu$m flux density with redshift for 
different types of galaxies ({\eg}Arp220, Mrk 231, type 1 and type 2 AGN/QSOs, 
and radio-loud galaxies) and found that any such galaxy would lie at {\it z} $>$ 2.5 if S$_{\rm 3.6~{\mu}m}$ $\leq$ 2.0 $\mu$Jy. 
Therefore, a conservative limit of $z$ $>$ 2.0 for our IFRSs with no redshift estimates, derived from the $K$-{\it z} relation 
is also consistent with their positions in S$_{\rm 1.4~GHz}$ versus S$_{\rm 3.6~{\mu}m}$ diagnostic plot. 
\par 
Table~\ref{table:IFRSSample} lists the redshifts of our IFRSs. In the SXDF, IFRSs 
are found to be distributed over $z$ $=$ 1.68 -- 4.32, while in the VLA-VVDS field, a lower limit of $z$ $>$ 2.0 is placed for all but 
one (at $z$ $=$ 2.45) IFRSs (see Table~\ref{table:IFRSSample}). We note that the redshift estimates of our sample IFRSs in
the two fields are consistent with previous studies suggesting IFRSs to be high redshift objects \citep[see][]{Garn08,Huynh10,Collier14}. 
Although, given the faintness of IFRSs in the optical and IR bands 
most of the previous attempts were limited to obtain the lower limit on redshifts based on SED modelling \citep[see][]{Garn08,Huynh10,Herzog14,Herzog15b}. 
Therefore, we emphasise that the deep optical and IR data available in the SXDF allow us to obtain the spectroscopic or 
photometric redshifts of all IFRSs in this field. 
In fact, our study of IFRSs in the SXDF provides the first instance where all the IFRSs found in a deep field have redshift estimates.
\\
However, unlike in the SXDF, the optical and IR data in the VLA--VVDS field are relatively less deep (see Table~\ref{table:MWData}), 
which, in turn, cause the lack of redshift measurements of our IFRSs. 
Indeed, the lack of redshift estimates for IFRSs is a very common issue. For instance, 
in a large sample of 1317 relatively bright IFRSs, \cite{Collier14} 
obtained  spectroscopic redshifts of only 19 IFRSs from the SDSS DR9, 
and all but one of these IFRSs are found be quasars located at 2 $<$ $z$ $<$ 2.99. 
Using VLT observations, \cite{Herzog14} attempted to estimate the redshifts of four IFRSs that are relatively bright in the optical 
(m$_{\rm r}$ $\sim$ 22 -- 24; Vega magnitude) in the ATLAS field and found redshifts for three brightest 
IFRSs at 1.84, 2.13, and 2.76. 
Therefore, in comparison to the previous studies which could find IFRSs at $z$ $\leq$ 2.99, 
our study reveals, for the first time, IFRSs at the highest redshift {\ie}$z$ $>$ 3.0. 
We also note that the deep data in the SXDF allow us to find redshifts for IFRSs at the faintest flux levels 
(see Section~\ref{sec:Comparison}).  

\section{Comparison of our IFRSs with previous IFRSs samples}
\label{sec:Comparison}
\begin{figure*}
\includegraphics[angle=0,width=8.6cm,trim={0.0cm 0.0cm 0.0cm 0.0cm},clip]{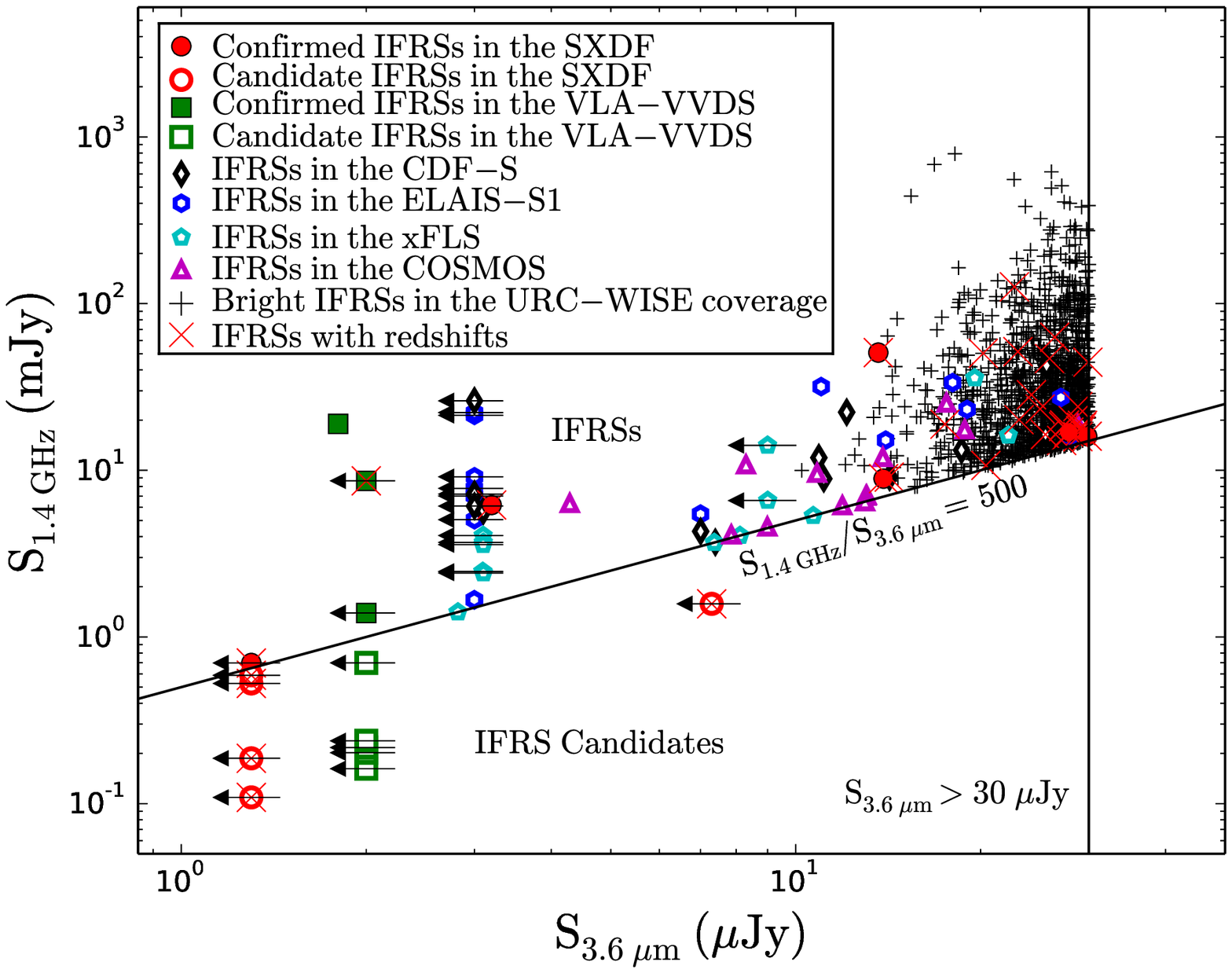}
{\includegraphics[angle=0,width=8.6cm,trim={0.0cm 0.0cm 0.0cm 0.0cm},clip]{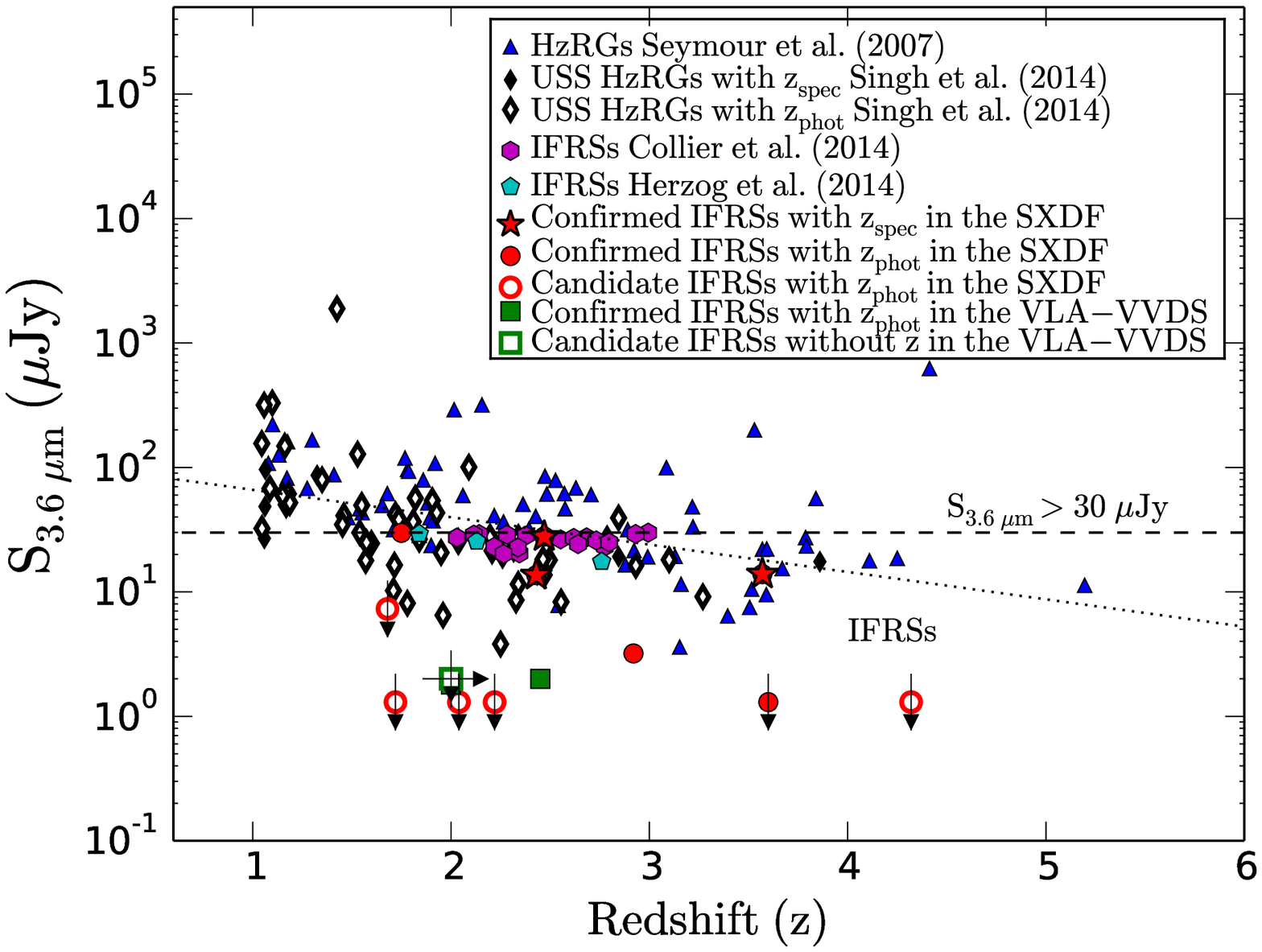}}
\caption{{\it Left panel} : S$_{\rm 1.4~GHz}$ versus S$_{\rm 3.6~{\mu}m}$ plot for IFRSs from different samples. 
The solid lines mark IFRSs selection criteria. The horizontal arrows depict upper limits on 3.6 $\mu$m fluxes. 
{\it Right panel} : S$_{\rm 3.6~{\mu}m}$ versus redshift plot for IFRSs and H$z$RGs from different samples. 
The horizontal dashed line marks one of the selection criteria (S$_{\rm 3.6~{\mu}m}$ $<$ 30 $\mu$Jy) for IFRSs. 
The dotted line represents the linear regression line for all confirmed IFRSs and H$z$RGs. 
Downward arrows represent the upper limits on 3.6 $\mu$m fluxes.     
For Collier et al. (2014) IFRSs sample 3.4 ${\mu}$m flux instead of 3.6 ${\mu}$m flux is available from the WISE survey.}
\label{fig:S36vsRedshift} 
\end{figure*}

In order to highlight the new parameter space investigated in our study, we compare our IFRSs with previous IFRSs samples 
using S$_{\rm 3.6~{\mu}m}$ versus S$_{\rm 1.4~GHz}$, and S$_{\rm 3.6~{\mu}m}$ versus $z$ plots (see Fig.~\ref{fig:S36vsRedshift}).   
Also, in Table~\ref{table:IFRSsComp}, we list details ({\ie}field, survey area, sensitivity in the radio and IR surveys, 
number of IFRSs, 1.4 GHz and 3.6 $\mu$m flux ranges with their median values, 
number of IFRSs with redshift estimates, and redshift ranges with their median values) 
of various IFRSs samples. 
\begin{table*}
\hspace{-3.5cm}
\begin{minipage}{140mm}
\caption{Comparison of various IFRSs samples}
\scalebox{0.8}{ 
\begin{tabular}{@{}ccccccccccc@{}}
\hline
 Reference  & Field        & Area       & S$_{\rm 1.4~GHz}$ limit  & S$_{\rm 3.6~{\mu}m}$ limit & N$_{\rm IFRSs}$ & S$_{\rm 1.4~GHz}$    & N$_{\rm det,~3.6~{\mu}m}$ & S$_{\rm 3.6~{\mu}m}$       & N$_{z}$  & $z$                     \\
            &              & (deg$^{2}$) & 5$\sigma$ ($\mu$Jy beam$^{-1}$) & 5$\sigma$ ($\mu$Jy beam$^{-1}$) &                 & range (mJy)          &                           & range (${\mu}$Jy)          &          & range                    \\ \hline
 1          & SXDF         &  0.8       & 100$^{a}$                & 1.3$^{sp}$                 & 6               & 0.7 -- 50.8 (16.0)  &  5                        & $<$ 1.3 -- 29.8 (13.9)    &  6       & 1.75 -- 3.6 (2.92)       \\
 1          & VLA-VVDS     &  1.0       & 80$^{b}$                 & 2.0$^{se}$                 & 3               & 1.4 -- 19.0 (8.6)   &  1                        &  $<$ 2.0 ($<$ 2.0)         &  1       & 2.45 -- $>$ 2.0 ($>$ 2.0)  \\
 2          & CDF-S        &  3.7       & 186                      & 3.1$^{sw}$                 & 14              & 3.7 -- 42.5 (11.9)  &  11                       & $<$ 3.0 -- 29.3 (11.1)    &  3       & 1.84 -- 2.76 (2.13)   \\
 3          & ELAIS-S1     &  3.6       & 160                      & 3.1$^{sw}$                 & 15              & 1.7 -- 33.6 (15.2)  &   9                       &  $<$ 3.0 -- 28.0 (3.0)    & ...      &  ...          \\
 4          & xFLS         &  3.1       & 105$^{c}$                & 9.0$^{s}$                  & 13              & 1.4 -- 35.8 (4.1)   &   6                       & $<$ 3.0 -- 22.2 ($<$ 7.3) &  ...     &  ...          \\
 5          & COSMOS       &  1.1       &  65$^{d}$                & 1.0$^{sc}$                 & 13              & 4.1 -- 25.3 (7.0)   &   13                      & 4.3 -- 28.8 (13.0)        &  ...     &  ...          \\ 
 6          & URC-SDSS     &  11000    & 1000                      & 80$^{w}$                   & 1317            & 7.9 -- 793.4 (25.5) &    1317                   &  10.3 -- 30.0 (25.94)     &   19     & 0.5 --  2.99 (2.38)    \\ \hline
 \end{tabular}}
\label{table:IFRSsComp} 
\\
Notes : Only confirmed IFRSs in the SXDF and VLA-VVDS are considered. 
The details on the IFRSs samples in deep fields (CDF-S, ELAIS-S1, xFLS and COSMOS) are based on \cite{Zinn11}.   
References - 
1 : This work; 2 : \cite{Norris06}; 3 : \cite{Middelberg08}; 4 : \cite{Garn08}; 5 : \cite{Zinn11}; 6 : \cite{Collier14}.
$^{a}$ : VLA BnC \citep{Simpson06}; $^{b}$ : VLA B \citep{Bondi03};  $^{c}$ : VLA B \citep{Condon03}; $^{d}$ : VLA AnC \citep{Schinnerer07}; 
$^{sp}$ : SpUDS \citep{Dunlop07}, $^{se}$ : SERVS \citep{Mauduit12}; $^{sw}$ : SWIRE \citep{Lonsdale03}; $^{s}$ : Spitzer \citep{Lacy05}; 
$^{sc}$ : S-COSMOS \citep{Sanders07}, $^{w}$ : WISE \citep{Wright10}. \\
N$_{\rm IFRSs}$ : number of IFRSs, N$_{\rm det,~3.6~{\mu}m}$ : number of IFRSs with detected 3.6~${\mu}$m counterparts, N$_{z}$ : number of IFRSs with redshift estimates. 
\end{minipage}
\end{table*}
We note that all IFRSs samples ({\ie}our IFRSs, IFRSs in other deep fields, and the bright IFRSs sample) 
follow same selection criteria as proposed by \cite{Zinn11}. 
From Fig.~\ref{fig:S36vsRedshift} (left panel) and Table~\ref{table:IFRSsComp} it is obvious that, in comparison to the previous IFRSs samples, 
our IFRSs sample reaches down to the faintest flux level in the radio (S$_{\rm 1.4~GHz}$ = 0.7 mJy) and 3.6 $\mu$m (S$_{\rm 3.6~{\mu}m}$ $<$ 1.3 $\mu$Jy) bands 
due to the depths of our data. Our 3.6 $\mu$m SpUDS data (5$\sigma$ = 1.3 $\mu$Jy beam$^{-1}$) in the SXDF are the second deepest 
{\it Spitzer} survey after the S-COSMOS (5$\sigma$ = 1.0 $\mu$Jy beam$^{-1}$). 
In the radio band also, our data are the second deepest data (5$\sigma$ = 80 -- 100 $\mu$Jy beam$^{-1}$) after 
the radio data in the COSMOS field. 
The relatively higher flux densities of IFRSs in the COSMOS and in other deep fields may also be due to slightly higher 
cut-off (SNR $\geq$ 10$\sigma$) imposed during the IFRSs sample selection \citep[see][]{Zinn11}. 
In comparison to the bright IFRSs sample reported in \cite{Collier14} our IFRSs sample includes nearly one order of magnitude fainter sources 
in both radio and 3.6 $\mu$m bands (see Table~\ref{table:IFRSsComp}).   
\par
Fig.~\ref{fig:S36vsRedshift} (right panel) shows S$_{\rm 3.6 {\mu}m}$ versus $z$ for IFRSs and H$z$RGs taken from different samples.  
In the literature, this diagram has been used to understand the nature of H$z$RGs and IFRSs \citep[see][]{Jarvis09,Norris11,Collier14}. 
In S$_{\rm 3.6~{\mu}m}$ versus $z$ diagram we show the positions of our IFRSs, the IFRSs from previous samples \citep{Collier14,Herzog14}, 
less powerful H$z$RGs (L$_{\rm 1.4~GHz}$ $>$ 10$^{24}$ W Hz$^{-1}$; \citealt{Singh14}), and powerful 
H$z$RGs (L$_{\rm 3~GHz}$ $>$ 10$^{26}$ W Hz$^{-1}$; \citealt{Seymour07}). 
For H$z$RGs we consider only sources with $z$ $\geq$ 1.0. 
From S$_{\rm 3.6 {\mu}m}$ versus $z$ plot (Fig.~\ref{fig:S36vsRedshift}; right panel) it is apparent that 
our study finds redshifts of IFRSs at much fainter flux limits ({\ie}S$_{\rm 3.6~{\mu}m}$ $<$ 1.3~$\mu$Jy), 
in comparison to the previous studies where redshift measurements were limited to relatively bright IFRSs 
({\ie}S$_{\rm 3.6~{\mu}m}$ $\sim$ 20 -- 30 $\mu$Jy). 
In fact, we find first spectroscopic redshift of an IFRS at S$_{\rm 3.6~{\mu}m}$ $<$ 15 $\mu$Jy, 
while previous studies provided spectroscopic redshifts of IFRSs at S$_{\rm 3.6~{\mu}m}$ $=$ 17 -- 30 $\mu$Jy 
\citep[see][]{Herzog14,Collier14}. 
Also, we obtain the first photometric redshift of an IFRS at S$_{\rm 3.6~{\mu}m}$ $<$ 1.3 $\mu$Jy (see Table~\ref{table:IFRSSample}).
\\
Furthermore, the availability of redshift estimates for faint IFRSs in our sample allows us to compare the nature of 
faint and bright IFRSs. 
In S$_{\rm 3.6~{\mu}m}$ versus $z$ diagram we find that powerful H$z$RGs exhibit a trend of decreasing 3.6 $\mu$m flux with 
the increase in redshift. 
And, the inclusion of less powerful H$z$RGs shows a similar trend but with a larger scatter. 
We note that all confirmed IFRSs, in our sample as well as in the previous samples, follow the trend shown by H$z$RGs, 
which can be understood as IFRSs are radio--loud AGN (see Section~\ref{sec:RadioProp}). 
The candidate IFRSs at $z$ $>$ 2, with no detected 3.6 $\mu$m counterparts, tend to deviate in S$_{\rm 3.6~{\mu}m}$ versus $z$ plot, 
as these are possibly weaker radio AGN. 
In general, powerful H$z$RGs are found to be hosted in massive galaxies that are bright in 3.6 $\mu$m band, in which 
stellar emission dominates \citep{Jarvis01c,Seymour07}. 
Therefore, relatively less powerful H$z$RGs at similar redshifts are
expected to be relatively fainter in the 3.6 $\mu$m band, which is apparent from the S$_{\rm 3.6~{\mu}m}$ -- $z$ plot. 
To probe the correlation between faint and bright IFRSs 
it may be useful to derive a correlation equation for IFRSs in S$_{\rm 3.6~{\mu}m}$ -- $z$ plot. 
However, we caution that the redshift estimates are available only for relatively bright sources 
spanning in a small range of S$_{\rm 3.6~{\mu}m}$ ({\ie}20 -- 30 $\mu$Jy; \citealt{Herzog14,Collier14}) along 
with a few IFRSs at fainter flux and higher redshift regime. 
Hence, S$_{\rm 3.6~{\mu}m}$ -- $z$ correlation equation derived from the present data points would be biased 
and non-representative to the full IFRSs population. 
Therefore, we present only a qualitative comparison between IFRSs and HzRGs in the S$_{\rm 3.6~{\mu}m}$ -- $z$ plot.    
\section{Radio Properties of our IFRSs}
\label{sec:RadioProp}
In this section, we discuss the radio properties of our IFRSs.
\subsection{1.4 GHz and 3.6 $\mu$m flux densities}
1.4 GHz radio flux densities of IFRSs in our sample are distributed over 0.1 mJy -- 50.8 mJy with a median of 0.7 mJy 
(see Table~\ref{table:IFRSSample} and Fig.~\ref{fig:S36vsRedshift}). 
The confirmed IFRSs are distributed over S$_{\rm 1.4~GHz}$ $=$ 0.68 mJy -- 50.8 mJy with the median of 8.9 mJy, 
while the candidate IFRSs span over S$_{\rm 1.4~GHz}$ $=$ 0.109 -- 1.58 mJy with the median of 0.238 mJy. 
Therefore, at radio wavelengths, the confirmed IFRSs in our sample are systematically brighter than the candidate IFRSs. 
The 3.6 $\mu$m flux densities of confirmed IFRSs range from $<$ 1.3 $\mu$Jy to 29.8 $\mu$Jy with the median of 3.2 $\mu$Jy, while all the candidate IFRSs 
show no detected 3.6 $\mu$m counterparts at 5$\sigma$ level. The candidate IFRSs can only be true IFRSs if their 3.6 $\mu$m flux densities are on 
average 1.2 -- 6.25 times lower than their upper limits. 
The visual inspection of the 3.6 $\mu$m image cut-outs shows no detection even at the faintest level. Thus, if we 
assume upper limit of 3.6 $\mu$m flux at 1$\sigma$ level (0.26 $\mu$Jy in the SpUDS and 0.4 $\mu$Jy in the SERVS) 
then all but two candidate IFRSs will 
be confirmed IFRSs. We perform the median stacking of 3.6 $\mu$m SERVS and SpUDS image cut-outs at the radio positions of 
13 IFRSs with no detected 3.6 $\mu$m counterparts, and we do not get any detection in the stacked image with the 
average rms noise of $\sim$ 0.09 $\mu$Jy beam$^{-1}$.
Therefore, on average 3.6 $\mu$m fluxes of our IFRSs with undetected 3.6 $\mu$m counterparts is $<$ 0.27 $\mu$Jy (assuming 3$\sigma$ of the median stacked image), 
and thus suggesting that, most if not all, candidate IFRSs are likely to be true IFRSs. 
Also, the candidate IFRSs with relatively higher ratio of radio-to-IR flux 
({\ie}$\frac{\rm S_{1.4~GHz}}{\rm S_{3.6~{\mu}m}}$ $>$ 350) are brighter in the radio and have higher probability of being true IFRSs. 
The candidate IFRSs with lower ratio of radio-to-IR flux can represent an extreme IFRS population which contains either powerful radio sources at 
much higher redshifts ($z$ $\geq$ 5.0) or less powerful radio sources at moderate redshifts ($z$ $\sim$ 2.0 -- 3.0). 
We note that the IFRSs in our sample are distributed over a wide range of radio flux densities, IR flux densities, 
and the ratios of radio-to-IR, which is consistent with the previous IFRS samples. 
For example, \cite{Zinn11} presented a sample of 55 IFRSs (with SNR cut-off $\geq$ 13$\sigma$) from four deep fields ({\ie}ELAIS-S1, CDFS, xFLS and COSMOS) 
and found them to be distributed over S$_{\rm 1.4~GHz}$ $\sim$ 1.14 -- 42.2 mJy with the median of 7.8 mJy. 
For our IFRSs sample, we use a lower value of SNR cut-off ({\ie}8$\sigma$) which allows us to find IFRSs in the sub-mJy regime.
\subsection{Radio morphologies, sizes and luminosities}
The radio morphologies of IFRSs can provide us clues in understanding their nature. 
Therefore, we investigate radio morphologies of our IFRSs using both 1.4 GHz and 325 MHz observations. 
The morphologies at two frequencies can be used in a complementary fashion as 
1.4 GHz VLA observations are more useful in resolving the structures due to relatively higher resolution, while low-frequency 325 MHz 
observations can be more effective in detecting diffuse relic emission. 
In figure~\ref{fig:IFRSDetected} we show 1.4 GHz  radio contours of IFRSs overlaid on to their 3.6 $\mu$m images. 
Using 1.4 GHz observations we find that 14/19 IFRSs in our sample are unresolved point sources, 
while five IFRSs show extended radio emission. 
There are three confirmed IFRSs (J021839-044149, J022420-042544 and J022526-043454) and one candidate IFRS (J022631-042453) 
which show distinct double-lobe radio morphology, and thus confirming them to be a typical radio galaxy. 
A candidate IFRS J022631-042453 ($\frac{S_{\rm 1.4~GHz}}{S_{\rm 3.6~{\mu}m}}$ $>$ 350) also shows 
extended morphology with marginally resolved double-lobe like structure. 
Thus, with the 1.4 GHz radio morphology itself, it is evident that at least a fraction of IFRSs are radio galaxies at higher redshifts. 
The 325 MHz radio morphologies seem less informative due to their relatively lower resolution ({\ie}9$\arcsec$.0) and lower 
sensitivity (5$\sigma$ $\simeq$ 750 $\mu$Jy beam$^{-1}$). 
At 325 MHz, some of the IFRSs with double-lobe morphologies seen in 1.4 GHz, appear as an unresolved point source ({\ie}J021839-044149), 
blended with a nearby source ({\ie}J021740-045157), and are undetected due to their inverted/flat radio spectra ({\ie}J021859-050837 and J021801-044200). 
\par
\begin{figure}
\includegraphics[angle=0,width=9.0cm,trim={0.0cm 0.0cm 0.0cm 0.0cm},clip]{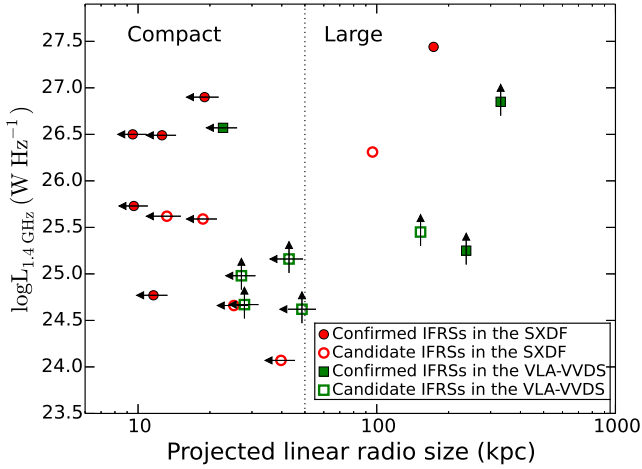}
\caption{Radio luminosity versus projected linear radio size plot for our IFRSs. 
The vertical dotted line separates compact and large, extended sources.}
\label{fig:RadioSizeVsLumin} 
\end{figure}

 
We try to constrain the 1.4 GHz radio sizes of our IFRSs irrespective of their apparent radio morphology. 
For extended sources, we obtain their projected linear sizes by measuring the distance between the 
two extremes along the direction of extension. 
For unresolved point sources we fit them with a single Gaussian component using the {\em JMFIT} task in the Astronomical 
Image Processing System (AIPS\footnote{http://www.aips.nrao.edu/index.shtml}). 
For unresolved sources, we consider the upper limit on their radio size as 
$\sqrt{\rm (max){\times}(min)}$, where `max' and `min' are the upper limits on 
major and minor axis, respectively, of the deconvolved Gaussian fit obtained from the {\em JMFIT}. 
If the upper limit on the deconvolved minor axis is `0', we set 
the upper limit on the minor axis equal to the upper limit on the major axis. 
\\
It is evident that all the unresolved sources are compact with the upper limit on their sizes ranging from 10 kpc to 50 kpc, 
while the linear projected size of extended sources ranges from 96 kpc to 330 kpc (see Table~\ref{table:IFRSSample} 
and figure~\ref{fig:RadioSizeVsLumin}). 
Among the nine confirmed IFRSs, six are compact sources with their radio sizes $\leq$ 23 kpc and three are extended double-lobe sources 
with the projected linear sizes of 173 kpc, 237 kpc and 331 kpc.  
The double-lobe morphology of some of our IFRSs is consistent with the previous hypotheses of IFRSs being radio galaxies 
at high-{\it z}. 
The typical radio size of an evolved radio galaxy is $\sim$ 100 kpc or larger \citep{Pentericci2000}, 
and therefore, our IFRSs having smaller radio sizes may be young radio galaxies. 
%
\cite{Garn08} reported the upper limits on the radio sizes of IFRSs with unresolved morphologies 
to be $\leq$ 20 kpc which is similar to our results. 
Very Long Baseline Interferometry (VLBI) observations of relatively bright IFRSs revealed only 
a compact radio core with upper limit on their radio sizes limited only to a few hundred parsec 
\citep{Norris07,Middelberg08,Herzog15a}. 
We also note that the resolution of VLA observations ({\ie}6$\arcsec$.0) used in our study is better than the ATLAS observations 
({\ie}11$\arcsec$.0 $\times$ 5$\arcsec$.0; \citealt{Norris06}), and therefore, we are likely to find more double-lobe radio sources.
\par
The radio luminosities of IFRSs can also be used to characterise their nature. 
We estimate 1.4 GHz rest-frame radio luminosities of our IFRSs that are $K$-corrected by assuming a 
power law radio spectrum (S$_{\nu}$ $\propto$ ${\nu}^{-{\alpha}}$) and the spectral index measured between 
325 MHz and 1.4 GHz (${\alpha}_{\rm 325~MHz}^{\rm 1.4~GHz}$). 
Thus, the radio luminosity of an IFRS at redshift $z$ and luminosity-distance d$_{\rm L}$ is given 
by L$_{\nu}$ = 4 $\pi$ d$_{\rm L}^{2}$ S$_{\nu}$ $\rm (1+{\it z})^{-(1+{\alpha})}$.  
We find that the 1.4 GHz radio luminosities of our IFRSs are distributed over 1.2 $\times$ 10$^{24}$ W Hz$^{-1}$ to 
2.7 $\times$ 10$^{27}$ W Hz$^{-1}$ with the median of 3.8 $\times$ 10$^{25}$ W Hz$^{-1}$ (see Table~\ref{table:IFRSSample}). 
For confirmed IFRSs, the 1.4 GHz radio luminosities range from 5.8 $\times$ 10$^{24}$ W Hz$^{-1}$ 
to 2.7 $\times$ 10$^{27}$ W Hz$^{-1}$ with the median of 3.2 $\times$ 10$^{26}$ W Hz$^{-1}$. 
While the radio luminosities of candidate IFRSs span over 1.2 $\times$ 10$^{24}$ W Hz$^{-1}$ 
to 2.0 $\times$ 10$^{26}$ W Hz$^{-1}$ with the median of 1.2 $\times$ 10$^{25}$ W Hz$^{-1}$. 
Given the high 1.4 GHz radio luminosities all our IFRSs can only be AGN as galaxies with L$_{\rm 1.4~GHz}$ $\geq$ 10$^{24}$ W Hz$^{-1}$ 
cannot be powered by star formation or star-burst alone \citep{Afonso05,Mauch07}. 
Therefore, high radio luminosities of our IFRSs suggest them to be radio-loud AGN in which 
radio emission is due to non-thermal synchrotron radiation from relativistic jets \citep{Rafter09}. 
The radio luminosities of our IFRSs straddle across the radio luminosity break between FR I/FR II radio galaxies ({\ie}L$_{\rm 1.4~GHz}$ 
$=$ 10$^{25.6}$ W Hz$^{-1}$ derived from L$_{\rm 178~MHz}$ $=$ 10$^{26}$ W Hz$^{-1}$ assuming a typical spectral index of -0.7; 
\citealt{Fanaroff74}). Thus, a fraction of our IFRSs with L$_{\rm 1.4~GHz}$ $\geq$ 10$^{25.6}$ W Hz$^{-1}$ are powerful FR II radio galaxies. 
Furthermore, FR I/FR II break radio luminosity is a function of optical luminosity of the host galaxy \citep{Ghisellini01}, and 
therefore, optically faint IFRSs ({\ie} m$_{\rm r}$ $\geq$ 25 at {\it z} $\geq$ 2.0 for IFRSs corresponds to M$_{\rm r}$ $>$ -21) 
with lower radio luminosities (L$_{\rm 1.4~GHz}$ $\sim$ 10$^{24}$ $-$ 10$^{25}$ W Hz$^{-1}$) can also be of FR II type radio galaxies. 
Indeed, the double-lobe radio morphologies seen in some of our IFRSs suggest them to be FR II like radio galaxies. 
In fact, \cite{Garn08} have shown that IFRSs can be modelled as less luminous FR II radio galaxies at high redshifts.  
\par
To understand the nature of our IFRSs we also plot the radio size versus radio luminosity (see Fig.~\ref{fig:RadioSizeVsLumin}). 
As mentioned earlier that the IFRSs can be grouped into two categories namely compact sources with the upper limit on their radio sizes 
ranging from $\leq$ 10 kpc to $\leq$ 50 kpc, and the extended sources with the linear projected size ranging 
from 96 kpc to 330 kpc. 
From Fig.~\ref{fig:RadioSizeVsLumin}, it is evident that all the extended sources are radio powerful 
with L$_{\rm 1.4~GHz}$ $=$ 1.8 $\times$ 10$^{25}$ W Hz$^{-1}$ -- 2.7 $\times$ 10$^{27}$ W Hz$^{-1}$, 
while the compact sources are distributed across moderate to high radio luminosities 
(L$_{\rm 1.4~GHz}$ $=$ 1.2 $\times$ 10$^{24}$ W Hz$^{-1}$ to 7.9 $\times$ 10$^{26}$ W Hz$^{-1}$). 
This suggests that the IFRSs population consists of powerful double-lobe radio galaxies as well as less powerful compact radio-loud AGN 
(possibly young radio galaxies).
\subsection{Radio spectra}
\begin{figure*}
\includegraphics[angle=0,width=8.5cm,trim={0.0cm 0.0cm 0.0cm 0.0cm},clip]{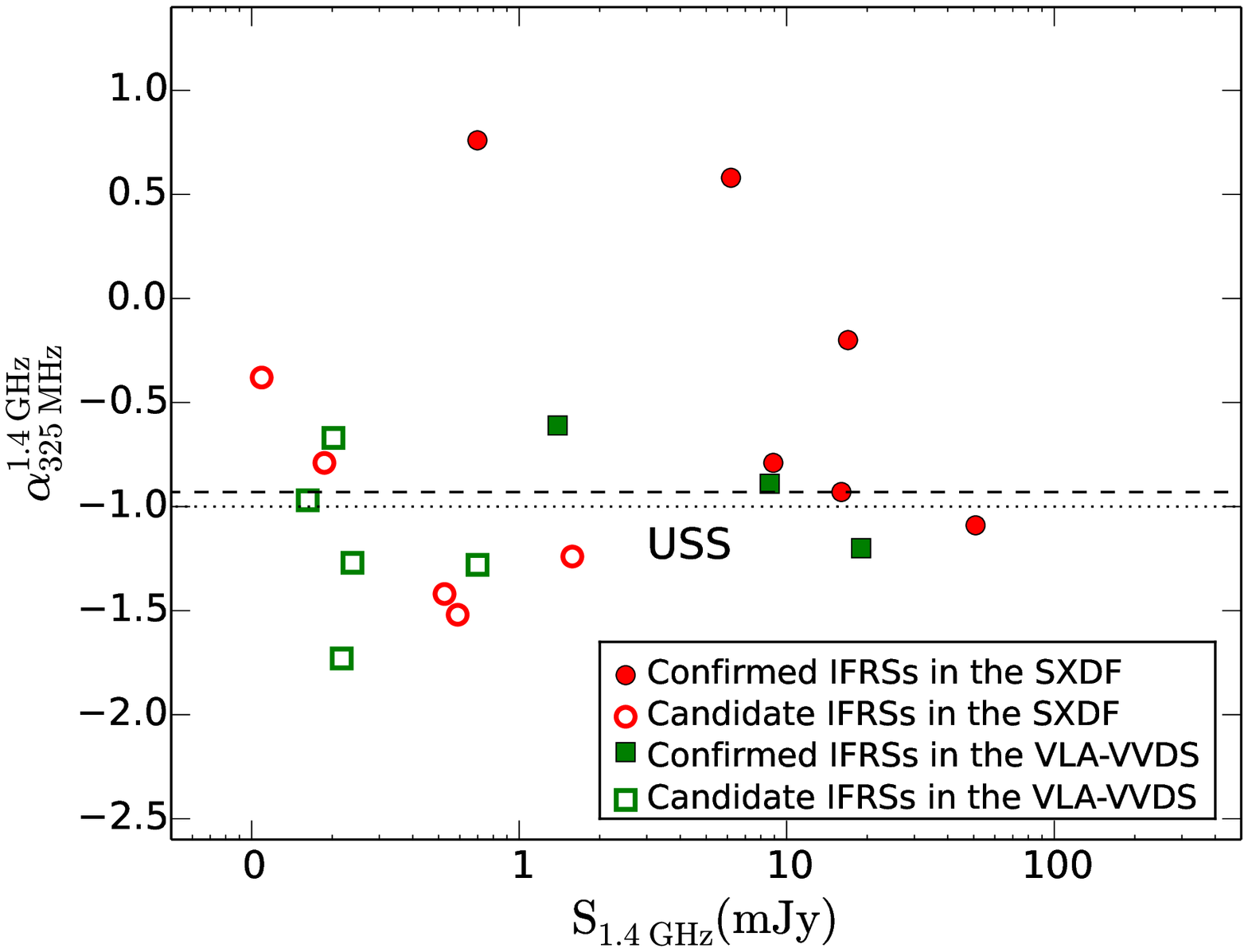}
{\includegraphics[angle=0,width=8.5cm,trim={0.0cm 0.0cm 0.0cm 0.0cm},clip]{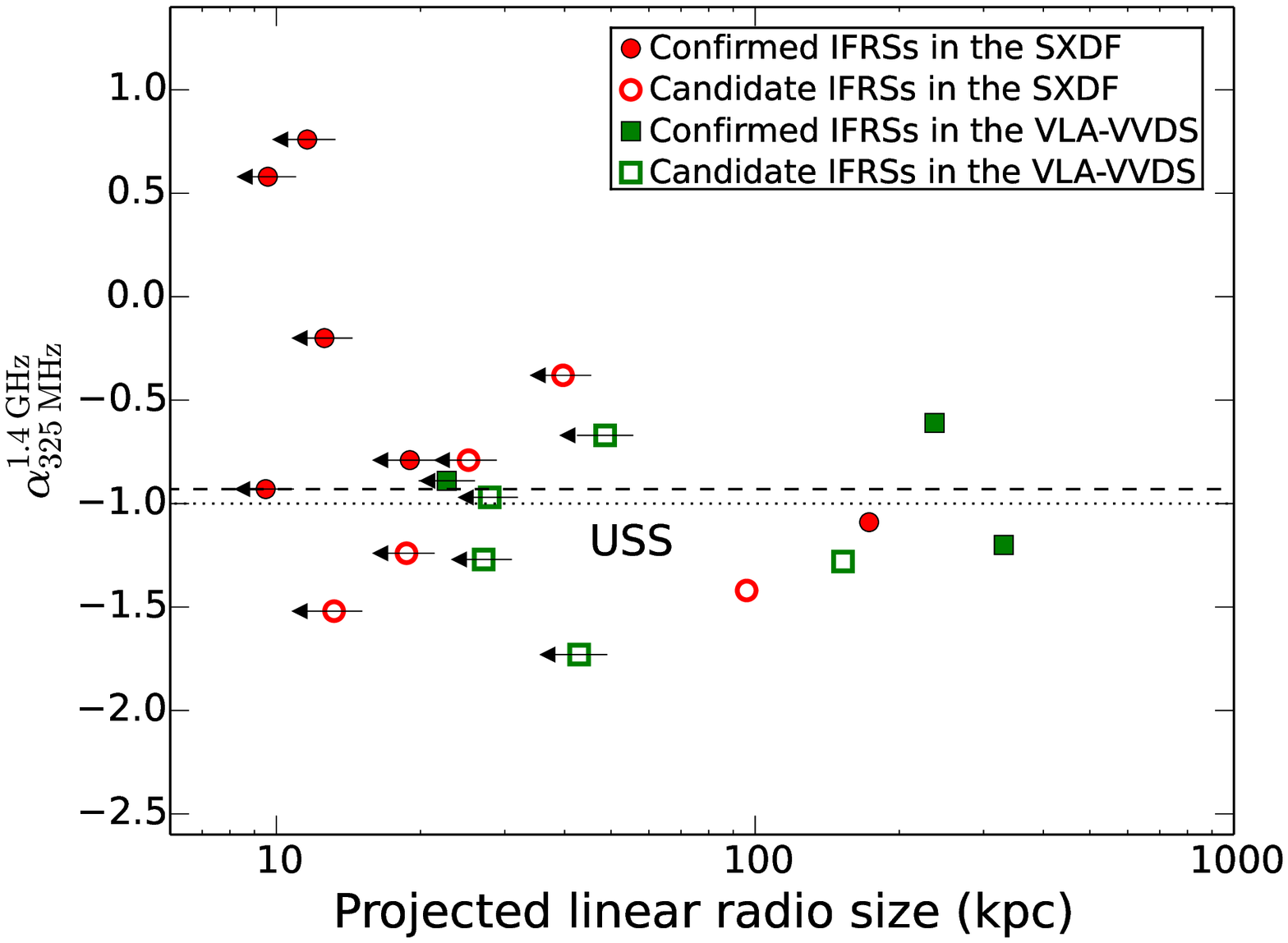}}
\caption{{\it Left panel}:  ${\alpha}_{\rm 325~MHz}^{\rm 1.4~GHz}$ versus S$_{\rm 1.4~GHz}$ plot. 
{\it Right panel} :  ${\alpha}_{\rm 325~MHz}^{\rm 1.4~GHz}$ versus radio size plot.  
Error bars are small and are of the order of the sizes of symbols. The horizontal dashed and dotted lines 
represent the median value (${\alpha}_{\rm 325~MHz}^{\rm 1.4~GHz}$ = -0.93) of the spectral index distribution 
and the USS limit (${\alpha}_{\rm 325~MHz}^{\rm 1.4~GHz}$ = -1.0), respectively.}
\label{fig:SpectralIndexHist} 
\end{figure*}

To examine the nature of radio spectra of our IFRSs we measure spectral index between 1.4 GHz and 325 MHz (see Table~\ref{table:IFRSSample}). 
In the SXDF we are limited to two point (1.4 GHz -- 325 MHz) spectral index, while in the VLA-VVDS 
we obtain three point (1.4 GHz -- 610 MHz -- 325 MHz) spectral index using the least square fit method. 
The resolutions of 1.4 GHz VLA, 610 MHz GMRT, and 325 MHz GMRT observations are not too different 
{\ie}6$\arcsec$.0, 6$\arcsec$.0, and 9$\arcsec$.0, respectively, and therefore, our 
spectral index estimates are unlikely to be affected by the resolution bias. 
For our full sample, the radio spectral indices span over +0.76 to -1.73 with the median of -0.93 (see Fig.~\ref{fig:SpectralIndexHist}). 
The radio spectral indices (${\alpha}_{\rm 325~MHz}^{\rm 1.4~GHz}$) of our confirmed IFRSs range from +0.76 to -1.2 with the 
median of -0.79. While the spectral indices of the candidate IFRSs are distributed over -0.38 to -1.73 with the median of -1.24. 
From S$_{\rm 1.4~GHz}$ versus ${\alpha}_{\rm 325~MHz}^{\rm 1.4~GHz}$ plot  (see Fig.~\ref{fig:SpectralIndexHist}, left panel) 
it is evident that the candidate IFRSs are not only systematically fainter but also exhibit systematically steeper spectra. 
Six out of eleven candidate IFRSs can be classified as Ultra Steep Spectrum (USS) radio 
sources ({\ie}${\alpha}_{\rm 325~MHz}^{\rm 1.4~GHz}$ $\geq$ -1.0). While only two out of nine confirmed IFRSs fall into the USS category. 
In general, USS sources are believed to be high-{\it z} sources owing to the {\it z} -- $\alpha$ correlation \citep{Miley08,Ker12}. 
In fact, the {\it z} -- $\alpha$ correlation is used to search for high--{\it z} radio galaxies \citep[see][]{Ishwara-Chandra10,Singh14}. 
Therefore, given the {\it z} -- $\alpha$ correlation all the candidate IFRSs with steep radio spectra are likely to be high-{\it z} sources. 
Indeed, all the USS candidate IFRSs in the SXDF have redshifts over 1.68 -- 4.32. 
Therefore, the assumption of candidate IFRSs with no IR counterparts being at $z$ $\geq$ 2.0 (see Section~\ref{sec:Redshifts}) is further supported 
by the fact that they are USS sources. 
However, we caution that all USS sources are not guaranteed to be high-{\it z} sources and vice-versa \citep{Afonso11}. \\
Apart from the IFRSs with steep spectra there are two confirmed IFRSs in the SXDF 
with positive spectral indices (${\alpha}_{\rm 325~MHz}^{\rm 1.4~GHz}$ $=$ 0.58 and 0.76) and these 
may possibly be Giga-hertz Peaked Spectrum (GPS) sources. 
The GPS radio sources show a turnover in their radio spectra at 500 MHz or higher frequencies 
in the rest-frame \citep{Fanti09}. 
Both the IFRSs with positive spectral indices between 325 MHz and 1.4 GHz can have 
turnover frequency at $\geq$ 1.4 GHz in the observed-frame, that corresponds to $\geq$ 357 MHz 
and $\geq$ 304 MHz in their respective rest-frames at redshifts 2.92 and 3.6, respectively. 
Moreover, we need high-frequency ($\geq$ 1.4 GHz) radio observations to confirm if these IFRSs are GPS sources.
\par      
It is interesting to note that the radio spectral indices of both the confirmed as well as candidate IFRSs are widely distributed 
and deviate from the typical value of spectral index ($\alpha$ $\sim$ -0.7) for the general radio population 
(see figure~\ref{fig:SpectralIndexHist}). 
This suggests that our IFRSs are unlikely to consist of SFGs and/or radio-quiet AGN.  
The distribution of radio spectral indices of our sample IFRSs is in agreement with the results from previous studies \citep{Middelberg11,Herzog16} 
which found IFRSs exhibiting primarily steep spectra. 
For instance, \cite{Middelberg11} studied the radio spectra over 1.4 GHz to 8.4 GHz for 18 IFRSs and 
reported that the IFRSs spectra at higher frequencies are remarkably steep with the median 
of ${\alpha}_{\rm 1.4~GHz}^{\rm 2.4~GHz}$ $\sim$ -1.4, and lack  sources with spectral indices 
(${\alpha}_{\rm 1.4~GHz}^{\rm 2.4~GHz}$) higher than -0.7. 
The steeper high-frequency radio spectra of IFRSs reported in \cite{Middelberg11}, 
in comparison to the relatively less steep low-frequency radio spectra of our IFRSs, can be explained if IFRSs spectra 
steepen towards higher frequencies.  
Indeed, based on the comparison of the resolution-matched spectral indices between 1.4 GHz and 2.4 GHz (${\alpha}_{\rm 1.4~GHz}^{\rm 2.4~GHz}$) 
and between 4.8 GHz and 8.6 GHz (${\alpha}_{\rm 4.8~GHz}^{\rm 8.6~GHz}$), \cite{Middelberg11} reported that 
the radio spectra of IFRSs generally steepen towards higher frequencies. 
However, \cite{Herzog16} reported that the radio spectra of relatively bright IFRSs 
can be characterised by a single power law over a large frequency range of 150~MHz -- 34~GHz, and, 
on average, IFRSs show steep spectra ($\alpha$ $<$ -0.8), but also include GPS sources exhibiting turnover at $\sim$ 1 GHz. 
\par
Fig.~\ref{fig:SpectralIndexHist} (right panel) shows that the powerful extended sources in our sample 
exhibit steep spectral indices which is one of the 
characteristic features of H$z$RGs. 
While, the compact sources in our sample are widely distributed in spectral indices and include steep, flat and inverted spectra, 
which can be characterised as Compact Steep Spectrum (CSS) sources and GPS sources. 
Both CSS and GPS sources are believed to represent the initial phase of the evolutionary path of radio galaxies \citep{Tinti06,Fanti09}. 
Thus, we find that our IFRSs constitute a diverse population of radio-loud AGN widely distributed over 
radio luminosities, radio-sizes and spectral indices.  
Therefore, it seems likely that our IFRSs are H$z$RGs in different evolutionary phases.   
\section{Multiwavelength counterparts of our IFRSs}
\label{sec:MWCounterparts}
To understand the nature of our IFRSs we also attempt to find their multiwavelength 
({\ie}optical, NIR, MIR, FIR and X-ray) counterparts using the existing deep data in both the fields. 
The counterparts of our IFRSs in different bands are searched by considering the closest match within a circle 
centred at the radio position and having radius equal to the one-third of the larger beam size in the two cross-matching bands.
The search radii used to find counterparts in different bands are similar to those derived from the cross-matching of 
large samples at different wavelengths \citep[\eg][]{Ivison07}. 
We note that our IFRSs are the subset of radio sources detected in the 1.4 GHz VLA surveys in both the fields. The optical counterparts 
for all 1.4 GHz radio sources are searched using likelihood ratio method, and only reliable optical counterparts are considered 
{\ie}there is $\geq$ 90 per cent probability that the identified optical counterpart is associated with the radio source 
\citep[see][]{Simpson06,McAlpine12}. In order to estimate the fraction of chance matches we shifted the positions of radio 
sources by 30${\arcsec}$ to 45${\arcsec}$ in random directions, and thereafter, cross-matched the radio and optical catalogues using the previously opted search radius. 
The cross-matching of radio sources with shifted positions and optical sources resulted only 2 -- 4 per cent of the total radio sources 
\citep[see][]{Singh14}. Therefore, the probability of an optical counterpart being merely a chance match is limited only to a few per cent. 
The lower surface source density in other bands ({\ie}NIR, MIR, FIR, and X-ray) results in an even lower fraction of chance matches.
\\
Table~\ref{table:MWCounterpart} lists the magnitudes/fluxes of the multiwavelength counterparts of our IFRSs. 
We note that the detection rate is highest in the optical band in which all but two IFRSs show counterparts. 
The r-band magnitudes of our sample IFRSs are widely distributed and range from 23.5 to 28.6 with the median of 26. 
The deep optical data available in both the fields allow us to obtain much higher detection rate  
than that reported in previous studies \cite[{\eg}][]{Garn08,Collier14}.
We note that the radio-bright IFRSs are also bright in the optical and NIR bands.     
Moreover, both radio-bright and radio-faint IFRSs are present at similar redshifts suggesting that the IFRSs constitute a 
diverse population. 

\begin{figure*}
\includegraphics[angle=0,width=8.5cm,trim={0.0cm 0.0cm 0.0cm 0.0cm},clip]{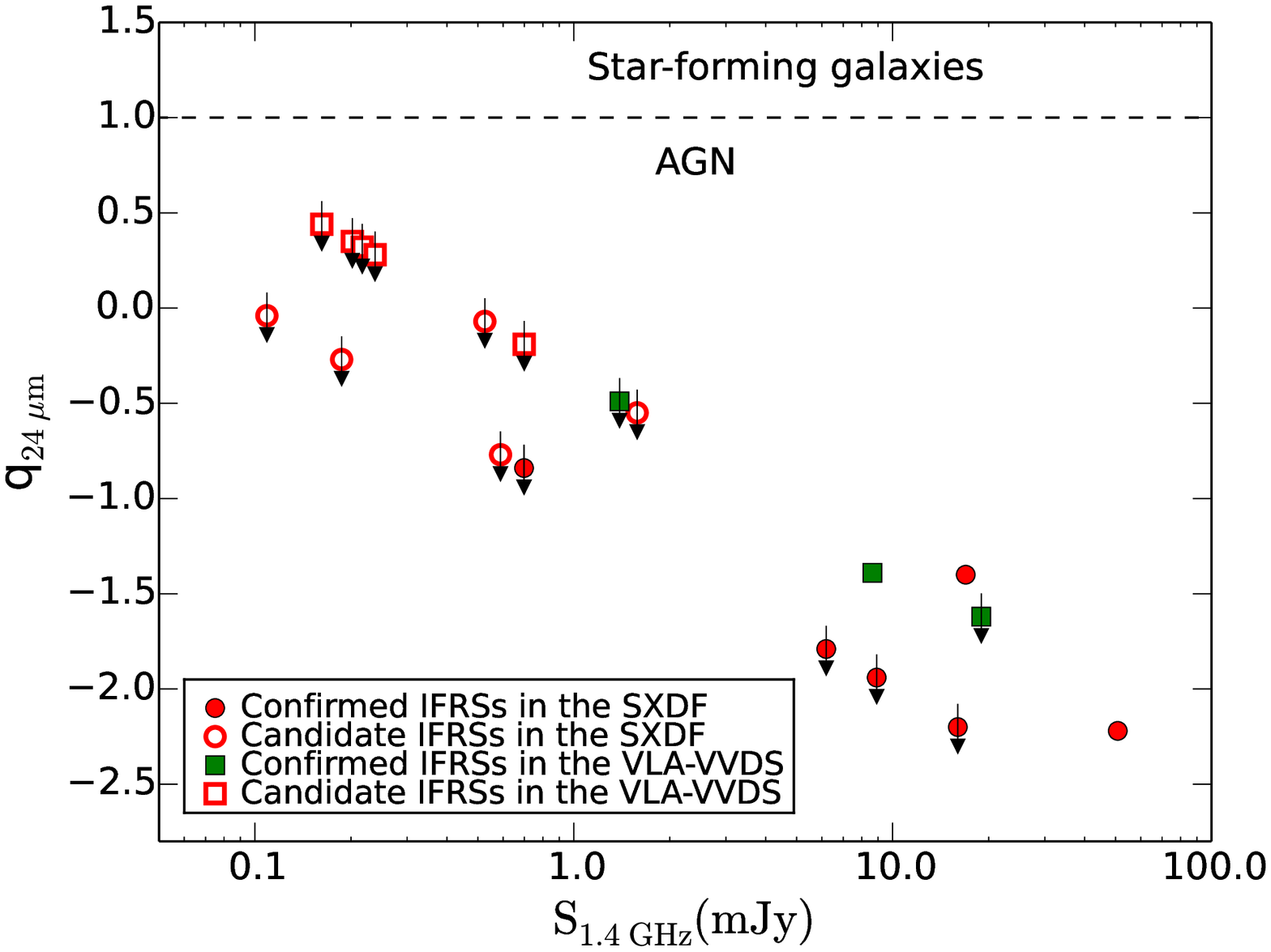}
{\includegraphics[angle=0,width=8.5cm,trim={0.0cm 0.0cm 0.0cm 0.0cm},clip]{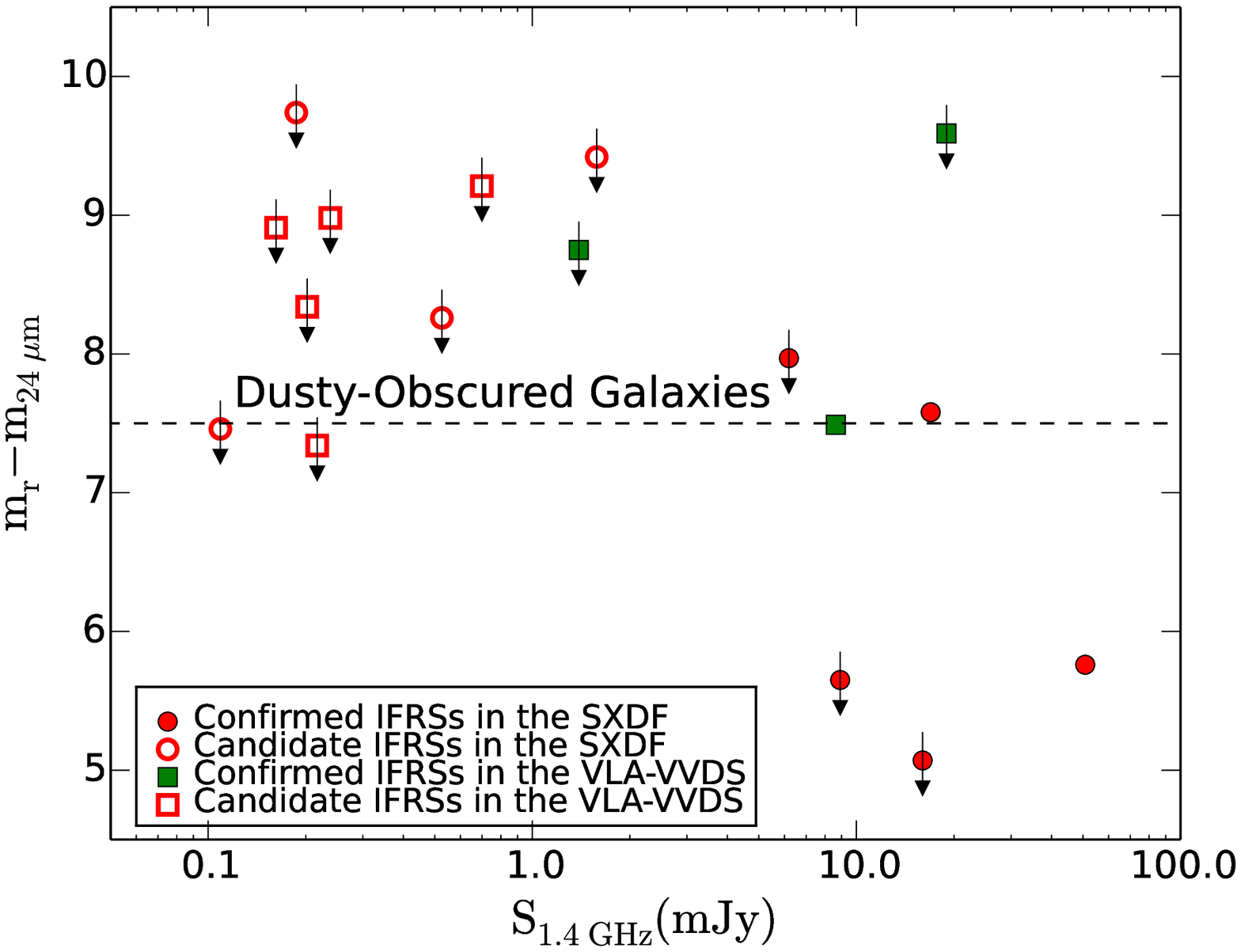}}
\caption{{\it Left panel} : q$_{\rm 24}$ = log($\frac{\rm S_{24~{\mu}m}}{\rm S_{1.4~GHz}}$) versus S$_{\rm 1.4~GHz}$ plot. 
{\it Right panel} :  Optical-to-24 $\mu$m colour (m$_{\rm r}$ - m$_{\rm 24~{\mu}m}$)versus S$_{\rm 1.4~GHz}$ plot.}
\label{fig:RadioVsq} 
\end{figure*}

\begin{table*}
\hspace{-2.5cm}
\centering
\begin{minipage}{140mm}
\caption{Multiwavelength counterparts of our IFRSs}
\scalebox{0.9}{
\begin{tabular}{@{}ccccccccc@{}}
\hline
  RA          &   DEC          &   m$_{\rm R}$   &  m$_{\rm K}$     & S$_{\rm 24~{\mu}m}$ & m$_{\rm R}$ - m$_{\rm 24 {\mu}m}$ & q$_{\rm 24~{\mu}m}$     &  S$_{\rm 250~{\mu}m}$ & S$_{\rm 2.0~-~10~keV}$ \\
 (h m s)      &   (d m s)      &                 &                  &      (mJy)          &             &               & (mJy)           & (ergs s$^{-1}$ cm$^{-2}$) \\   \hline  
   SXDF       &                &                 &                  &                     &             &               &               &                           \\
  02 18 39.55 & -04 41 49.4    & 23.45$\pm$0.01  & 18.98$\pm$0.01   & 0.306$\pm$0.023     &  5.76       &  -2.22        & ...           &        ...                 \\  
  02 17 52.12 & -05 05 22.4    & 26.87$\pm$0.10  & 21.34$\pm$0.03   &  $<$  0.1           & $<$ 7.97    & $<$ -1.79     & ...           &  6.38$\pm$2.06 $\times$ 10$^{-15}$ \\
  02 18 53.63 & -04 47 35.6    & 24.42$\pm$0.02  & 19.26$\pm$0.01   & 0.671$\pm$0.026     &  7.57       &  -1.40        &  13.2$\pm$1.3 &        ...                \\
  02 18 51.38 & -05 09 01.6    & 23.97$\pm$0.01  & 18.65$\pm$0.01   &  $<$  0.1           & $<$ 5.07    & $<$ -2.20     & ...           &        ...                \\
  02 18 03.41 & -05 38 25.5    & 24.55$\pm$0.02  & $>$ 15.85        &  $<$ 0.1            & $<$ 5.65    & $<$ -1.95     &               &                           \\
  02 18 38.24 & -05 34 44.2    & 26.69$\pm$0.10  & $>$ 15.85        &  $<$ 0.45           & $<$ 9.45    & $<$ -0.55     &               &                            \\ 
  02 17 40.69 & -04 51 57.3    & 25.53$\pm$0.03  & 22.32$\pm$0.06   &  $<$ 0.45           & $<$ 8.26    & $<$ -0.07     & ...           &        ...                \\   
  02 17 45.84 & -05 00 56.4    &  $>$ 27.7       & $>$ 25.3         &   $<$ 0.1           & $<$ 8.80    & $<$ -0.77     & ...           &        ...                \\  
  02 18 01.23 & -04 42 00.8    & 26.36$\pm$0.06  & 22.10$\pm$0.07   &   $<$ 0.1           & $<$ 7.46    & $<$ -0.04     &  ...          &        ...                \\  
  02 18 30.13 & -05 17 17.4    & 28.64$\pm$0.15  & $>$ 25.3         &   $<$ 0.1           & $<$ 9.74    & $<$ -0.27     & ...           &        ...                \\  
  02 18 59.19 & -05 08 37.8    &  $>$ 27.7       & $>$ 25.3         &   $<$ 0.1           & $<$ 8.8     & $<$ -0.84     & ...           &        ...                 \\
  VLA-VVDS    &                &                 &                  &                     &             &               &               &                           \\
  02 27 48.26 & -04 19 05.3    & 26.18$\pm$0.18  &  $>$ 23.5        &    $<$ 0.45         & $<$ 8.91    & $<$ 0.44      & ...           &     ...          \\  
  02 25 02.13 & -04 40 26.9    & 25.61$\pm$0.10  & $>$ 23.5         &    $<$ 0.45         & $<$ 8.34    & $<$ 0.35      & ...           &     ...          \\
  02 26 58.10 & -04 18 14.9    & 24.61$\pm$0.04  & $>$ 23.5         &    $<$ 0.45         & $<$ 7.34    & $<$ 0.32      &  ...          &     ...          \\  
  02 27 09.90 & -04 23 44.8    & 26.25$\pm$0.21  & $>$ 23.5         &    $<$ 0.45         & $<$ 8.98    & $<$ 0.27      & ...           &     ...           \\  
  02 26 31.12 & -04 24 53.3    & 26.48$\pm$0.24  & $>$ 23.5         &    $<$ 0.45         & $<$ 9.21    & $<$ 0.19      & ...           &     ...          \\
  02 25 26.14 & -04 34 54.4    & 26.02$\pm$0.14  & $>$ 23.5         &    $<$ 0.45         & $<$  8.57   & $<$ -0.49     &  ...          &     ...          \\ 
  02 26 09.09 & -04 33 34.7    & 25.04$\pm$0.25  & 21.80$\pm$0.09   &  0.347$\pm$0.022    &  7.49       &  -1.39        & ...           &     ...          \\  
  02 24 20.96 & -04 25 44.6    & 26.86$\pm$0.32  & $>$ 23.5         &    $<$ 0.45         & $<$ 9.59    & $<$ -1.62     &  ...          &     ...          \\  
 \hline
\end{tabular}}
\label{table:MWCounterpart} 
\\
Notes - In the VLA-VVDS field the r-band magnitudes are from the CFHTLS-D1. 
The K-band magnitudes in the SXDF are based on the UDS DR 11 and the lower limits are set equal 
to the median depth (5$\sigma$) of the UDS DR 11. 
Two IFRSs in the SXDF lie outside UDS footprint and the lower limits (m$_{\rm K}$ $>$ 15.85) are based on the 2MASS. 
The 24 $\mu$m counterparts are searched from the SpUDS and SWIRE surveys, while 250 $\mu$m fluxes are based on the HerMES level 4 data. 
The X-ray counterparts are searched from the SXDS and XMDS surveys (see Table~\ref{table:MWData}).    
\end{minipage}
\end{table*} 
In 24 $\mu$m MIR band, only three IFRSs show counterparts and for rest of the sources we put an upper limit based on the 
sensitivity limit of SpUDS (5$\sigma$ $=$ 0.1 mJy beam$^{-1}$) and SWIRE (5$\sigma$ $=$ 0.45 mJy beam$^{-1}$) surveys. 
To get further insight into the nature of our IFRSs we estimate the ratio of 24 $\mu$m MIR to 1.4 GHz radio flux densities, which is 
defined as q$_{\rm 24~{\mu}m}$ = log$_{\rm 10}$($\frac{\rm S_{24~{\mu}m}}{\rm S_{1.4~GHz}}$).   
The q$_{\rm 24~{\mu}m}$ parameter is conventionally used to segregate the populations of AGN and SFGs 
owing to the fact that SFGs exhibit tight correlation between MIR and radio emission, while AGN tend to deviate 
due their radio excess \citep{Appleton04}. 
We find that q$_{\rm 24~{\mu}m}$ values of our IFRSs are in the range of $\leq$ 0.44 to -2.22 and show a large radio excess from the typical value of q$_{\rm 24~{\mu}m}$ $\simeq$ 1.0 for SFGs, and thus, reinforcing 
the conclusion that IFRSs belong to the AGN population.    
Given the non-detection of 24 $\mu$m counterparts for most of our IFRSs, in particular, in the radio-faint regime it is unclear 
if q$_{\rm 24~{\mu}m}$ values for radio-faint and radio-bright IFRSs are similar. 
The confirmed IFRSs clearly stand out with much lower values/limits of q$_{\rm 24~{\mu}m}$ than that for candidate IFRSs 
(see Fig.~\ref{fig:RadioVsq}, left panel). 
\\
We also investigate the optical-to-MIR colours of our sample sources to examine if our IFRSs are hosted in Dust-Obscured Galaxies (DOGs).
The DOGs at higher redshifts are faint in the UV and optical bands but become relatively bright in the MIR and FIR bands. 
In the literature, DOGs have been identified with colour cut of m$_{\rm r}$ - [24] $\geq$ 7.5 (AB mag) and 
S$_{\rm 24~{\mu}m}$ $>$ 100 $\mu$Jy \citep[see][]{Calanog13}. 
We obtain the optical-to-MIR colours (m$_{\rm r}$ - [24]) for our sample sources, and get the lower limits on the colours 
for sources that do not show detected counterparts at 24 $\mu$m. 
Using the optical-to-MIR colours (m$_{\rm r}$ - [24]) we find that two of three IFRSs that show 24 $\mu$m counterparts can be classified as DOGs 
(see Table~\ref{table:MWCounterpart}). From radio flux density versus optical-to-MIR colour plot 
(figure~\ref{fig:RadioVsq}, right panel) it is evident that a significant fraction of our IFRSs can belong to DOGs even if their 
24 $\mu$m counterparts are one-to-two order of magnitudes fainter than the SpUDS and SWIRE sensitivity limits at 5$\sigma$. 
However, due to the lack of detected counterparts in 24 $\mu$m band for majority of IFRSs we only have upper limits on 
the optical-to-MIR colours (m$_{\rm r}$ - [24]) which prevent us from obtaining the exact fraction of IFRSs hosted in DOGs, 
and the variation of optical-to-MIR colour across the redshift. 
We perform image stacking to get average 24 ${\mu}$m flux of the IFRSs population 
that remained undetected in 24 $\mu$m Spitzer surveys (SpUDS and SWIRE). 
The stacking method involves making the 24 $\mu$m image cut-outs of 60$\arcsec$ $\times$ 60$\arcsec$ centred at the radio positions of IFRSs 
and then obtaining a stacked image by combining all the image cut-outs on a pixel-by-pixel basis. We prefer to take the median rather than the mean, which can be affected by the presence of extreme outliers. 
The flux density is obtained from the value of the central pixel in the stacked image. 
For all our 16 IFRSs with no detected 24 $\mu$m counterparts, we stack 24 $\mu$m image cut-outs 
and find that the median stacked image does not show any detection with the rms noise of $\simeq$ 0.014 $\mu$Jy beam$^{-1}$. 
The non-detection in the stacked 24 $\mu$m image suggests that a significant fraction of IFRSs are fainter than the typical DOGs 
(S$_{\rm 24{\mu}m}$ $>$ 100 $\mu$Jy).    
\par
Since dusty galaxies are more probable to be detected in the FIR bands we also search for the FIR counterparts of our IFRSs 
using the HerMES data. 
We find that only one IFRS ({\ie}J021853-044735) is detected in the {\it Herschel}/SPIRE bands with the flux of 
13.2$\pm$1.3 mJy, 13.9$\pm$1.3 mJy and 6.2$\pm$1.9 mJy in the 250 $\mu$m, 350 $\mu$m, and 500 $\mu$m band, respectively. 
For all 18 IFRSs with no 250 $\mu$m counterparts we obtain median stacked image using the method similar to 
the one performed for 24 $\mu$m band. 
We find that the stacked 250 $\mu$m HerMES image does not show any detection with the rms noise of $\simeq$ 0.52 mJy beam$^{-1}$. 
The non-detections of all but one IFRSs in the {\it Herschel}/SPIRE bands may not be surprising 
owing to the fact that nearly half of DOGs population remained undetected in the 250 $\mu$m band at 3$\sigma$ level 
in the HerMES survey of the COSMOS deep field \citep[see][]{Calanog13}. 
As expected, the DOGs population undetected at 250 $\mu$m is on average relatively less luminous in the IR and has higher dust temperature.   
The non-detection of most of our sample IFRSs in the FIR bands is consistent with previous studies. 
For example, \cite{Herzog15b} found that all six IFRSs in their sample remained undetected in all five bands of {\it Herschel} 
(with stacking limits $\sigma$  = 0.74 mJy beam$^{-1}$, 2.68 mJy beam$^{-1}$ and 3.45 mJy beam$^{-1}$ 
at 100 $\mu$m, 250 $\mu$m, and 500 $\mu$m, respectively). 
By comparing the SED templates of different type of galaxies on to the upper limits of fluxes at 3.6 $\mu$m, 24 $\mu$m 
and FIR bands \cite{Herzog15b} concluded that the non-detections of radio-bright IFRSs (7 -- 25 mJy) can be only be explained if IFRSs are:
(i) H{\it z}RGs at very high redshifts ({\it z} $\geq$ 10.5), (ii) low-luminosity variants of H{\it z}RGs with additional 
dust obscuration at moderate redshifts, (iii) scaled or unscaled versions of Cygnus A at any redshift, and (iv) 
scaled and dust-obscured radio-loud quasars or CSS.
We further note that all our IFRSs are not likely to be hosted in DOGs. For instance, 
\citep{Zinn11} demonstrated that a fraction of IFRSs are possibly similar to 3C 48 placed at high redshifts 
and such IFRSs will fall below the detection limits of the SWIRE and the {\it Herschel} bands. 
We note that, based on the optical spectra, one of our IFRSs J021839-044149 is classified as a narrow line AGN, 
while two other IFRSs {\ie}J021853-044735 and J021803-053825 are also possible narrow line AGN (see Section~\ref{sec:Redshifts}). 
\par 
We also search for the X-ray counterparts of our IFRSs using the deepest available X-ray surveys {\ie} the SXDS and XMDS (see Table~\ref{table:MWData}). 
All but one IFRSs do not show X-ray counterparts. 
IFRS J021752-050522 is detected in the SXDS with the X-ray flux of 
S$_{\rm 0.5~-~2.0~keV}$ $\simeq$ 2.5$\pm$1.7 $\times$ 10$^{-16}$ ergs s$^{-1}$ cm$^{-2}$ and 
S$_{\rm 2.0~-~10~keV}$ $\simeq$ 6.38$\pm$2.06 $\times$ 10$^{-15}$ ergs s$^{-1}$ cm$^{-2}$; 
where fluxes are estimated from the count rates assuming photon index ($\Gamma$) $=$ 1.7 of the X-ray spectrum 
and conversion factors given in \cite{Ueda08}. 
The estimated X-ray luminosities in the soft and hard X-ray bands are 
L$_{\rm 0.5~-~2.0~keV}$ $=$ 1.87$\pm$1.27 $\times$ 10$^{43}$ ergs s$^{-1}$ and 
 L$_{\rm 2.0~-~10~keV}$ $=$ 4.77$\pm$1.54 $\times$ 10$^{44}$ ergs s$^{-1}$, respectively. 
The hard X-ray luminosity is similar to the X-ray bright quasars in the local Universe \citep{Ballo14}. 
The non-detection of most of our sample IFRSs is consistent with the previous studies \citep[{\eg}][]{Zinn11}. 
For example, \cite{Huynh10} used {\it Chandra} 2Ms source catalogue and did not find the X-ray counterpart of any IFRS 
in the CDFS field. 
The non-detection of our IFRSs in the X-ray band further reinforces the notion that IFRSs are hosted in dusty 
galaxies and the X-ray emission is absorbed by the dust present in the host galaxy.
In fact, it is proposed that IFRSs can be part of the missing population of Compton-thick AGN predicted by the models of 
cosmic X-ray background emission \citep[see][]{Zinn11}.
\section{Summary and Conclusions}
\label{sec:Summary}

In this paper, we identify and investigate the nature of IFRSs in the
SXDF and VLA-VVDS field using one of the deepest set of
multiwavelength data currently available. The key results of our study
are outlined below.

\begin{itemize}
\item 
With the selection criteria of : (i) $\frac{\rm S_{1.4~GHz}}{\rm S_{3.6~{\mu}m}}$ $\geq$ 500, and (ii) 
S$_{\rm 3.6~{\mu}m}$ $\leq$ 30 ${\mu}$Jy, we identify nine confirmed IFRSs (six in the SXDF and three in the VLA-VVDS) 
over 1.8 deg$^{2}$ in the two deep fields. 
We also identify ten candidate IFRSs (five in each of the two fields) 
with the selection criteria of : (i) a radio source with no 3.6 $\mu$m counterpart, 
and (ii) the lower limit on $\frac{\rm S_{1.4~GHz}}{\rm S_{3.6~{\mu}m}}$ is less than 500. 
For candidate IFRSs, the visual inspection of 3.6 $\mu$m image cut-outs centred at the radio positions does 
not show any emission at the faintest level, thus inferring them to be true IFRSs if their 
3.6 $\mu$m fluxes are below 1$\sigma$ detection limit. 
\item 
The availability of deep radio (S$_{\rm 1.4~GHz}$ $\sim$ 80 -- 100 $\mu$Jy beam$^{-1}$ at 5$\sigma$), 
optical (m$_{\rm r}$ $\sim$ 26 -- 27.7 at 5$\sigma$), 
and NIR (S$_{\rm 3.6~{\mu}m}$ $\sim$ 1.3 -- 2.0 $\mu$Jy beam$^{-1}$ at 5$\sigma$) data allow us to find IFRSs in the faintest flux regime. 
Also, in comparison to previous studies, we obtain the highest optical identification rate of IFRSs in the SXDF.
\item 
In the SXDF, all our IFRSs have spectroscopic or photometric redshifts that are distributed over $z$ = 1.68 -- 4.3. 
In the VLA-VVDS, all but one (at $z$ = 2.45) IFRSs have only a lower limit $z$ $>$ 2.0, on their redshifts. 
We emphasise that by using deep multiwavelength data our study reveals, for the first time, IFRSs at $z$ $>$ 3.0. 
Also, in comparison to previous studies, we find redshift estimates for IFRSs lying in 
the faintest 3.6 ${\mu}$m flux regime. For instance, 
an IFRS with spectroscopically measured redshift ($z$ = 2.43) at S$_{\rm 3.6~{\mu}m}$ $=$ 13.6 $\mu$Jy, 
and an IFRS with photometrically measured redshift ($z$ = 4.3) at S$_{\rm 3.6~{\mu}m}$ $<$ 1.3 $\mu$Jy. 
Hitherto, all the previous attempts of measuring the redshifts of IFRSs were limited to the relatively bright flux regime 
(S$_{\rm 3.6~{\mu}m}$ $=$ 20 -- 30 $\mu$Jy) and at lower redshifts ($z$ $\leq$ 2.99).
%
\item 
The 1.4 GHz VLA radio images show that 14/19 sources in our sample are unresolved point sources, 
while five sources exhibit extended double-lobe morphology. Therefore, 
radio morphology itself reveals that at least a fraction of IFRSs are typical radio galaxies. 
The upper limits on the radio sizes of unresolved sources range from $\leq$ 10 kpc to $\leq$ 50 kpc, 
while the projected linear radio sizes of the extended sources range from 96 kpc to 330 kpc. 
\item 
The radio luminosity distribution of our IFRSs suggests them to be radio-loud AGN with 1.4 GHz radio luminosity spanning from  
1.2 $\times$ 10$^{24}$ W Hz$^{-1}$ to 2.7 $\times$ 10$^{27}$ W Hz$^{-1}$ with the median of 3.8 $\times$ 10$^{25}$ W Hz$^{-1}$. 
We note that high radio luminosities (L$_{\rm 1.4~GHz}$) $\geq$ 10$^{24}$ W Hz$^{-1}$ cannot be 
powered by star-formation or star-burst alone. 
Also, all IFRSs with extended radio morphology show high radio luminosities 
(L$_{\rm 1.4~GHz}$ $\simeq$ 1.8 $\times$ 10$^{25}$ W Hz$^{-1}$ -- 2.7 $\times$ 10$^{27}$ W Hz$^{-1}$) that are typical of 
FR-II radio galaxies. On the other hand, IFRSs having compact radio morphologies are found to be widely distributed from moderate 
(L$_{\rm 1.4~GHz}$ $=$ 1.2 $\times$ 10$^{24}$ W Hz$^{-1}$) to high radio luminosity (L$_{\rm 1.4~GHz}$ $=$ 7.9 $\times$ 10$^{26}$ W Hz$^{-1}$). 
\item 
The radio spectral indices measured between 1.4 GHz to 325 MHz (${\alpha}_{\rm 325~MHz}^{\rm 1.4~MHz}$) for our IFRSs 
predominately show steep spectra with the median spectral index of -0.93, however, our sample also contains sources with flat/inverted spectra. 
In comparison to the confirmed IFRSs (median ${\alpha}_{\rm 325~MHz}^{\rm 1.4~MHz}$ $=$ -0.79), the 
candidate IFRSs show systematically steeper spectra (median ${\alpha}_{\rm 325~MHz}^{\rm 1.4~MHz}$ $=$ -1.24) 
which suggests them to be at higher redshifts owing to the ${z}$--${\alpha}$ empirical correlation. 
The systematically steep radio spectra of our IFRSs also indicate that deeper low-frequency radio surveys can be more useful in 
unveiling IFRSs at higher redshifts.   
\item 
Using the optical-to-MIR colours (m$_{\rm r}$ - m$_{\rm 24~{\mu}m}$) we find that two out of three IFRSs detected in the 
24 ${\mu}$m band are hosted in DOGs, while upper limits on the optical-to-MIR colours for rest of the IFRSs suggest 
that a large fraction of IFRSs is likely to be hosted in DOGs, even if their 24 ${\mu}$m fluxes are one-to-two order of magnitude lower than 
their upper limits. 
The non-detection of all but one IFRSs in the deep X-ray surveys further supports the possibility of IFRSs  
being AGN hosted in an obscured environment.
\item
The IFRSs in our sample are widely distributed in a variety of parameters {\ie}optical magnitude, IR flux, radio flux density, 
radio size, radio power, radio spectral index, and redshift. 
Therefore, it is likely that our IFRSs consist of 
a diverse population of radio-loud AGN ranging from powerful radio galaxies to less powerful radio AGN.   
\end{itemize}
Our study based on the two small-area deep fields also acts as a test-bed to explore large population of IFRSs by using 
data from upcoming large-area deep optical and IR surveys {\eg}the Large Synoptic Survey Telescope (LSST; \citealt{Ivezic08}), 
the James Webb Space Telescope (JWST; \citealt{Gardner06}), and the deep radio surveys 
from the upgraded GMRT (uGMRT), the Evolutionary Map of the Universe (EMU; \citealt{Norris11b}) and the Square Kilometre Array 
(SKA; \citealt{Dwedney09}).       
\section*{Acknowledgements}
We thank the two anonymous referees for useful comments and suggestions that helped in improving the quality of this publication. 
We also thank Dr. Aveek Sarkar for proof-reading the manuscript. 
VS acknowledges the support from Square Kilometre Array (SKA) consortium during his tenure at UKZN, South Africa. 
YW thanks IUCAA for hosting him on sabbatical when a substantial part of this work was completed. 
YW and AB acknowledge support from the Indo-French Center for the Promotion of Advanced Research 
(Centre Franco-Indien pour la Promotion de la Recherche Avanc$\acute{e}$e under program No. 4404-3 
during which 325 MHz GMRT observations of the XMM-LSS were carried out. 
We thank the staff of GMRT who have made these observations possible. GMRT
is run by the National Centre for Radio Astrophysics of the Tata Institute of
Fundamental Research. We thank Marco Bondi for providing 1.4 GHz and
610 MHz radio images of the VLA-VVDS field. We thank Chris Simpson for providing
1.4 GHz VLA radio image of the SXDF field. This work
is based on observations made with the {\it Spitzer} Space Telescope, which is operated
by the Jet Propulsion Laboratory (JPL), California Institute of Technology
(Caltech), under a contract with NASA. This work used the CFHTLS data products,
which are based on observations obtained with MegaPrime/MegaCam, a
joint project of CFHT and CEA/DAPNIA, at the CFHT which is operated by the
National Research Council (NRC) of Canada, the Institut National des Sciences
de l$^{\prime}$Univers of the Centre National de la Recherche Scientifique (CNRS) of
France, and the University of Hawaii. This work is based in part on data products
produced at TERAPIX and the Canadian Astronomy Data Centre as part of
the CFHTLS, a collaborative project of NRC and CNRS. This research uses data
from the VIMOS VLT Deep Survey, obtained from the VVDS database operated
by Cesam, Laboratoire d$^{\prime}$Astrophysique de Marseille, France.
This research has made use of data from HerMES project (http://hermes.sussex.ac.uk/). 
HerMES is a Herschel Key Programme utilising Guaranteed Time from the SPIRE instrument team, ESAC scientists 
and a mission scientist. The HerMES data was accessed through the {\it Herschel} Database in Marseille 
(HeDaM - http://hedam.lam.fr) operated by CeSAM and hosted by the Laboratoire d'Astrophysique de Marseille.
\\
\bibliographystyle{mnras}
\bibliography{RadioXMMLSS}
\appendix
\section{IFRSs images}

\begin{figure*}
\includegraphics[angle=0,width=8.0cm,trim={2.0cm 1.75cm 2.5cm 1.5cm},clip]{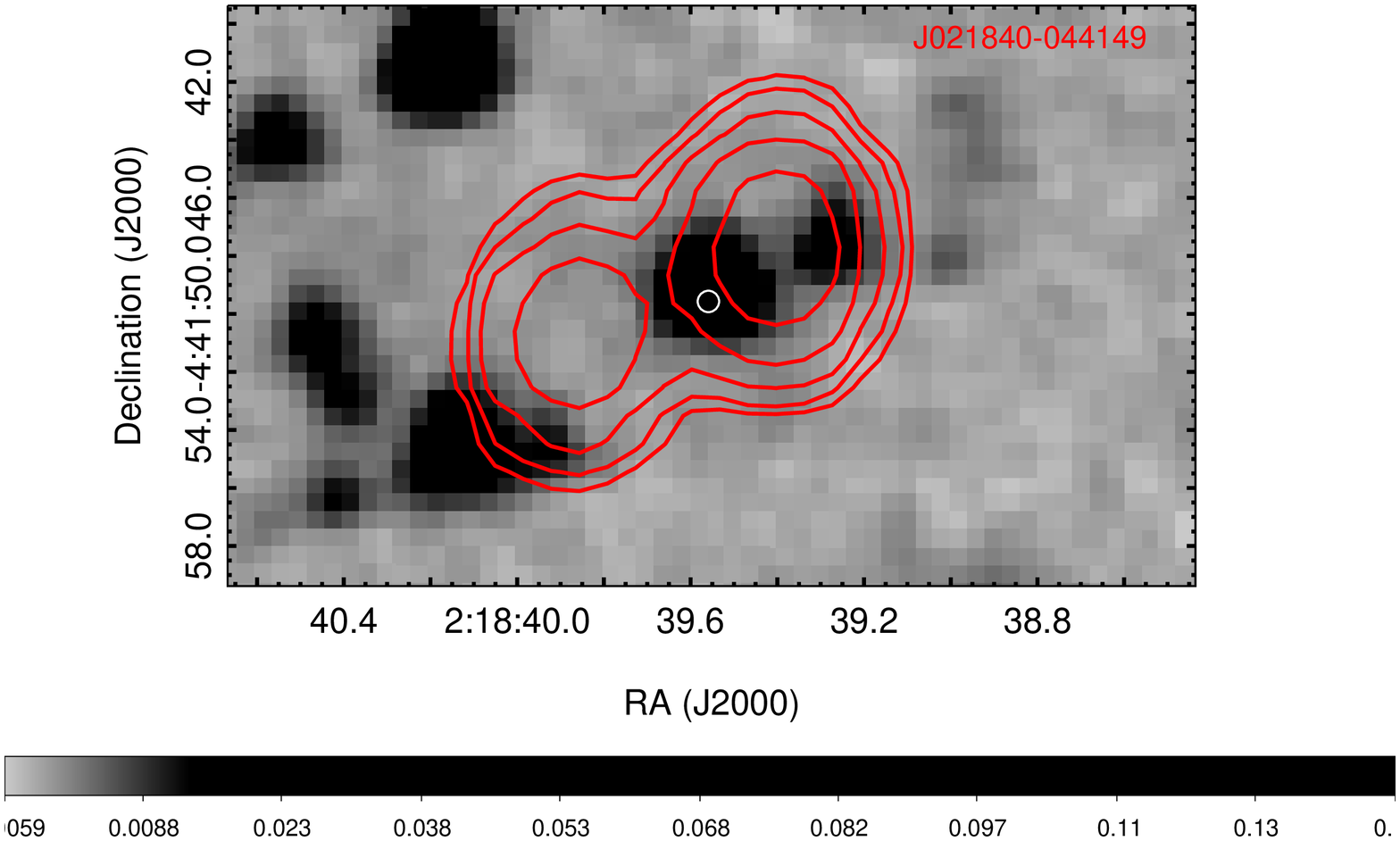}
{\includegraphics[angle=0,width=8.0cm,trim={1.5cm 1.75cm 2.5cm 1.5cm},clip]{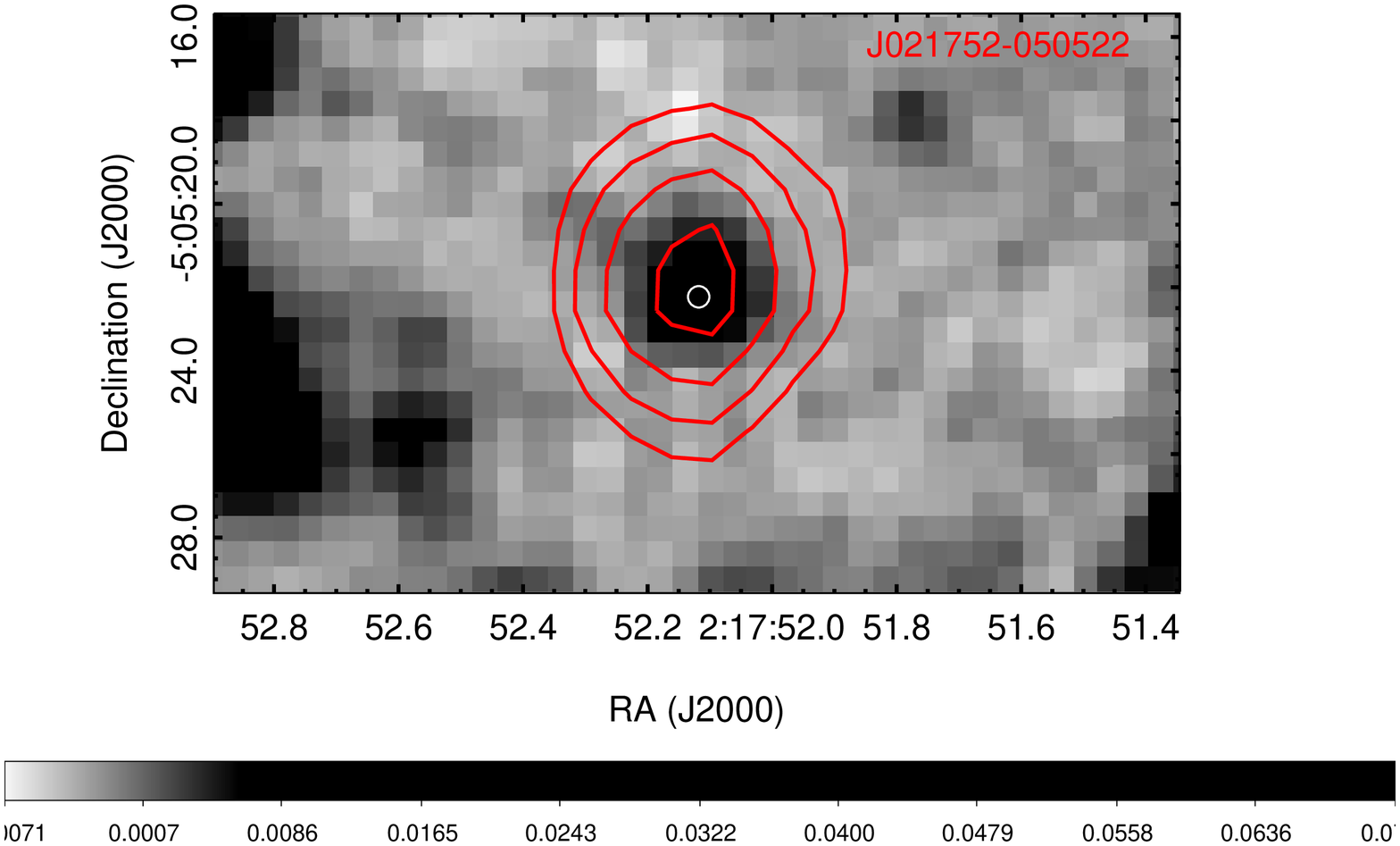}}
\includegraphics[angle=0,width=8.0cm,trim={1.0cm 1.75cm 2.5cm 1.5cm},clip]{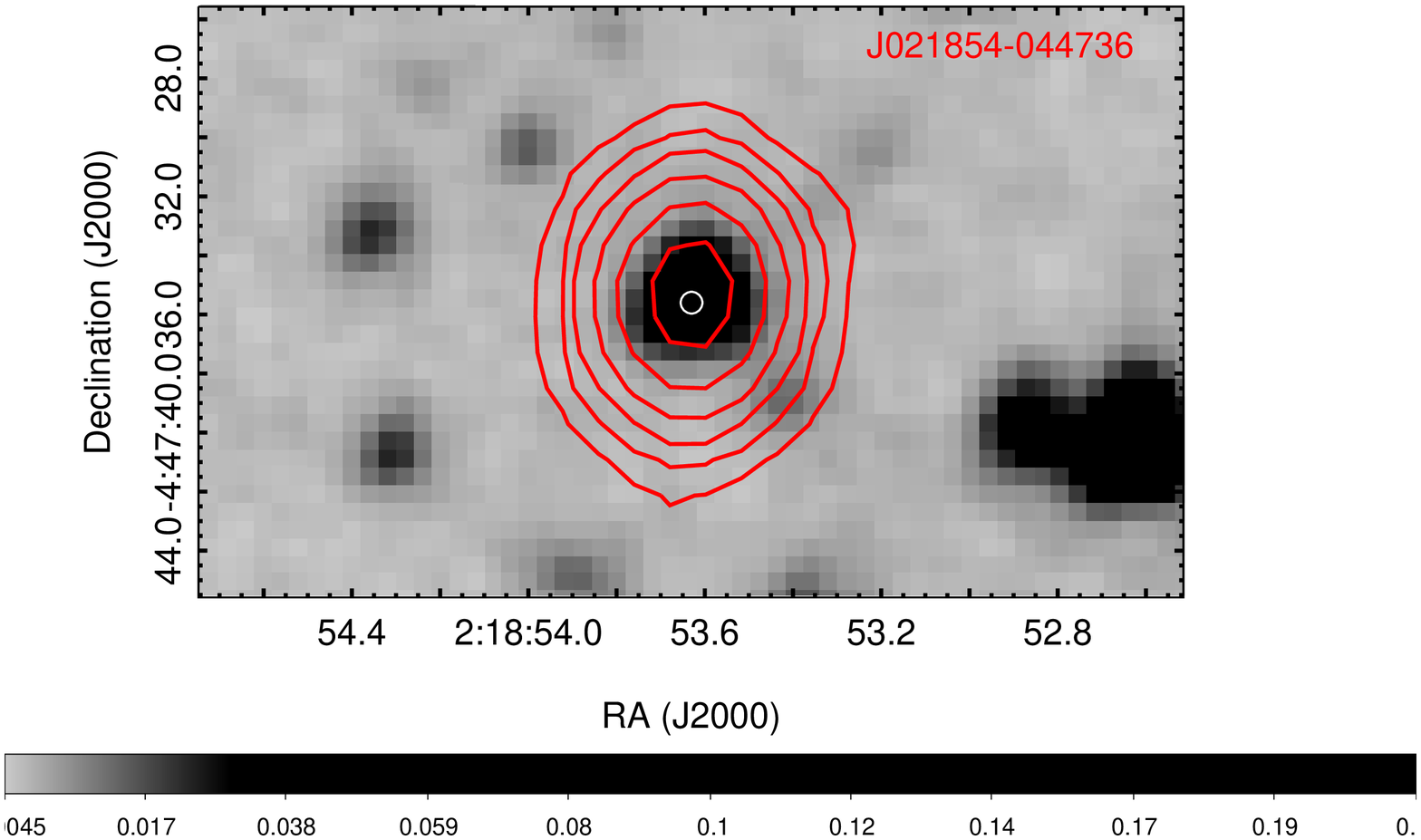}
{\includegraphics[angle=0,width=8.0cm,trim={1.5cm 1.75cm 2.5cm 1.5cm},clip]{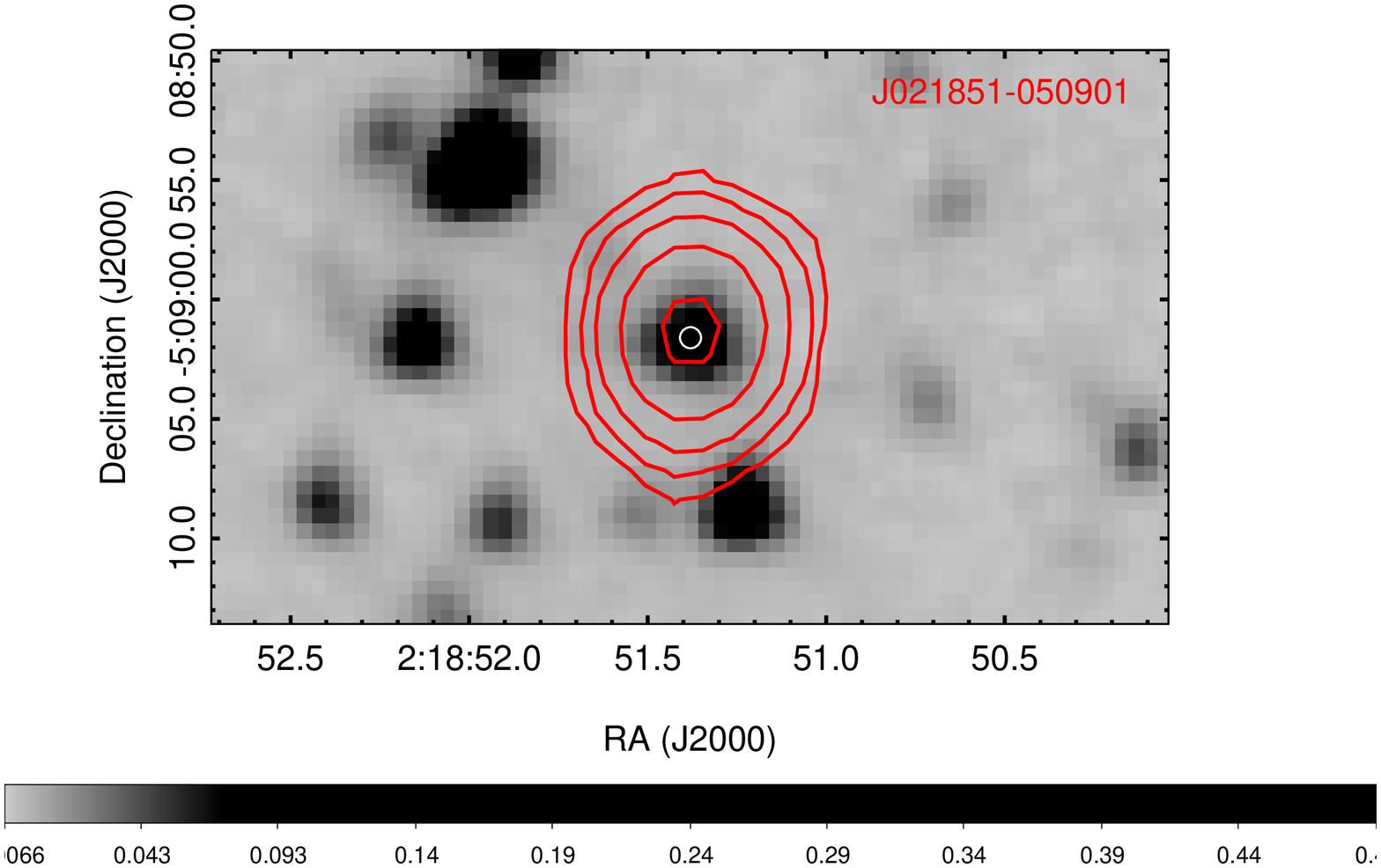}}
\includegraphics[angle=0,width=8.0cm,trim={2.0cm 2.5cm 2.5cm 0.5cm},clip]{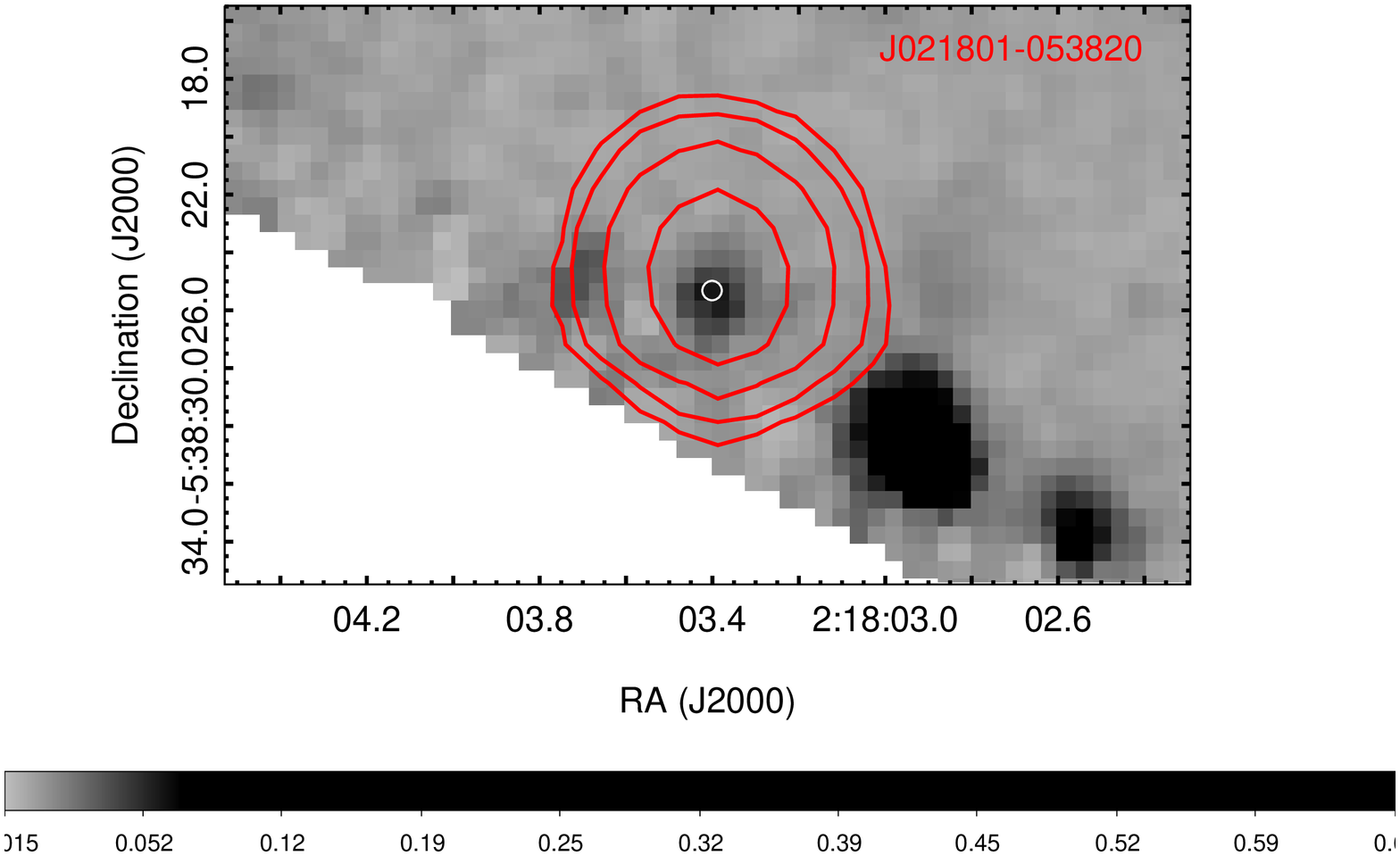}
{\includegraphics[angle=0,width=8.0cm,trim={2.0cm 1.75cm 3.0cm 2.5cm},clip]{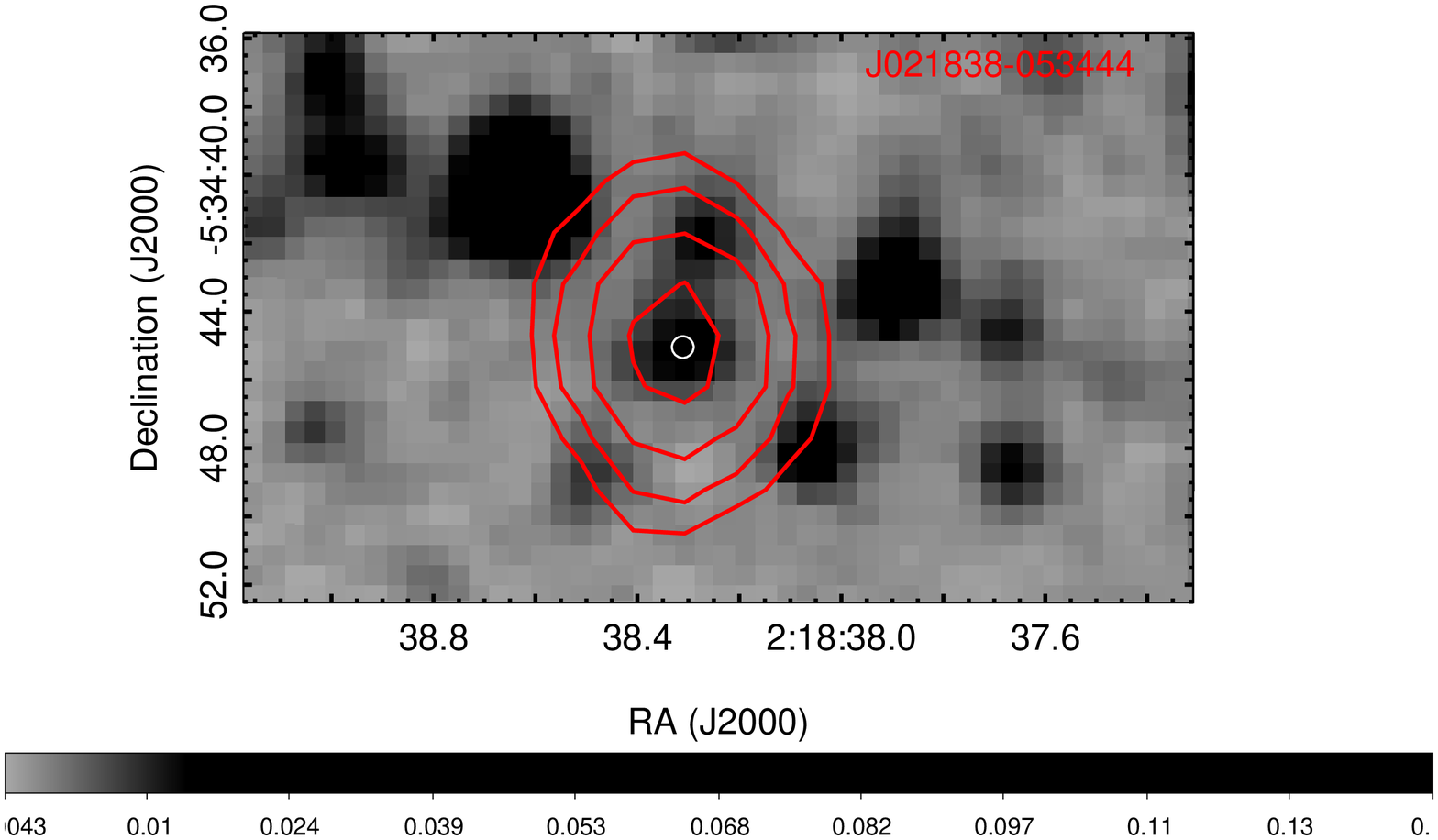}}
\includegraphics[angle=0,width=8.0cm,trim={1.5cm 1.65cm 2.5cm 2.25cm},clip]{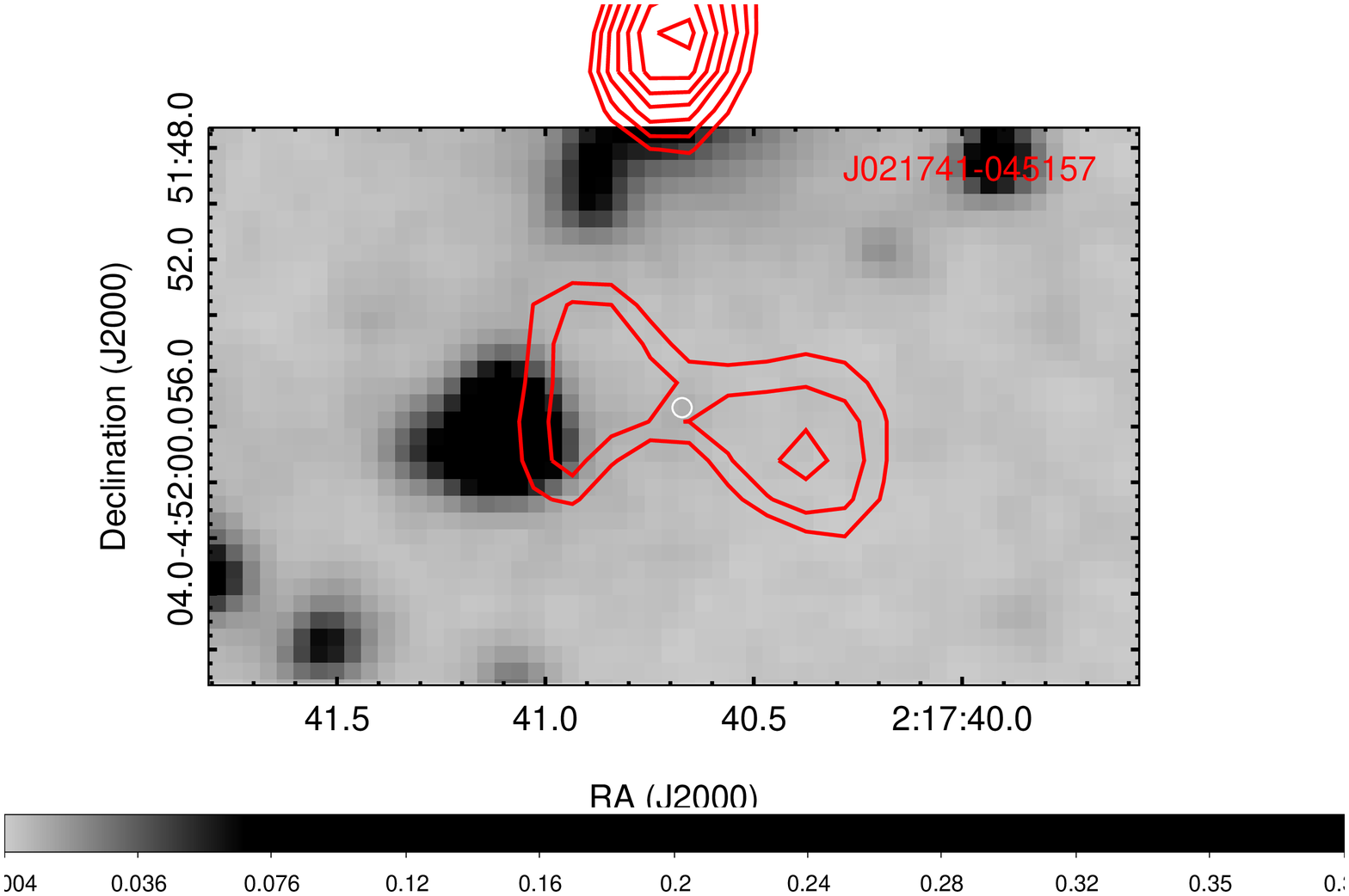}
{\includegraphics[angle=0,width=8.0cm,trim={3.0cm 2.0cm 4.5cm 2.0cm},clip]{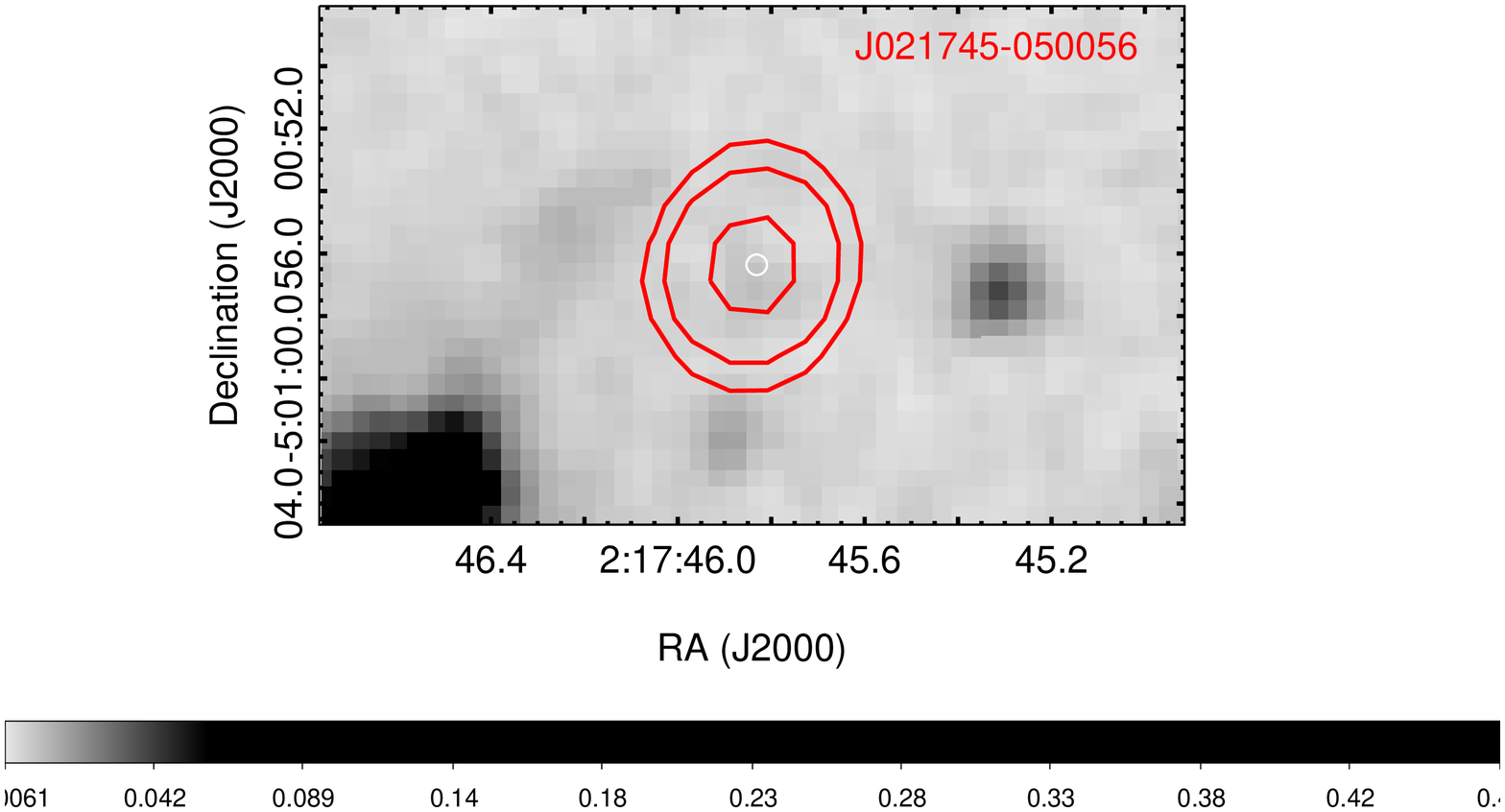}}
\caption{{\it ...continue.}}
\label{fig:IFRSDetected} 
\end{figure*}
\addtocounter{figure}{-1}
\begin{figure*}
\includegraphics[angle=0,width=8.0cm,trim={2.0cm 1.75cm 2.5cm 1.5cm},clip]{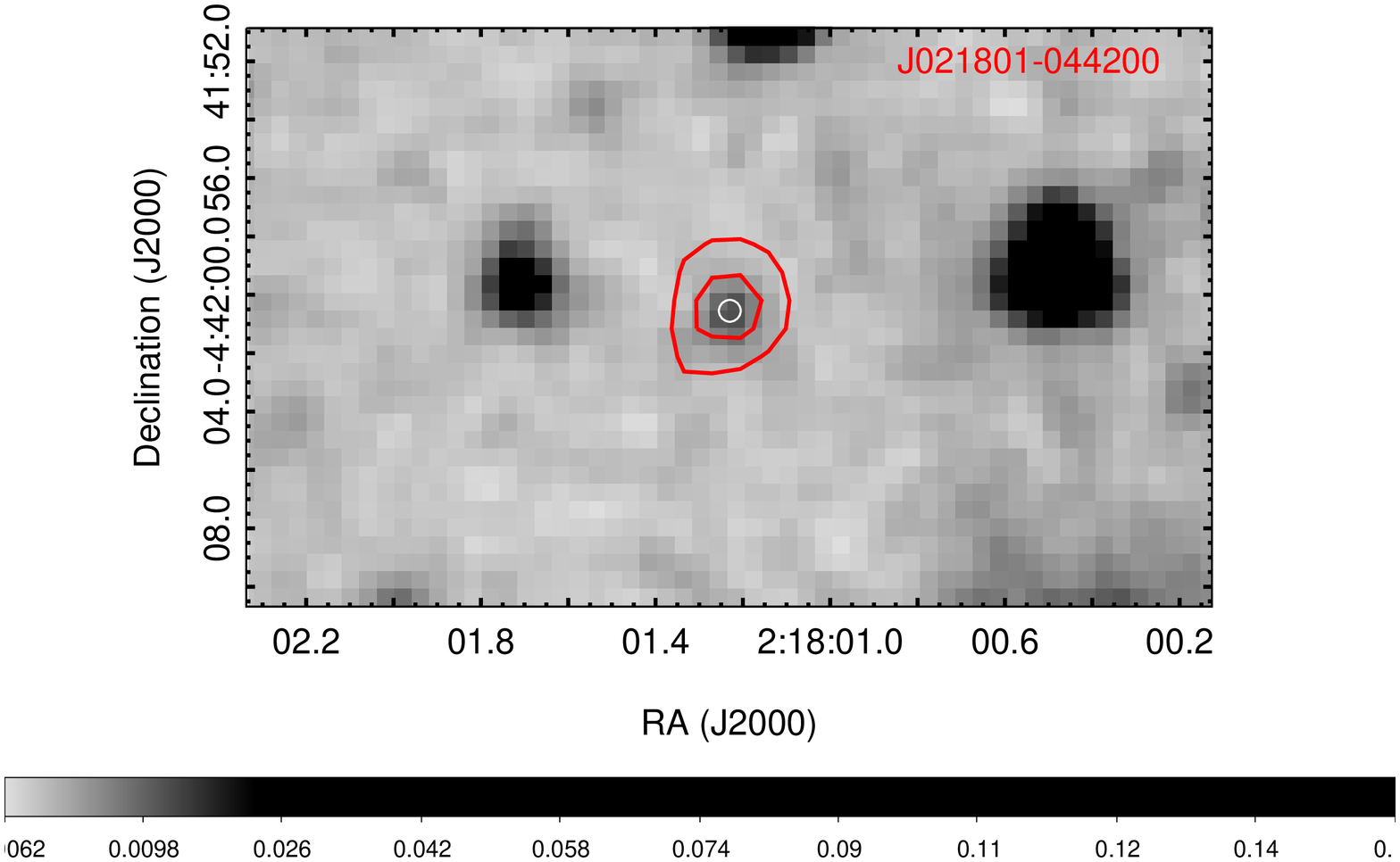}
{\includegraphics[angle=0,width=8.0cm,trim={2.0cm 1.75cm 2.5cm 2.0cm},clip]{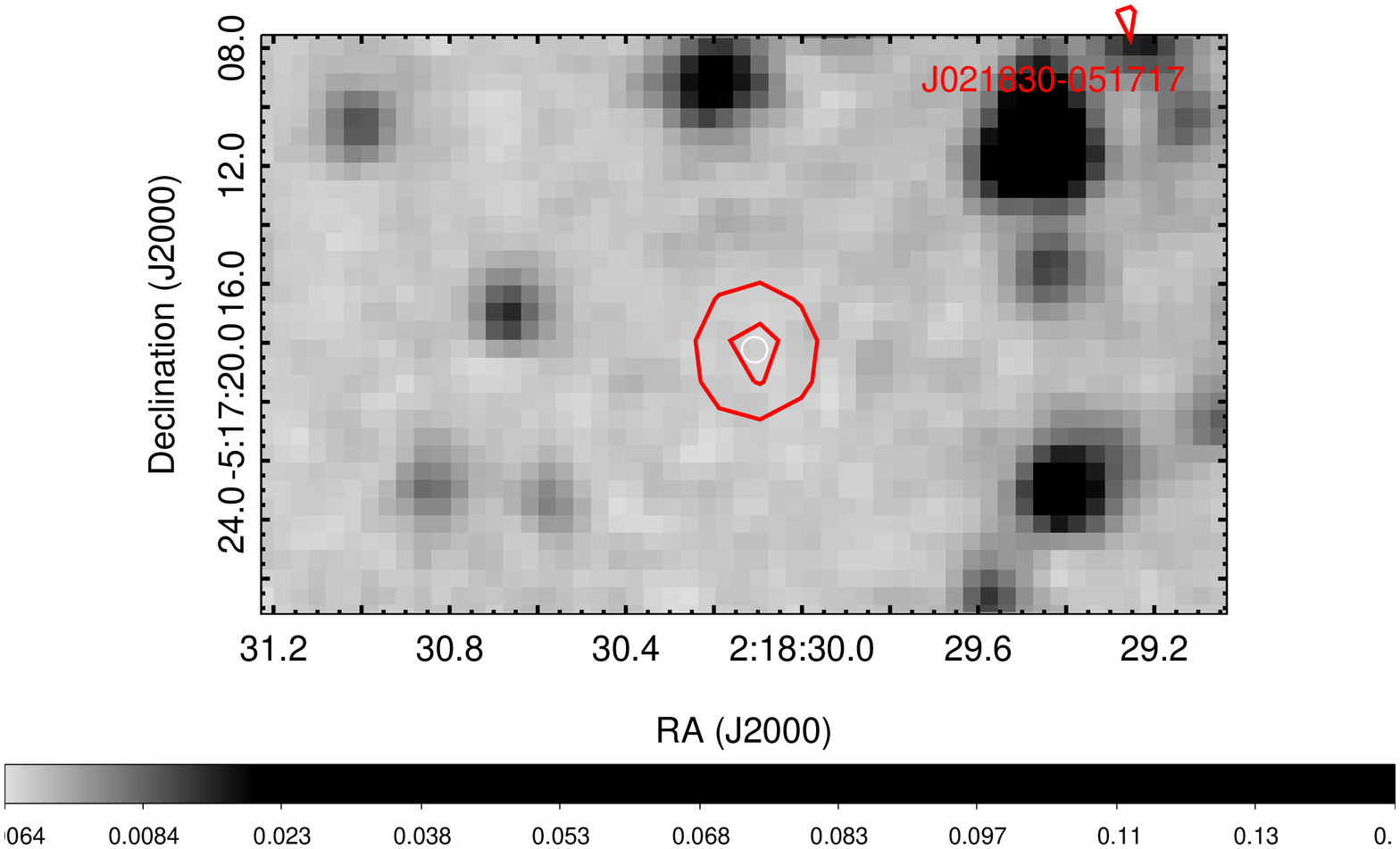}}
\includegraphics[angle=0,width=8.5cm,trim={3.0cm 2.0cm 4.0cm 2.0cm},clip]{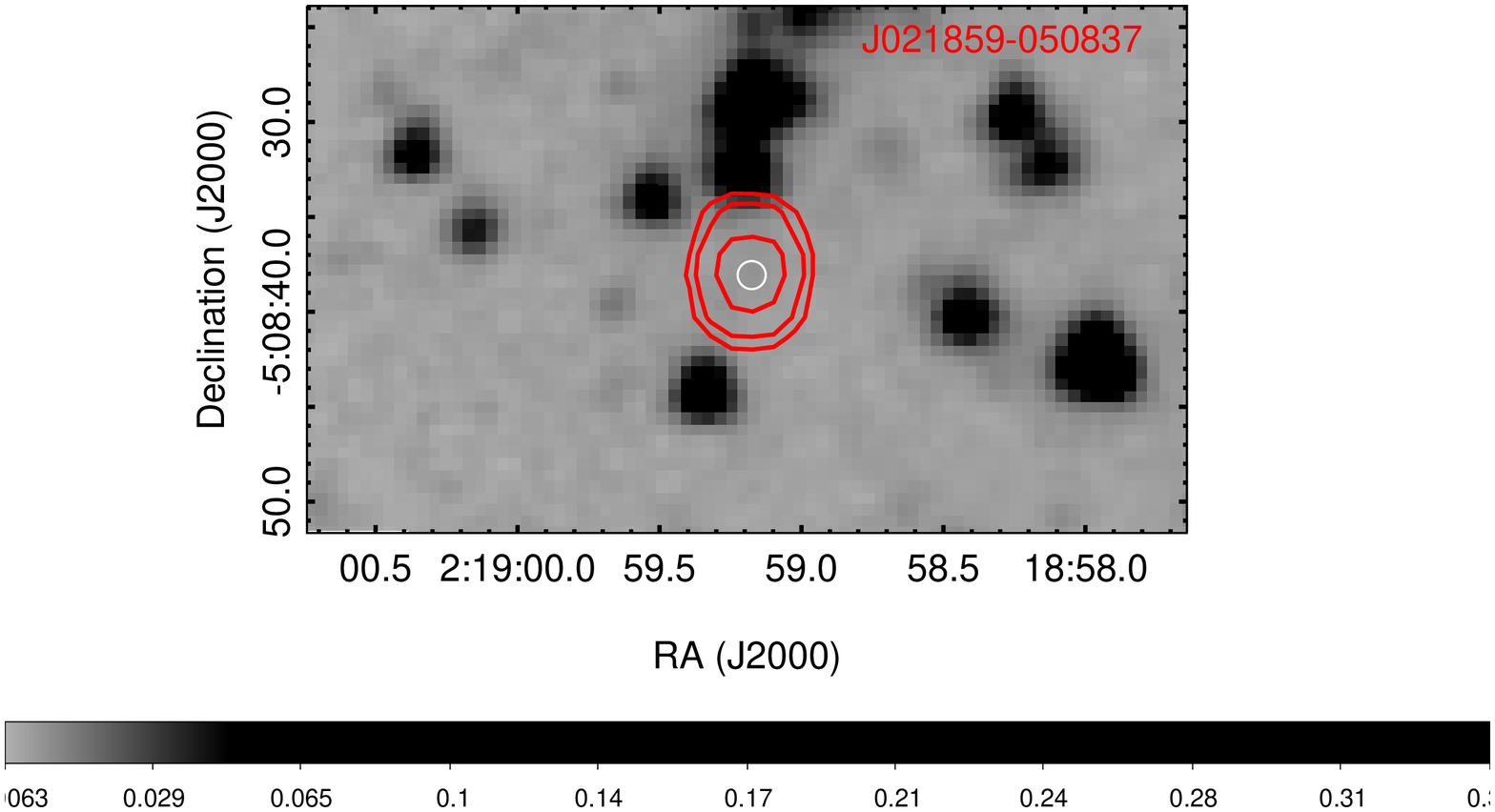}
{\includegraphics[angle=0,width=8.5cm,trim={3.0cm 2.25cm 4.0cm 2.5cm},clip]{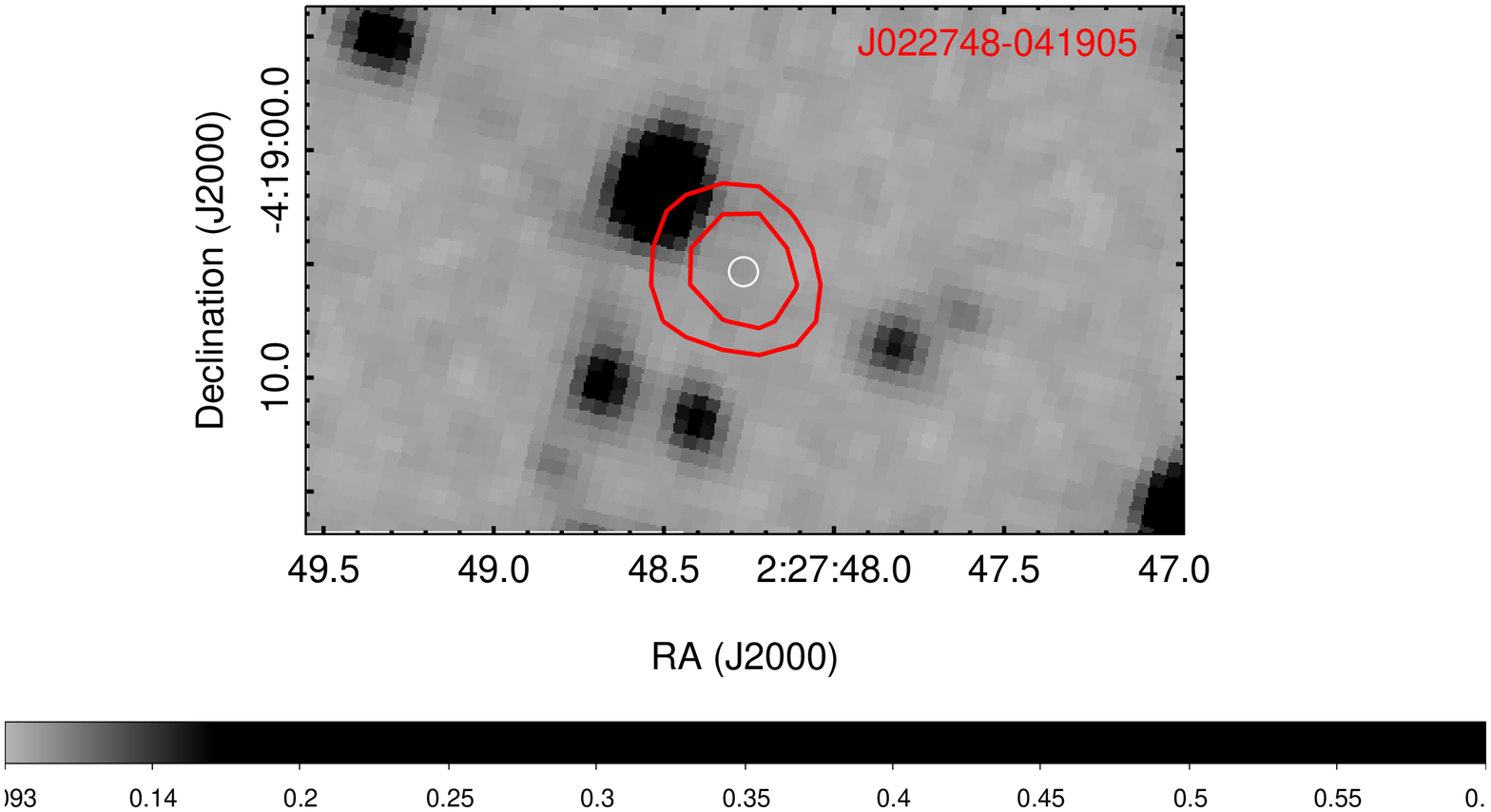}}
\includegraphics[angle=0,width=8.5cm,trim={2.0cm 2.25cm 4.0cm 1.5cm},clip]{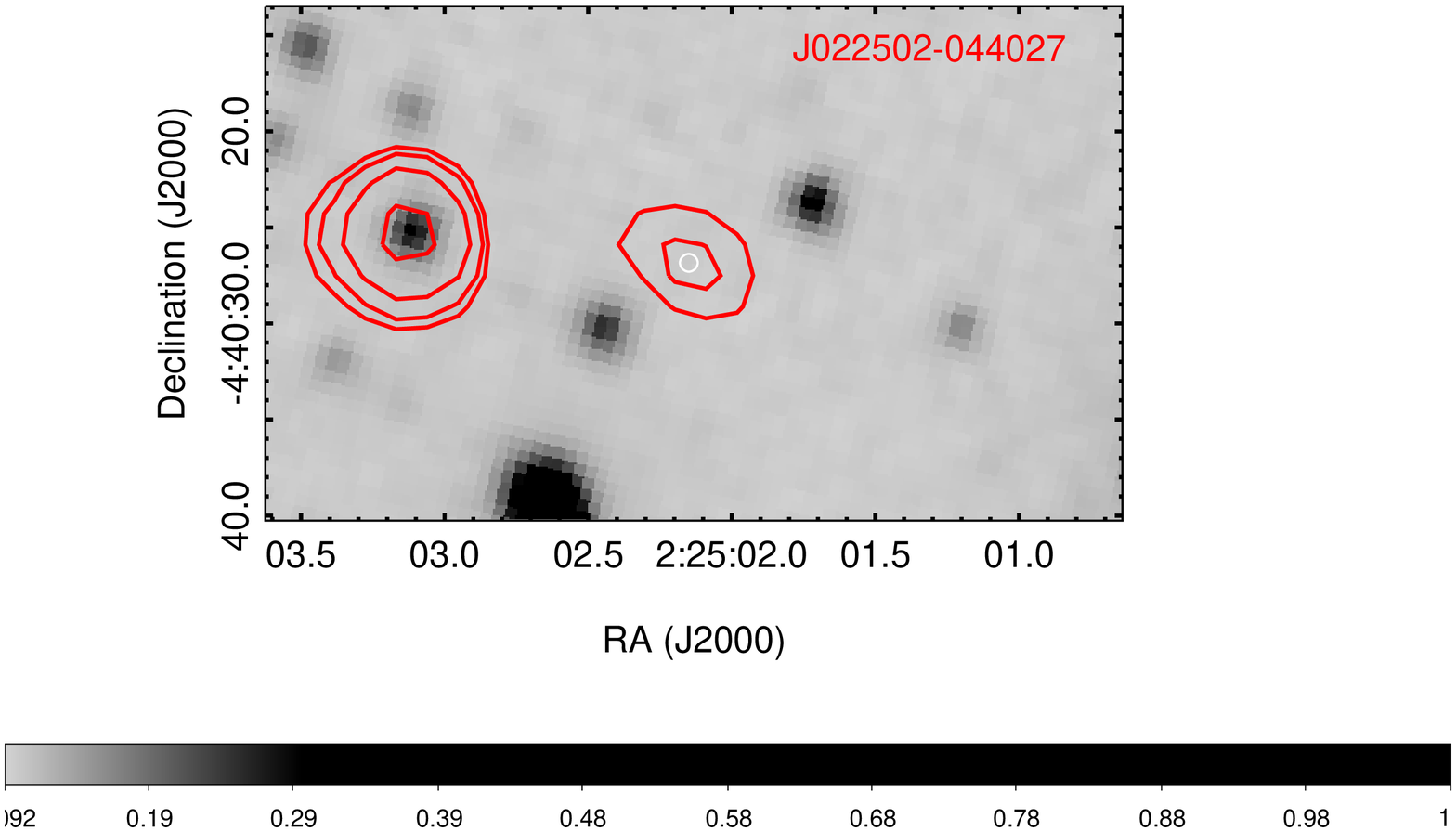}
{\includegraphics[angle=0,width=8.5cm,trim={1.8cm 1.75cm 2.0cm 1.5cm},clip]{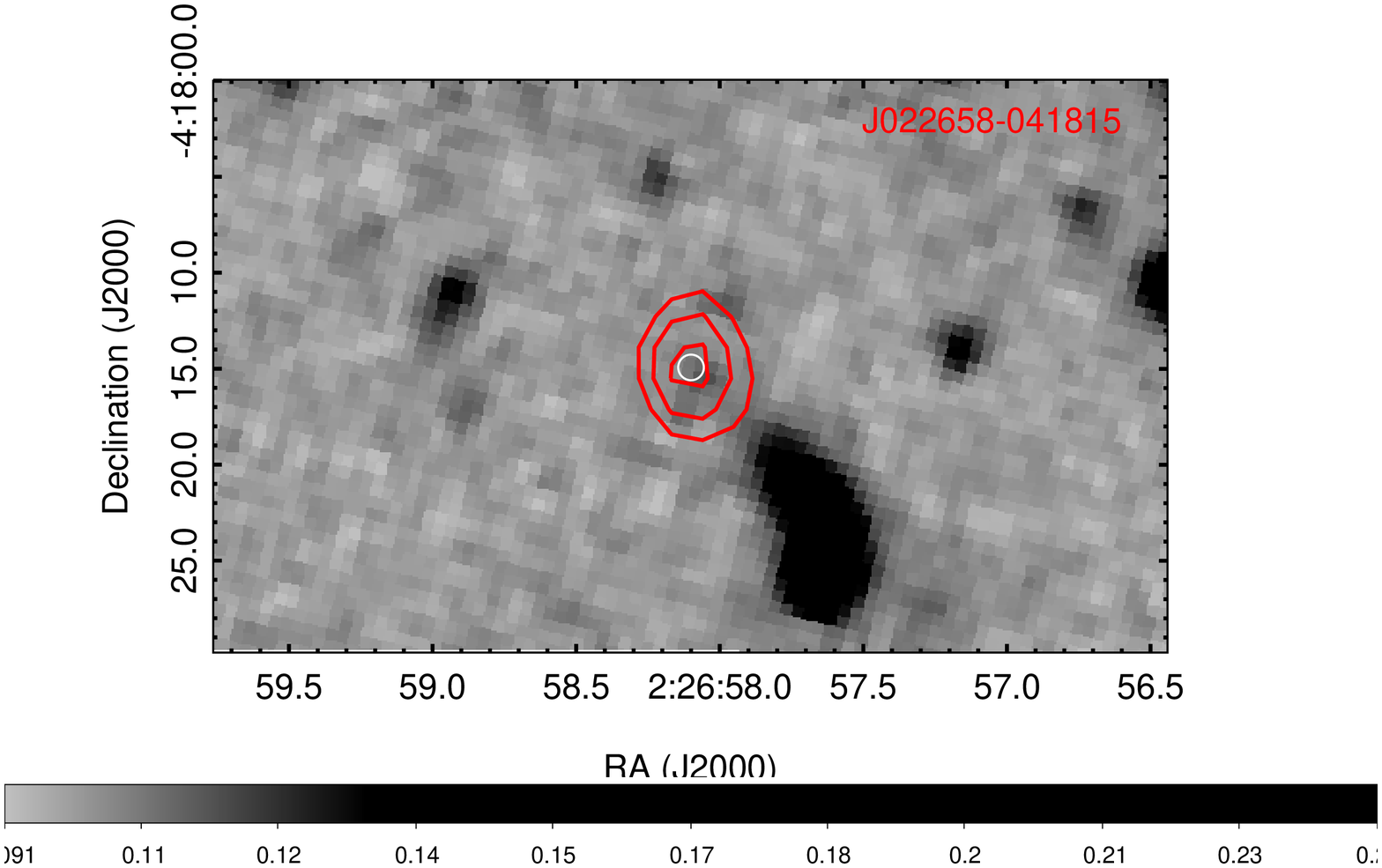}}
\includegraphics[angle=0,width=8.5cm,trim={1.0cm 1.75cm 2.0cm 1.5cm},clip]{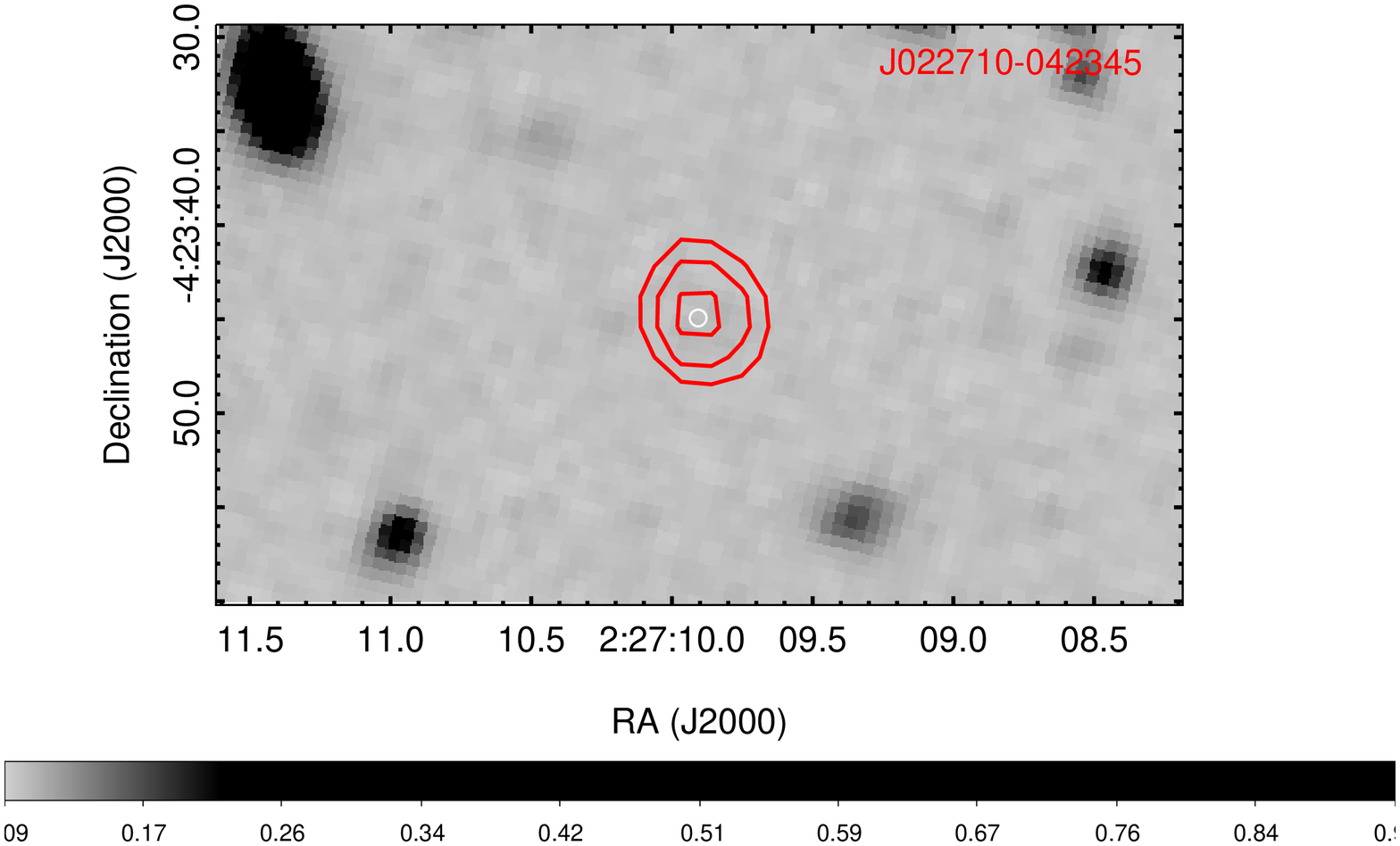}
{\includegraphics[angle=0,width=8.5cm,trim={2.0cm 1.75cm 2.0cm 1.5cm},clip]{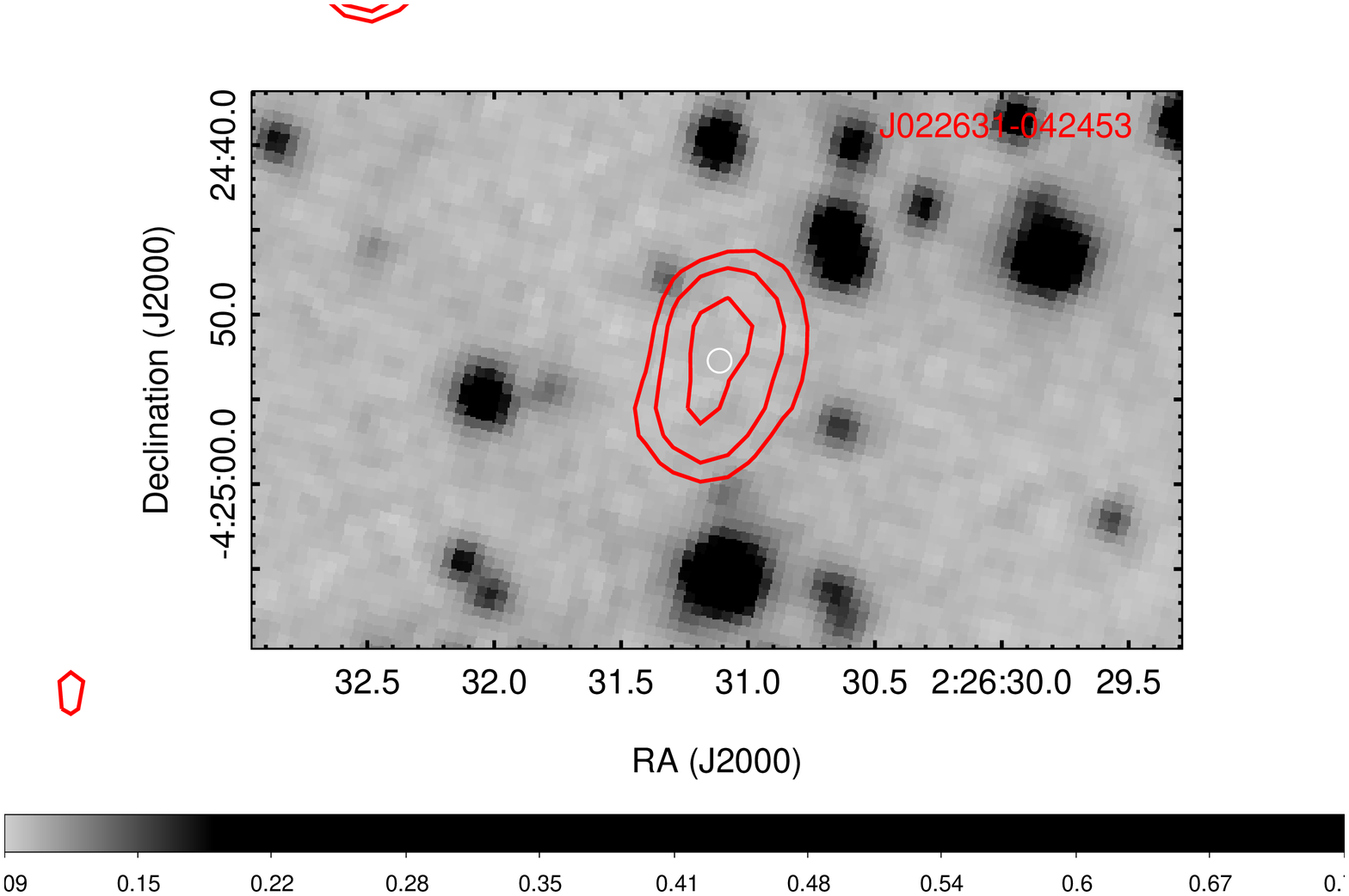}}
\caption{{\it ...continue.}}
\label{fig:IFRSDetected} 
\end{figure*}
\addtocounter{figure}{-1}
\begin{figure*}
\includegraphics[angle=0,width=8.5cm,trim={2.0cm 1.75cm 2.0cm 1.5cm},clip]{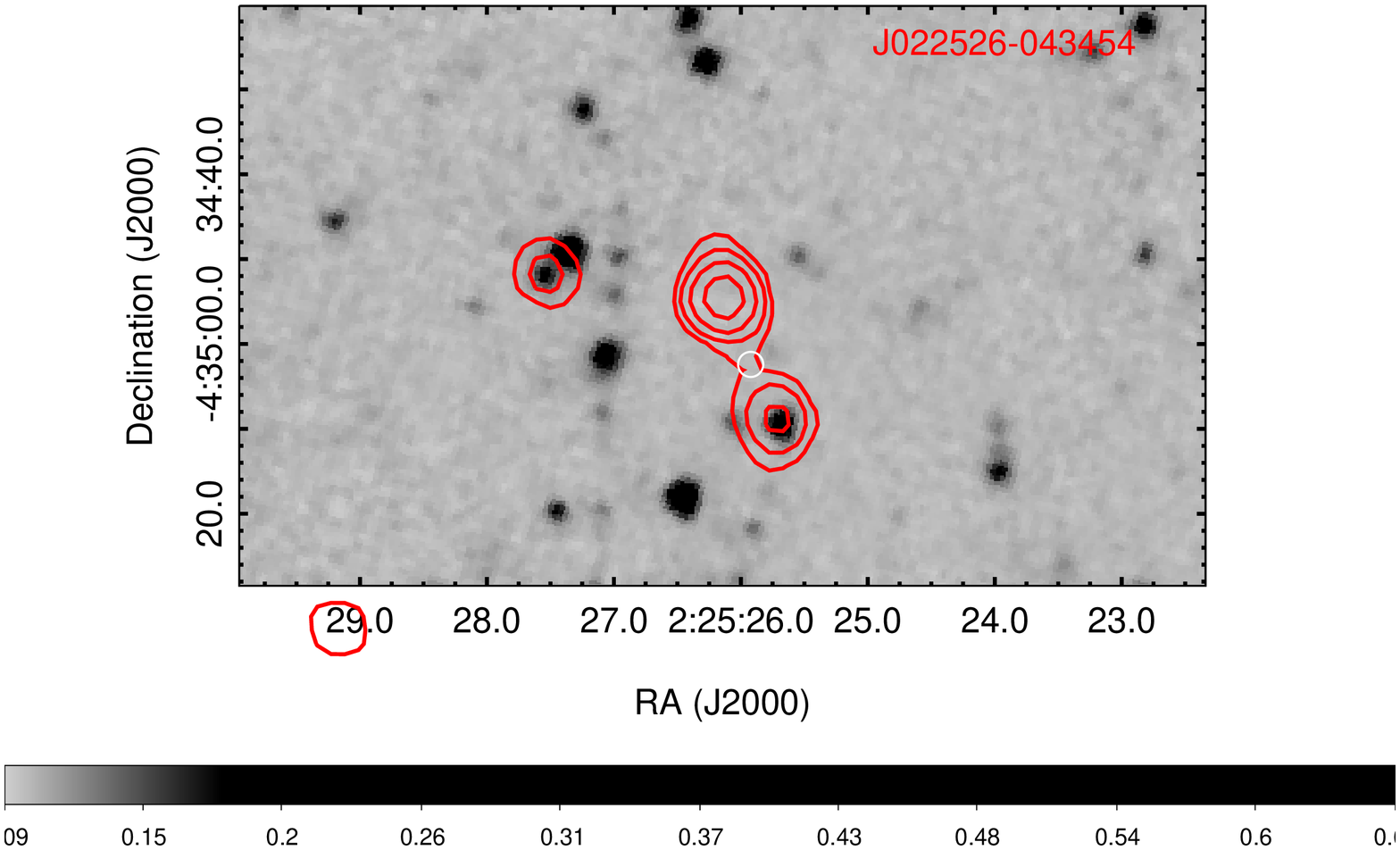}
{\includegraphics[angle=0,width=8.5cm,trim={3.0cm 2.25cm 4.5cm 3.15cm},clip]{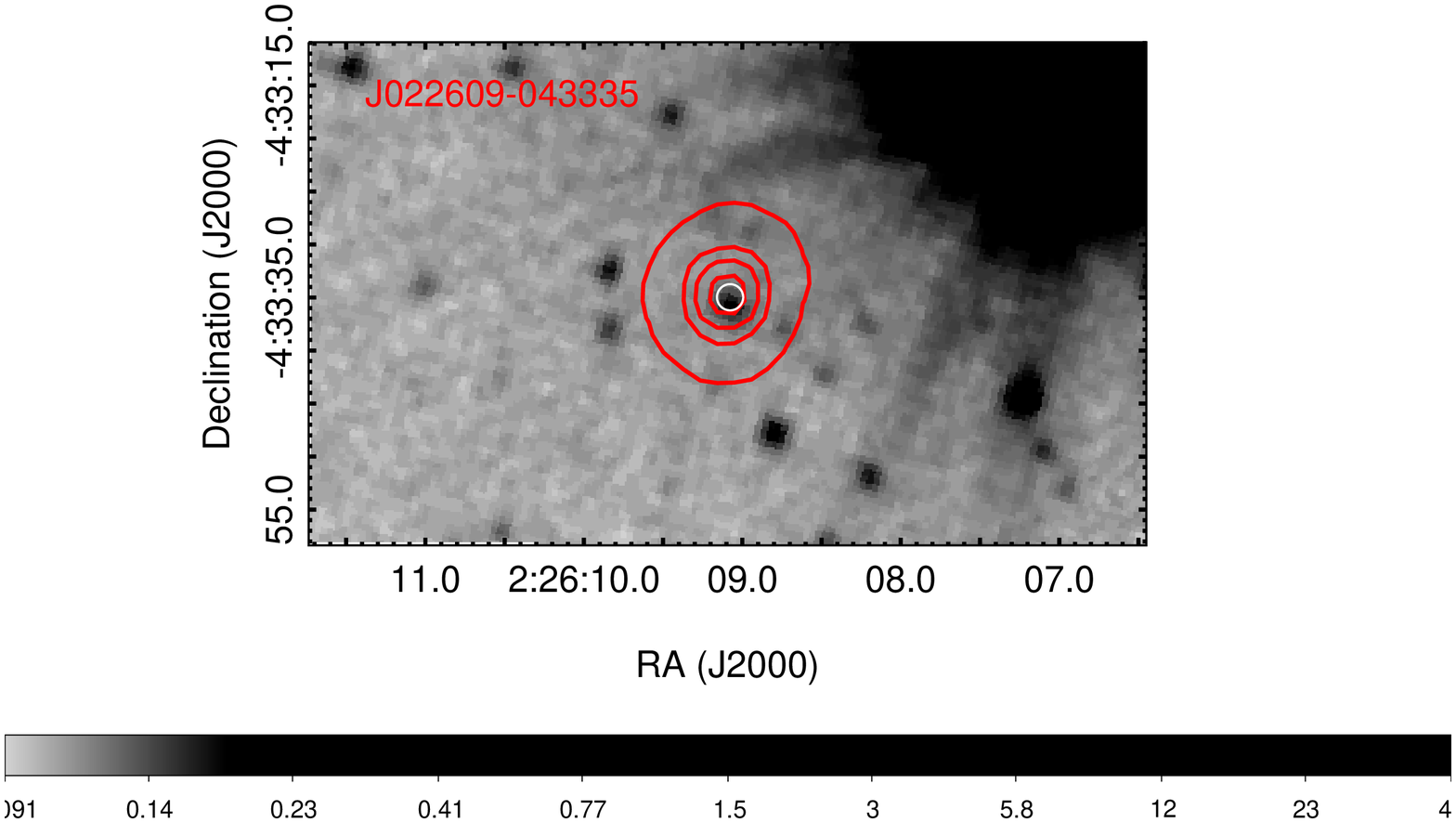}}
\includegraphics[angle=0,width=8.5cm,trim={2.0cm 1.75cm 2.0cm 1.5cm},clip]{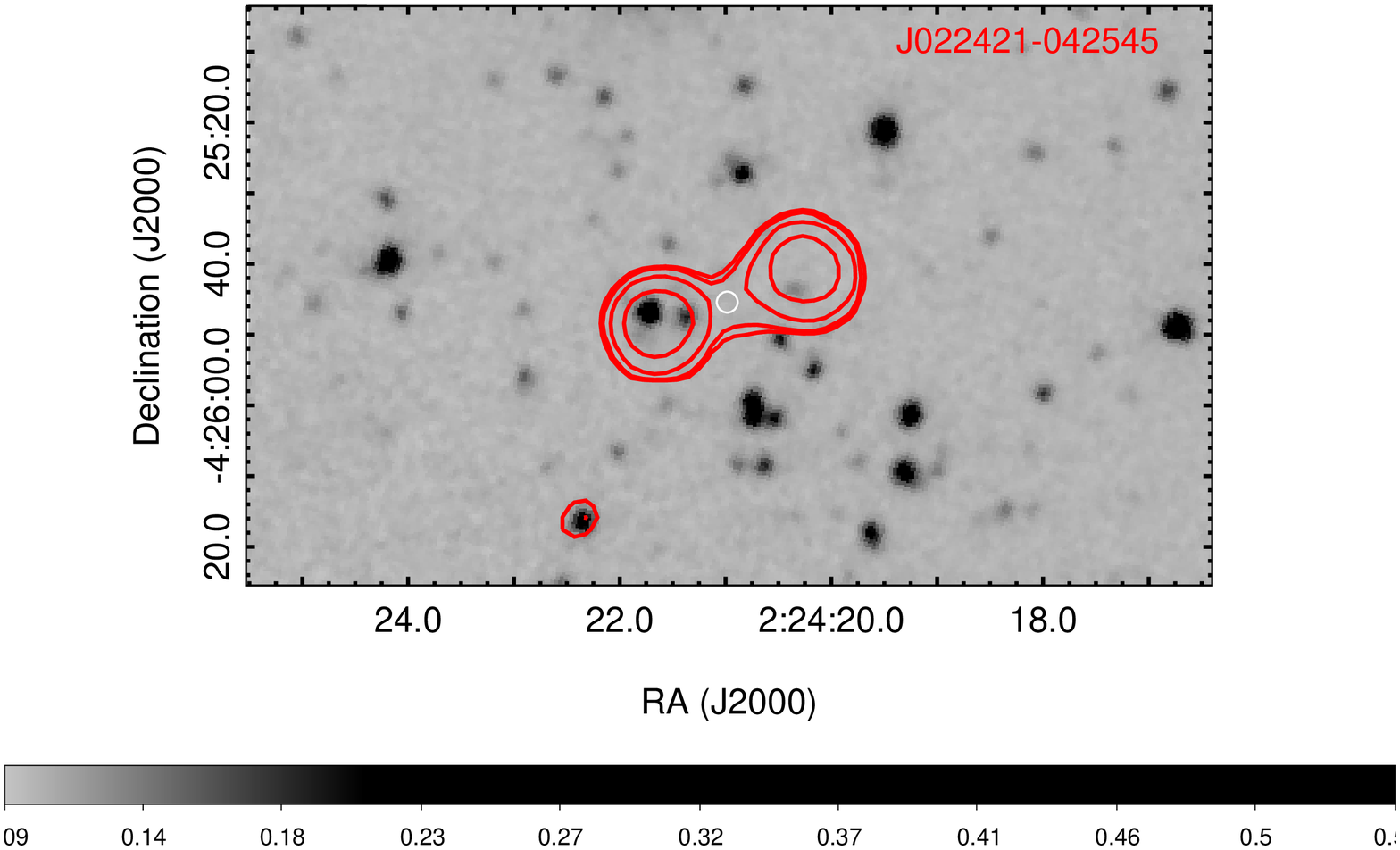}
\caption{1.4 GHz radio contours of our IFRSs over-plotted on to 3.6 $\mu$m grey-scale images. 
The outermost first radio contour begins at 5$\sigma$ with successive contour levels increased by $\sqrt{2}$ factor. 
White circles represent the radio positions of the IFRSs.}
\label{fig:IFRSUnDetected} 
\end{figure*}

\bsp	
\label{lastpage}
\end{document}